\documentclass[aps,prd,10pt,notitlepage,nofootinbib,superscriptaddress,showkeys,showpacs]{revtex4-1}

\usepackage{amsfonts, amsmath, amssymb, amsthm, bm, bbm}
\usepackage{cancel}
\usepackage{hyperref}

\usepackage{color}

\usepackage{graphicx}

\usepackage[caption=false]{subfig}

\usepackage[left = 2.0cm, right = 2.0cm, top = 2.5cm, bottom = 2.5cm]{geometry}

\newcommand{\be}{\begin{equation}}
\newcommand{\ee}{\end{equation}}

\DeclareMathOperator{\tr}{tr}

\newcommand{\cB}{\mathcal{B}}

\newcommand{\cC}{\mathcal{C}}

\newcommand{\cO}{\mathcal{O}}

\newcommand{\C}{\mathbbm{C}}

\newtheorem{remark}{Remark}
\newtheorem{corollary}{Corollary}

\newcommand{\vk}{\boldsymbol{k}}
\newcommand{\vn}{\boldsymbol{n}}
\newcommand{\ve}{\boldsymbol{e}}
\newcommand{\vlam}{\boldsymbol{\lambda}}
\newcommand{\vmu}{\boldsymbol{\mu}}

\begin{document}

\title{\Large \bf Blobbed topological recursion for the quartic melonic tensor model}

\author{{\bf Valentin Bonzom}}\email{bonzom@lipn.univ-paris13.fr}
\affiliation{LIPN, UMR CNRS 7030, Institut Galil\'ee, Universit\'e Paris 13, Sorbonne Paris Cit\'e,
99, avenue Jean-Baptiste Cl\'ement, 93430 Villetaneuse, France, EU}

\author{{\bf St\'ephane Dartois}}\email{stephane.dartois@outlook.com}
\affiliation{LPTM, UCP - CNRS, Universit\'e Cergy-Pontoise,
2 avenue A. Chauvin, 95302 Pontoise, France, EU}

\date{\small\today}

\begin{abstract}
\noindent 
Random tensor models are generalizations of random matrix models which admit $1/N$ expansions. In this article we show that the topological recursion, a modern approach to matrix models which solves the loop equations at all orders, is also satisfied in some tensor models. While it is obvious in some tensor models which are matrix models in disguise, it is far from clear that it can be applied to others. Here we focus on melonic interactions for which the models are best understood, and further restrict to the quartic case. Then Hubbard-Stratonovich transformation maps the tensor model to a multi-matrix model with multi-trace interactions. We study this matrix model and show that after substracting the leading order, it satisfies the blobbed topological recursion. It is a new extension of the topological recursion, recently introduced by Borot and further studied by Borot and Shadrin. Here it applies straightforwardly, yet with a novelty as our model displays a disconnected spectral curve, which is the union of several spectral curves of the Gaussian Unitary Ensemble. Finally, we propose a way to evaluate expectations of tensorial observables using the correlation functions computed from the blobbed topological recursion.
\end{abstract}

\medskip


\maketitle

\section*{Introduction}

Random tensor models are generalizations of random matrix models. They were introduced as a way to generalize the successful relationship between random matrices and two-dimensional quantum gravity (see the classics \cite{MatrixModels} and the more modern \cite{EynardBook}) to higher dimensions. Indeed, random matrix models are integrals over matrices of size $N\times N$ which admit some $1/N$ expansions. At each order $2g$ in the $1/N$ expansion, the matrix integrals give rise to Feynman diagrams which are ribbon graphs, or equivalently combinatorial maps, and represent discrete surfaces of genus $g$. The original motivation behind tensor models for a tensor with $d \geq 2$ indices is that their Feynman diagrams represent gluings of $d$-dimensional building blocks.

The original proposals of tensor models \cite{AmbjornTensors, GrossTensors, SasakuraTensors} were plagued with difficulties. The Feynman diagrams represent gluings of simplices which can be quite singular \cite{Lost}. Moreover, no tensor models were known to admit a $1/N$ expansion (where $N$ is the range of each index) for twenty years. A reason which explains the absence of major development in tensor models for such a long time might have been the impossibility to generalize random matrix techniques. For instance, tensors cannot be diagonalized which implies no reduction to eigenvalues and therefore no orthogonal polynomials. As a consequence, methods to solve matrix models, such as the one considered in the present article, were not applicable as they rely on the existence of a $1/N$ expansion.

Tensor models have come a long way in the past six years. We now have at our disposal a family of building blocks known as \emph{bubbles} \cite{Uncoloring} which generalize the unitary-invariant observables (matrix traces) of matrix models. Matrix models use those invariant observables as potential: $V(M, M^\dagger) = \sum_{n} t_n \tr(MM^\dagger)^n$, in the case of single-trace interactions, where the term of order $n$ represents a $2n$-gon to be used to form surfaces.

Similarly, bubbles are unitary-invariant polynomials in the tensor entries, which also represent simplicial building blocks: instead of $n$-gons which are gluings of $n$ triangles, one can have arbitrary \emph{colored} gluings of $d$-simplices with a boundary. Using those polynomials as potential in the tensor integrals gives Feynman diagrams which are colored triangulations and which are classified according to some $1/N$ expansion. The fact that colored gluings of simplices emerges in modern tensor models is the reason why $1/N$ expansions exist and was the missing ingredient in the original tensor models (see Gurau's original $1/N$ expansion articles for the combinatorics of colored triangulations \cite{Gurau1/N}).

In contrast with matrix models for which the $1/N$ expansion is the genus expansion for any choice of potential, the $1/N$ expansions in tensor models depend on the choice of bubbles \cite{SigmaReview}. Gurau's $1/N$ expansion, with respect to Gurau's degree, exists for all tensor models but leads to trivial large $N$ limits when the bubbles used as potential do not fall in the class of melonic bubbles \cite{Uncoloring}. In the case of melonic bubbles, Gurau's degree is the only way to get a $1/N$ expansion and it leads to very strong result, see Gurau's universality theorem \cite{Universality} and its extension to almost-melonic bubbles \cite{New1/N}.

For tensor models whose potential consist in non-melonic bubbles, it is then possible to recourse to other $1/N$ expansions, which depend on the choice of bubbles \cite{SigmaReview, New1/N}. Some models for instance resemble matrix models and admit a genus-like expansion \cite{MelonoPlanar}. Recent efforts in tensor models have thus been organized in two directions:
\begin{itemize}
\item Finding $1/N$ expansions and large $N$ limits for tensor models with non-melonic potentials,
\item Solving tensor models to all order in their $1/N$ expansions.
\end{itemize}
The first aspect is new compared to matrix models, but the second aspect is a typical problem in matrix models. Therefore, one can try and extend methods developed for matrix integrals to tensor integrals in order to solve tensor models at all orders in their $1/N$ expansions. Doing so however requires a good understanding of the large $N$ limit first, i.e. the first non-trivial order of the $1/N$ expansion.

Progress in identifying and solving large $N$ tensor models with non-melonic potentials is very recent and still limited. Some strategies and applications have been presented in the review \cite{SigmaReview} and in \cite{MelonoPlanar, StuffedWalshMaps, Octahedra}.

In this article, we will be interested in solving some tensor models at all orders in their $1/N$ expansions and we will therefore focus on cases which are better understood like the melonic case. One of the preferred strategies to do so in matrix models is by using Schwinger-Dyson equations, also known as loop equations in that context. Schwinger-Dyson equations are derived from the matrix integrals and form a sequence of equations which relate the expectation values of the unitary-invariant polynomials $\tr (MM^\dagger)^n$ for different values of $n$.

Loop equations can be rearranged as equations on the generating functions of products of traces, $\prod_{k=1}^p \tr (MM^\dagger)^{n_k}$, known as the $p$-point functions. Solving the loop equations at all orders in the $1/N$ expansion is then a difficult problem. A fascinating development known as the topological recursion has been put forward for the past ten years as it provides a recursive solution to the loop equations in a beautiful mathematical way \cite{EynardBook, TROverview, BorotEynardOrantinAbstract}.

The topological recursion was developed in the context of matrix models and enumeration of combinatorial maps by Eynard \cite{TR}. It has since been found to be applicable to a much larger set of situations, leading to fascinating results in combinatorics, enumerative geometry \cite{HurwitzTR, GromovWittenTR}, quantum topology \cite{KnotTR} and theoretical physics \cite{TROverview, BorotEynardOrantinAbstract}. The existence of a solution to a system of equations satisfying the topological recursion furthermore comes with additional, interesting properties, such as symplectic invariance and integrability.

Tensor models are based on tensor integrals which generalize matrix integrals. It is thus not unexpected that tensor models also give rise to Schwinger-Dyson equations which generalize those of matrix models \cite{BubbleAlgebra}. They form a system of equations which relate the expectation values of unitary invariant polynomials labeled by different bubbles. Just like the Schwinger-Dyson equations of matrix models can be recast as a system of differential constraints forming (half) a Virasoro algebra, those of tensor models also form an algebra. A set of generators is formed by bubbles with a marked simplex and the commutator is related to the gluing of bubbles forming larger bubbles.

The Schwinger-Dyson equations of tensor models have been used to solve melonic tensor models in the large $N$ limit \cite{SDLargeN}, i.e. to calculate the large $N$ expectation of polynomials for melonic bubbles in tensor models with melonic potentials. In this case, the Schwinger-Dyson equations simplify drastically and allow to rederive Gurau's universality theorem. Also in the case of melonic potentials, the Schwinger-Dyson equations have been solved to next-to-leading and next-to-next-to-leading orders and in the double-scaling limit \cite{DoubleScaling}. A non-melonic example of Schinwger-Dyson equations, which combine melonic bubbles and other bubbles which are reminiscent of matrix models, has been given in \cite{MelonoPlanar}. In fact, the equations then reduce at large $N$ to those of matrix models with double-trace interactions \cite{DoubleTrace}.

In spite of this progress, solving the Schwinger-Dyson equations for tensor models remains extremely challenging. Even for one of the simplest non-trivial models involving melonic bubbles with four simplices, going beyond a few finite orders in the $1/N$ expansion is still out of reach. However, we know some tensor models can be completely solved at all order in the $1/N$ expansion and in fact satisfy the topological recursion. This is because those tensor models are matrix models in disguise. From the tensor $T_{a_1 a_2 a_3 a_4}$ with four indices, one can built a matrix $M_{A B}$ with multi-indices $A=(a_1 a_2), B=(a_3 a_4)$ so that $M_{AB} = T_{a_1 a_2 a_3 a_4}$ and then write a matrix model for $M, M^{\dagger}$. Such a tensor model obviously satisfy the topological recursion since its Schwinger-Dyson equations are loop equations of a matrix model. Although quite trivial, this statement can be seen as a proof of principle that tensor models can be solved using modern matrix model techniques.

Instead of facing directly the Schwinger-Dyson equations of tensor models, most of the recent progress in tensor models made use of a correspondence between tensor models and some multi-matrix models with multi-trace interactions. The correspondence is described through a bijection between the Feynman graphs of both sides in \cite{StuffedWalshMaps}. In the case of quartic interactions, it is in fact the Hubbard-Stratonovich transformation which transforms the tensor model into a matrix model \cite{QuarticModels}. This transformation was used in \cite{DoubleScalingDartois} to find the double-scaling limit with melonic quartic bubbles.

The correspondence between tensor models and matrix models offers the opportunity to study tensor models indirectly, by using matrix model techniques on the corresponding matrix models. The matrix models arising in the correspondence are however quite complicated as they involve multiple matrices which do not commute. Nevertheless, the matrix model obtained by the Hubbard-Stratonovich transformation applied to the quartic melonic tensor model is simpler and can be used to test the use of matrix model techniques in that context.

This program has been initiated in \cite{IntermediateT4}. It was shown that all eigenvalues fall into the potential well without spreading at large $N$. This is in contrast with ordinary matrix models for which the Vandermonde determinant causes the eigenvalues to repel each other and prevent them from simultaneously falling in the potential well. However, in this matrix model, the Vandermonde determinant is a sub-leading correction at large $N$. This result was in fact expected from the direct analysis of the Schwinger-Dyson equations of the quartic melonic tensor model at large $N$ \cite{SDLargeN}.

It was further shown in \cite{IntermediateT4} that after substracting the leading value of the eigenvalues and appropriately rescaling the corrections, another matrix model is obtained, involving $d$ matrices $M_1, \dotsc, M_d$. The matrix $M_c$ is called the matrix of color $c=1, \dotsc, d$. Although the model couples those matrices all together, it only does so at subleading orders in the $1/N$ expansion. At large $N$, it reduces to a GUE for each color independently, with their expected semi-circle eigenvalue distributions. The matrix model action however contains corrections to the GUE at arbitrarily high order in the $1/N$ expansion. It is a natural question to ask whether the $1/N$ expansion of this matrix model satisfies the topological recursion. If so, it would give another example of a topological recursion for a tensor model, with a different origin than the matrix model with multi-indices mentioned above.

To answer that question, a quick look at the action which will be presented below shows that it is a multi-matrix model with matrices $M_1, \dotsc, M_d$ for $d\geq 3$ and multi-trace interactions of the form $\prod_{c=1}^d \tr M_c^{a_c}$. Interestingly and fortunately, the topological recursion for matrix models with multi-trace interactions was presented by Borot very recently \cite{BlobbedTR}. In that situation, it is necessary to supplement the ordinary topological recursion with an infinite set of ``initial conditions'' (which depend on the model under consideration) called the blobs. They give rise to the notion of blobbed topological recursion whose properties have been analyzed in details in \cite{BlobbedTR2} by Borot and Shadrin. The blobbed topological recursion splits the correlation functions into two parts, a polar part which is singular along a cut and is evaluated using the ordinary topological recursion, and the blobs which bring in a holomorphic part of the correlation functions.

We thus find that the quartic melonic tensor model, after some transformation to a matrix model, satisfies the blobbed topological recursion. It is a fairly straightforward application of \cite{BlobbedTR} and \cite{BlobbedTR2}, with one notable difference: the spectral curve is not connected. This novelty is a direct consequence of the multi-matrix aspect. Indeed, without the multi-trace interactions, the matrices $M_1, \dotsc, M_d$ of different colors would not be coupled. Then, correlation functions for any fixed color would be obtained from the ordinary topological recursion with a one-matrix spectral curve. Moreover, the connected correlators involving different colors would vanish. However, upon turning on the multi-trace interactions which couple the $d$ colors all together, it becomes necessary to package the fixed-color spectral curves into a global spectral curve which is the union of the fixed-color ones.

Once the correct spectral curve is found, the blobbed topological recursion can be applied fairly generically for all $d\geq 2$. However, we find that there are properties which depend on $d$. In particular, the matrix model admits a topological expansion, i.e. an expansion in powers of $1/N^2$ for $d=4\delta +2$ and $\delta$ a non-negative integer, so $d = 2, 6, 10, \dotsc$ For other values of $d$, the $1/N$ expansion is not topological and is actually an expansion in powers of $1/\sqrt{N}$ (some orders of the expansion can vanish: for instance, at $d=4$, it really is an expansion in $1/N$). Therefore, when expanding the correlation functions, we might distinguish the contribution which comes from the usual topological expansion in powers of $1/N^2$ and the others. This is not what we will do. Indeed, the non-topological orders come from the fact that the coupling constants of the model have $1/\sqrt{N}$ expansions. We can thus perform all calculations of the correlation functions as in the topological cases, find them as functions of the coupling constants and eventually expand the latter. This is however very awkward and since it is obvious that it can be done in principle, we will instead focus on the topological case $d=6$.

The organization of the paper is as follows. In Section \ref{sec:Quartic}, we recall the definition of the quartic melonic tensor model and summarize the analysis of \cite{IntermediateT4}. It leads to the matrix model \eqref{MatrixModel} which is the one the topological recursion is then applied to. In Section \ref{sec:LoopEqs}, we define the correlation functions. A key point of our article is that we distinguish the local observables which are labeled by sets of colors from $1$ to $d$ and global observables obtained by packaging the local ones appropriately and which are defined globally on the spectral curve. The loop equations are then derived for the local observables. To find the spectral curve, we solve the loop equations in the planar limit for the 1-point and 2-point functions in Section \ref{sec:LargeN}. In the same section, we further show that non-topological corrections do not have the singularities one expects for topological expansions.

The spectral curve is described explicitly in Section \ref{sec:SpectralCurve}, which then offers the possibility of writing the loop equations on the spectral curve for the global observables. Those loop equations take exactly the same form as in the multi-matrix models of \cite{BlobbedTR} except for the disconnected spectral curve and the fact that the correlation functions do not have a topological expansion for generic $d$.

We then focus on the six-dimensional case in Section \ref{sec:TR}. We derive the linear and quadratic loop equations in Sections \ref{sec:LLE} and \ref{sec:QLE}, following \cite{BlobbedTR, BlobbedTR2} which lead to the blobbed topological recursion in Section \ref{sec:BlobbedTR}. The most important equations are the split of correlation functions into polar and holomorphic parts \eqref{SplitPolarHol}, the recursion for the polar part \eqref{eq:Polarpart} which takes the form of the ordinary topological recursion, and Borot's formula \cite{BlobbedTR} for the holomorphic part \eqref{eq:Holomorphicpart}. We then apply the results of \cite{BlobbedTR2} to obtain a graphical expansion of the correlation functions at fixed number of points and genus and a graphical expansion of the blobs which are the completely holomorphic parts of the correlation functions.

Finally, in Section \ref{sec:TensorObs} we propose a way to use the correlation functions of the matrix model \eqref{MatrixModel} in order to evaluate expectation values of tensorial observables in the quartic melonic tensor model.

\section{From the quartic melonic tensor model to a matrix model} \label{sec:Quartic}

Consider a tensor $T$ with entries $T_{a_1 \dotsb a_d}$, i.e. an $d$--dimensional array of complex numbers where all indices range from $1$ to $N$. We denote its complex conjugate $\overline{T}$ with entries $\overline{T}_{a_1 \dotsb a_d}$. Tensor models are based on tensorial invariants, i.e. polynomials in the tensor entries which are invariant under the natural action of $U(N)^d$ on $T$ and $\overline{T}$. The simplest tensorial invariant is the quadratic contraction of $T$ with $\overline{T}$,
\begin{equation}
T\cdot \overline{T} = \sum_{a_1, \dotsc, a_d = 1}^N T_{a_1 \dotsb a_d}\ \overline{T}_{a_1 \dotsb a_d}.
\end{equation}
The quartic melonic polynomial of color $c$, denoted $B_c(T, \overline{T})$ for $c=1, \dotsc, d$, is defined by
\begin{equation}
B_c(T, \overline{T}) = \sum_{\substack{a_1, \dotsc, a_d\\ b_1, \dotsc, b_d}} T_{a_1 \dotsb a_c \dotsb a_d} \overline{T}_{a_1 \dotsb a_{c-1} b_c a_{c+1} \dotsb a_d} T_{b_1 \dotsb b_c \dotsb b_d} \overline{T}_{b_1 \dotsb b_{c-1} a_c b_{c+1} \dotsb b_d},
\end{equation}
where each sum runs from $1$ to $N$. In other words, one can define the matrix $A_c(T, \overline{T})$ to be the contraction of $T$ and $\overline{T}$ on all their indices except those in position $c$, i.e.
\begin{equation} \label{MatrixA}
\bigl(A_c(T, \overline{T})\bigr)_{a b} = \sum_{a_1, \dotsc, a_{c-1}, a_{c+1}, \dotsc, a_d} T_{a_1 \dotsb a \dotsb a_d} \overline{T}_{a_1 \dotsb a_{c-1} b a_{c+1} \dotsb a_d}
\end{equation}
Then, the melonic polynomial of color $c$ simply is $B_c(T, \overline{T}) = \tr A_c A_c^\dagger$.

The quartic melonic tensor model has the partition function
\begin{equation}\label{eq:partfunction}
Z_{\text{tensor}} (\{ g_c\}, N) = 
\int dT\,d\overline{T}\ \exp -N^{d-1}\Bigl( T\cdot \overline{T} + \frac{1}{2} \sum_{c=1}^d g_c^2 B_c(T, \overline{T}) \Bigr),
\end{equation}
where the $g_c^2$ are the coupling constants of the model. One can perform a Hubbard--Stratonovich transformation on this integral, \cite{IntermediateT4}
\begin{equation}\label{eq:HubbardSplit}
\exp - N^{d-1} B_c(T, \overline{T}) = \int dX_c\ \exp -N^{d-1}\Bigl(\frac12 \tr X_c^2 - i g_c \tr A_c(T, \overline{T}) X_c \Bigr).
\end{equation}
Here $X_c$ is a $N\times N$ Hermitian matrix and the matrix $A_c(T, \overline{T})$ is defined by \eqref{MatrixA}. Introducing the $d$ matrices $X_1, \dotsc, X_d$ that way, the integral over $T$ and $\overline{T}$ becomes Gaussian and can thus be performed explicitly. The result is an integral over the matrices $X_1, \dotsc, X_d$. It is convenient to introduce $Y_c = \mathbbm{1}^{\otimes (c-1)} \otimes X_c\otimes \mathbbm{1}^{\otimes (d-c)}$ and rewrite the integral as
\begin{equation}
Z_{\text{tensor}} (\{g_c\}, N) = \int \prod_{c=1}^d dX_c\ \exp \Bigl[-\frac12 \sum_{c=1}^d \tr Y_c^2 - \tr \ln \Bigl(\mathbbm{1}^{\otimes d} + i  \sum_{c=1}^d g_c Y_c\Bigr) \Bigr]
\end{equation}
The matrices $Y_1, \dotsc, Y_d$ can be diagonalized simultaneously, in terms of the eigenvalues of $X_1, \dotsc, X_d$. In \cite{IntermediateT4}, only the symmetric model was investigated where all coupling constants $g_c^2$ are equal, $g_c^2=\lambda$. The saddle point analysis at large $N$ \cite{IntermediateT4} then reveals that the eigenvalues all fall into a potential well, without the usual spreading of the eigenvalue density. This is due to the fact that the Vandermonde determinant is here negligible at large $N$. The eigenvalue density is thus $\rho(\omega) = \delta(\omega - \alpha)$ where $\alpha$ is the extremum of the potential where all the eigenvalues condensate,
\begin{equation} \label{alpha}
\alpha = \frac{\sqrt{1+4d\lambda}-1}{2id\sqrt{\lambda}}.
\end{equation}

That result was indirectly known from the analysis of the tensor model at large $N$ in \cite{SDLargeN}, where a resolvent was proposed (in spite of the model not having eigenvalues since it is based on tensors which cannot be diagonalized) and found to be $1/(\omega - \alpha)$ (it is the resolvent associated with the density $\rho(\omega) = \delta(\omega - \alpha)$).

In the non-symmetric case the conclusion is quite the same as indeed the Vandermonde is still negligible at large $N$. However the eigenvalue density for the matrix $X_c$ of color $c$ now depends on the color,
\begin{equation}\label{eq:alphac}
\alpha_c = g_c \frac{\sqrt{1+4\sum_q g_q^2}-1}{2i\sum_q g_q^2}.
\end{equation} 
In \cite{IntermediateT4} this result further led the authors to substract to the matrix $X_c$ the amount $\alpha \mathbbm{1}$ in order to study the fluctuations. Their idea can be extended to the non-symmetric case and lead to the same type of conclusion. Making the following change of variable,
\begin{equation}
X_c = \alpha_c\,\mathbbm{1} + \frac{1}{N^{\frac{d-2}{2}}} M_c,
\end{equation}
the partition function rewrites
\begin{equation}
Z_{\text{tensor}} (\{g_c\}, N)=\frac{e^{-\frac{N^d}{2}\sum_{c}\alpha_c^2}}{(1+iG)^{N^d}}Z(\{\alpha_c\},N),
\end{equation}
where $G:=\sum_c g_c\alpha_c$ and the partition function $Z(\{\alpha_c\},N)$ is following matrix model partition function
\begin{equation}\label{MatrixModel}
Z(\{\alpha_c\},N)=\int \prod_{c=1}^d dM_c \exp \biggl(-\frac{N}{2}\sum_c\tr(M_c^2)+\tr\sum_{p\ge 2}\frac{N^{\frac{2-d}{2}p}}{p} \Bigl(\sum_c\alpha_c\mathcal{M}_c\Bigr)^p \biggr),
\end{equation}
with $\mathcal{M}_c=\mathbbm{1}^{\otimes (c-1)}\otimes M_c \otimes \mathbbm{1}^{\otimes (d-c)}$.

As was already the case in the symmetric model, the term of order at least 3 in the matrices $M_c$, i.e. $p\geq 3$ are subleading at large $N$. To see that, we assume that the trace of $M_c$ is of order $N$, we then use a power counting argument. The quadratic term of the action scales like $N^2$, as in regular matrix models.
Looking at the expansion of
\begin{multline}\label{eq:LogExpansion}
\tr\sum_{p\ge 2}\frac{N^{\frac{2-d}{2}p}}{p} \Bigl(\sum_c\alpha_c\mathcal{M}_c \Bigr)^p = \sum_c\frac{N}{2}\alpha_c^2\tr(M_c^2)+\sum_{c< c'}\alpha_c\alpha_{c'}\tr(M_c)\tr(M_{c'})  \\
+\sum_c\frac{\alpha_c^3}{3 N^{\frac{d-4}{2}}} \tr(M_c^3) + \sum_{c\neq c'} \frac{\alpha_c^2\alpha_{c'}}{N^{\frac{d-2}{2}}}\tr(M_c^2)\tr(M_{c'}) + \sum_{c<c'<c''} \frac{2}{N^{\frac{d}{2}}} \alpha_c \alpha_{c'} \alpha_{c''} \tr(M_{c}) \tr(M_{c'}) \tr(M_{c''}) + \dotsb
\end{multline}
we find that a typical term $\prod_{c=1}^d \tr M_c^{q_c}$ of order $p = \sum_{c=1}^d q_c$ in \eqref{eq:LogExpansion} scales like 
\begin{equation}
\frac1{N^{\frac{d-2}{2}p}} \prod_{c=1}^d \tr M_c^{q_c} \sim N^{d-\frac{d-2}{2}p} = N^{2 - \frac{d-2}{2}(p-2)} \leq N^2.
\end{equation}
The Gaussian term thus scales like $N^2$ as expected in matrix models, but all higher order terms for $p\geq 3$ are subleading.

\section{Loop equations for generic $d$--matrix model with multi-trace couplings} \label{sec:LoopEqs}

The matrix model \eqref{MatrixModel} is a multi--matrix model with $d$ matrices, $d$ coupling constants $\alpha_1, \dotsc, \alpha_d$ and some specific scalings with $N$ and multi-trace interactions which couple the matrices. It is not more difficult to write the loop equations for a generic $d$--matrix model with interactions $\prod_{c=1}^d \tr M_c^{q_c}$ than for our model. We therefore derive the loop equations for the generic action
\begin{equation}\label{}
S = N \sum_{c=1}^d \tr V_c(M_c) + N^{2-d} \sum_{a_1, \dotsc, a_d \geq 0} t_{a_1 \dotsb a_d} \prod_{c=1}^d \tr M_c^{a_c},
\end{equation}
whose partition function is $Z = \int \prod_c dM_c\ e^{-S}$ defined as a \emph{formal integral}. It is a formal series in $N$ and the couplings $t_{a_1 \dotsb a_d}$.

\subsection{Observables and notations}\label{sec:observablesnotations}

\subsubsection{Local observables} \label{sec:localobservablesnotations}

If $\mathcal{O}(M_1, \dotsc, M_d)$ is a function of the $d$ matrices, its expectation is defined by
\begin{equation}
\langle \mathcal{O}(M_1, \dotsc, M_d) \rangle = \frac1Z \int \prod_{c=1}^d dM_c\ \mathcal{O}(M_1, \dotsc, M_d)\ e^{-S}.
\end{equation}
The observables of the model are the expectations of products of traces of powers of $M_1, \dotsc, M_d$. Let $(k_1, \dotsc, k_d)\in\mathbbm{N}^d$ and introduce
\begin{equation}
\begin{aligned}
\overline{W}(p_1^{(1)}, \dotsc, p_{k_1}^{(1)}; p_1^{(2)}, \dotsc, p_{k_2}^{(2)}; \dotsc; p_1^{(d)}, \dotsc, p_{k_d}^{(d)}) &= \left\langle \prod_{c=1}^d \prod_{j_c=1}^{k_c} \tr M_c^{p^{(c)}_{j_c}} \right\rangle\\
&= \langle \underbrace{\tr M_1^{p_1^{(1)}} \dotsm \tr M_1^{p^{(1)}_{k_1}}}_{\text{$k_1$ times}} \underbrace{\tr M_2^{p_1^{(2)}} \dotsm \tr M_2^{p^{(2)}_{k_2}}}_{\text{$k_2$ times}} \dotsm \underbrace{\tr M_d^{p_1^{(d)}} \dotsm \tr M_d^{p^{(d)}_{k_d}}}_{\text{$k_d$ times}} \rangle
\end{aligned}
\end{equation}

We will often have lists of $d$ elements, one for each color, like the list $(k_1, \dotsc, k_d)$ above. It is convenient to think of such data as an integer--valued vector
\begin{equation}
\vk = k_1 \ve_1 + \dotsb + k_d \ve_d
\end{equation}
where $\ve_c = (0, \dotsc, 0, 1, 0, \dotsc, 0)\in\mathbbm{N}^d$ with a $1$ in position $c\in\{1, \dotsc, d\}$. This provides a natural addition on lists of that type. We will moreover have lists labeled by such vectors. A typical example is the list of exponents of the matrices $M_1, \dotsc, M_d$ in the above observables. The list is $(p^{(1)}_1, \dotsc, p^{(1)}_{k_1})$ for the color 1, and so on. It has $k_c$ elements of color $c$. We denote such a list
\begin{equation}
p_{\vk} = \bigl( (p^{(1)}_1, \dotsc, p^{(1)}_{k_1}), (p^{(2)}_1, \dotsc, p^{(2)}_{k_2}), \dotsc, (p^{(d)}_1, \dotsc, p^{(d)}_{k_d}) \bigr).
\end{equation}
This way, the observable defined above is simply written
\begin{equation}
\overline{W}(p_{\vk}) \equiv \overline{W}(p_1^{(1)}, \dotsc, p_{k_1}^{(1)}; p_1^{(2)}, \dotsc, p_{k_2}^{(2)}; \dotsc; p_1^{(d)}, \dotsc, p_{k_d}^{(d)}) \qquad \text{with $\vk = \sum_{c=1}^d k_c \ve_c$}.
\end{equation}

The \emph{correlation functions} are the generating functions of the above defined observables. They are therefore labeled by $\vk$. They are obtained by multiplying $\overline{W}(p_{\vk})$ with $(x^{(c)}_{j_c})^{-p^{(c)}_{j_c}-1}$, where $x^{(c)}_{j_c}\in \C, \: \forall j_c$, and then summing over all values of $p^{(c)}_{j_c}$.

We say that $x^{(c)}_{j_c}$ are the variables of color $c$ and there are $k_c$ of them. To emphasize the assignment of the variables to different colors, we introduce $d$ copies of $\C$, respectively denoted for convenience $\C_1,\ldots,\C_d$, so that a variable of color $c$ belongs to $\C_c$. The correlation functions are thus functions of the variables $x^{(1)}_{1}, \dotsc, x^{(1)}_{k_1}, \dotsc,  x_1^{(d)}, \dotsc, x^{(d)}_{k_d}\in \C_1^{k_1}\times \ldots \times \C_d^{k_d}$. These variables are packed into a list denoted $x_{\vk}$. We thus write the correlation functions as
\begin{equation}\label{eq:coloredbarW}
\overline{W}_{\vk}(x_{\vk}) = \sum_{p_{\vk}=(p^{(c)}_{j_c=1, \dotsc, k_c})_{c=1, \dotsc, d}} \frac{\overline{W}(p_{\vk})}{\prod_{c=1}^d \prod_{j_c=1}^{k_c} \Bigl(x^{(c)}_{j_c}\Bigr)^{p^{(c)}_{j_c}+1}} = \left\langle \prod_{c=1}^d \prod_{j_c=1}^{k_c} \tr \frac1{x^{(c)}_{j_c} - M_c} \right\rangle
\end{equation}
Note that it is symmetric under the permutations which preserve the colors of the variables, i.e. under the group $\mathfrak{S}_{k_1} \times \mathfrak{S}_{k_2} \times \dotsb\times \mathfrak{S}_{k_d}$ which acts like $x^{(c)}_{j_c} \mapsto x^{(c)}_{\sigma(j_c)}$ on the variable of color $c$, for $j_c =1, \dotsc, k_c$. As a consequence, it turns out that $\overline{W}_{\vk}$ depends on the variables of color $c$ as a set and not as a list.

\begin{remark}
$\overline{W}(p_{\vk})$ are by definition formal series in $N$ and $N^{-1}$ and the couplings $t_{a_1\ldots a_d}$. The correlations functions are also formal series in $N, N^{-1}$ and $t_{a_1\ldots a_d}$. At each order of the $1/N$ expansion, they are holomorphic functions at $\infty$ in their variables.  
\end{remark}

We use similar notations for the cumulants, or connected correlation functions 
\begin{equation}
W_{\vk}(x_{\vk}) = \left\langle \prod_{c=1}^d \prod_{j_c=1}^{k_c} \tr \frac1{x^{(c)}_{j_c} - M_c} \right\rangle_c
\end{equation}
which satisfy
\begin{equation} \label{Connected}
\overline{W}_{\vk}(x_{\vk}) = \sum_{\Lambda \vdash \vk} \prod_{j=1}^{|\Lambda|} W_{\vlam_j}(x_{\vlam_j}).
\end{equation}
Here $\vlam_1, \dotsc, \vlam_{|\Lambda|}$ are $d$-dimensional vectors with integral coefficients and we say that $\Lambda = \{\vlam_1, \dotsc, \vlam_{|\Lambda|}\}$ is a partition of $\vk$ if $\sum_j \vlam_j = \vk$ and each $\vlam_j\neq 0$ and we write $\Lambda \vdash \vk$.

To recover the observables from their generating functions, one can extract the coefficients in front of $1/\Bigl(x^{(c)}_{j_c}\Bigr)^{p^{(c)}_{j_c}+1}$. We will often have to extract the coefficient of the generating function in front of only a subset of variables. Assume that $\vk = \vk_L + \vk_R$, with $\vk_R = \sum_{c=1}^d k_{Rc} \ve_c$. One can then choose a splitting of the list of variables $x_{\vk}$ into a list $x_{{\vk}_L}$ and a list $x_{\vk_R}$. For instance, the former consists of the variables, say, $x^{(c)}_{j_c}$ for $j_c\in\{1, \dotsc, k_{Lc}\}$ and $c=1, \dotsc, d$, while the latter consists of the variables $\tilde{x}^{(c)}_{j_c} = x^{(c)}_{k_{Lc}+j_c}$ for $j_c\in\{1, \dotsc, k_{Lc}\}$ and $c=1, \dotsc, d$. Notice that there is no ordering between the variables of a given color due to the symmetry under color--preserving permutations. Therefore any splitting can be used for each color as long as it has length $k_{Lc}$ on one side and $k_{Rc}$ on the other side.

Assume that we want to extract the coefficient of $\overline{W}_{\vk}(x_{\vk})$ with respect to $x_{\vk_R}$ at order $(x^{(c)}_{j})^{-a^{(c)}_{j}}$. The list of exponents $a^{(1)}_{1}, \dotsc, a^{(1)}_{k_{R1}}, \dotsc, a^{(d)}_{1}, \dotsc, a^{(d)}_{k_{Rd}}$ can be stored as a list denoted $a_{\vk}$. Then we will use the shorthand notation
\begin{equation} \label{CoefficientExtraction}
\overline{W}_{\vk}( x_{\vk_L}, [x_{\vk_R}^{-a_{\vk_R}}]) = \Bigl[ \prod_{c=1}^d \prod_{j_c=1}^{k_{Rc}} \bigl(\tilde{x}^{(c)}_{j_c}\bigr)^{-a_{j_c}^{(c)}} \Bigr] \overline{W}_{\vk}(x_{\vk_L}, x_{\vk_R}) = \left\langle \prod_{c=1}^d \prod_{l_c=1}^{k_{Rc}} \tr M_c^{a^{(c)}_{l_c}-1} \prod_{j_c=1}^{k_{Lc}} \tr \frac1{x^{(c)}_{j_c} - M_c} \right\rangle
\end{equation}
where $[x^n]F(x) = F_n$ if $F(x) = \sum_n F_n x^n$.

\begin{remark}
In the matrix models literature (see for instance \cite{BlobbedTR}), these coefficients are often expressed as integrals around the cut of the planar resolvent. As we will see later, in our model there are several cuts and we can use the same integral representation provided we integrate over the corresponding cuts. We use both notations depending on the context.
\end{remark}

If the potential $V_c$ is the same for all $d$ matrices, and if $t_{a_1 \dotsb a_d}$ is symmetric, then the correlation functions are invariant under color permutations. This means that if $\sigma \in \mathfrak{S}_d$ acts on $\vk = \sum k_c \ve_c$ by permuting its elements as follows $\sigma \mathbf{k} = \sum_{c=1}^d k_{\sigma(c)} e_c$, then 
\begin{equation}
\overline{W}_{\vk}(x_{\vk})= \overline{W}_{\sigma \vk}(x_{\sigma\vk})\qquad \text{and} \qquad 
W_{\vk}(x_{\vk})=W_{\sigma \vk}(x_{\sigma\vk}).
\end{equation}
Here we abuse the notation a little bit and denote $x_{\sigma \vk}$ the list of variables where the sets $\{x^{(c)}_j\}_{j=1, \dotsc, k_c}$ have been permuted by $\sigma$.

\subsubsection{Global observables} \label{sec:globalobservablesnotations}

The correlation functions can be described in two ways.
\begin{itemize}
\item The $n$-point correlation functions $W_{\vn}(x_{\vn})$ we have just described are functions over $\C_1^{n_1} \times \dotsb \times \C_d^{n_d}$, i.e. one copy of $\C$ for each variable and keeping track of the color assigned to each variable (in fact, as expected, those functions have cuts, so it should read $\C_c$ minus some cut which depends on the color $c$). This will be the preferred choice to derive the loop equations and perform actual calculations. We call them \emph{local} observables, as they require a choice of color assigned to each variable.

Equivalently, one can package them, at fixed $n$, in a tensor $W_{c_1 \dotsc c_k}(x_1, \dotsc, x_n)$ where $c_i\in \{1, \dotsc, d\}$ indicates the color of the $i$-th variable, for $i=1, \dotsc, n$. Either way, the essential point is that each variable is assigned a color.

We will write the loop equations in terms of those correlation functions, in Equations \eqref{ConnectedDiscEq} and \eqref{FullLoopEq}, and use them to solve for the 1-point and 2-point functions in Section \ref{sec:LargeN}.

\item As an alternative, and this will be the better choice to understand the structure of the model, we can package all $n$-point functions $W_{\vn}(x_{\vn})$ into a single $n$-point function $W_n(x_1, \dotsc, x_n)$ by allowing the variables $x_1, \dotsc, x_n$ to live on any possible color. This means that the variable $x_j$ lives on the union of the $d$ copies of $\C$, or in fact $\C_c\setminus \Gamma_c$ where $\Gamma_c$ is the cut on the color $c$, as we will see after solving for the 1-point functions at large $N$, i.e.
\begin{equation}
\forall j=1, \dotsc, n, \quad x_j \in \bigcup_{c=1}^d \C_c\setminus \Gamma_c.
\end{equation}
Then the \emph{global} $n$-point function is defined as
\begin{equation}
W_n(x_1,\ldots,x_n)=\sum_{i_1,\ldots, i_n =1}^d \prod_{j=1}^n\mathbf{1}_{\hat{\C}_{i_j}\setminus \Gamma_{i_j}}(x_j) \ W_{\sum_{k=1}^n\ve_{i_k}}(x_1,\ldots,x_n),
\end{equation}
\end{itemize}

These global observables will be the key objects satisfying the topological recursion, see Section \ref{sec:TR}. Before that, we need to find the planar 1-point and 2-point functions using their local versions, which we do in Section \ref{sec:LargeN}. This allows us to identify the spectral curve and show that it is disconnected in Section \ref{sec:SpectralCurve}, leading to a precise definition of the global correlation functions in Equation \eqref{eq:globalWandw}. Then one can write the loop equations directly in terms of the global correlation functions, Equation \eqref{eq:GlobalLoopEq}

\subsection{Exact resolvent equations}

In regular matrix models, the disc equation is the loop equation obtained for all $n \in \mathbb{N}^*$ from
\begin{equation}
\frac{1}{Z}\int dM\ \sum_{a,b} \frac{\partial}{\partial M_{ ab}} \Bigl( (M)^n_{ab}\,e^{-S} \Bigr) = 0.
\end{equation}
In the model we investigate, these equations are replaced by a family of the form
\begin{equation}
\frac{1}{Z}\int \prod_{c=1}^d dM_c\ \sum_{a,b} \frac{\partial}{\partial (M_{i})_{ab}} \Bigl( (M_i^n)_{ab}\,e^{-S} \Bigr) = 0,\: \forall i \in \{1, \dotsc, d\}.
\end{equation}
Moreover the form of the multi-matrix interaction of the action gives rise to unusual terms in the loop equations. We then provide a detailed derivation of the corresponding set of discs equations. Computing the above derivative explicitly gives 
\begin{equation}
\sum_{k=0}^{n-1} \langle \tr M_i^k \tr M_i^{n-1-k} \rangle - N \langle \tr M_i^n\, V_i'(M_i) \rangle - N^{2-d} \sum_{\substack{a_i \geq 1\\ a_{j} \geq0, \,\forall j\neq i}} a_i t_{a_1 \dotsb a_d} \langle \tr M_i^{n+a_i-1} \prod_{c\neq i} \tr M_c^{a_c} \rangle = 0,
\end{equation} 
for all $i\in \{1, \cdots, d\}$, and $n\in \mathbb{N}^*$.

We turn those equations into an equation on generating functions by multiplying the equation for $n$ on matrix $M_i$ with $x^{-n-1}$ -- to emphasize that $x$ is the variable for the matrix $M_i$, we further write $x\in \C_i$. We then sum them over $n\geq0$. There are three types of contributions. The first term gives $\overline{W}_{2\ve_i}(x,x)$. To rewrite the others, we use the standard trick
\begin{equation}
\sum_{n\geq0} \tr M^{n+a} x^{-n-1} = x^a \tr \frac1{x-M} - \tr \frac{x^a - M^a}{x-M},
\end{equation}
where importantly the last term is a polynomial. This way, one finds for the second contribution
\begin{equation}
\sum_{n\geq 0} \langle \tr M_i^n\, V_i'(M_i) \rangle x^{-n-1} = V_i'(x) \overline{W}_{\ve_i}(x) - P_{\ve_i}(x),
\end{equation}
with $P_{\ve_i}(x) = \langle \tr \frac{V_i'(x) - V_i'(M_i)}{x - M_i} \rangle$. The last contribution is similar, with the insertion of $\prod_{c\neq 1} \tr M_c^{a_c}$ into the expectations,
\begin{equation}
\sum_{n\geq 0} \langle \tr M_i^{n+a_i-1} \prod_{c\neq i} \tr M_c^{a_c} \rangle x^{-n-1} = x^{a_i -1} \langle \tr \frac1{x-M_i}\prod_{c\neq i} \tr M_c^{a_c} \rangle - \langle \tr \frac{x^{a_i-1} - M_i^{a_i-1}}{x - M_i} \prod_{c\neq i} \tr M_c^{a_c} \rangle
\end{equation}
The quantity $\sum_{n\geq 0} \langle \tr M_i^{n+a_i-1} \prod_{c\neq i} \tr M_c^{a_c} \rangle x^{-n-1} = x^{a_i -1} \langle \tr \frac1{x-M_i}\prod_{c\neq i} \tr M_c^{a_c} \rangle$ can be rewritten as an extraction of coefficients in the generating function $\overline{W}_{\sum_{c=1}^d \ve_c}(x_1, x_2, \dotsc, x_d)$,
\begin{equation}
\langle \tr \frac1{x-M_i}\prod_{c\neq i} \tr M_c^{a_c} \rangle =\bigl[\prod_{c\neq i}^d \tilde{x}_c^{-a_c-1}\bigr] \overline{W}_{\sum_{c=1}^d \ve_c}(\tilde{x}_1, \tilde{x}_2, \dotsc, \tilde{x}_i=x,  \dotsc, \tilde{x}_d).
\end{equation}
We also introduce 
\begin{equation}
\overline{U}^{(a_i)}_{\ve_i;\sum_{c\neq i} \ve_c}(x,; \{x_c\}_{c\neq i}) = \langle \tr \frac{x^{a_i-1} - M_i^{a_i-1}}{x - M_i} \prod_{c\neq i}^d \tr \frac1{x_c - M_c} \rangle
\end{equation}
so that
\begin{equation}
\langle \tr \frac{x^{a_i-1} - M_i^{a_i-1}}{x - M_i} \prod_{c\neq i} \tr M_c^{a_c} \rangle = \bigl[\prod_{c\neq i} \tilde{x}_c^{-a_c-1}\bigr] \overline{U}^{(a_i)}_{\ve_i;\sum_{c\neq i} \ve_c}(x; \{\tilde{x}_c\}_{c\neq i}).
\end{equation}
The disc equations thus read, for all $i\in \{1, \dotsc, d\}$ and $x\in \C_i$, 
\begin{multline} \label{DiscEq}
0 = \overline{W}_{2\ve_i}(x,x) - N V_i'(x) \overline{W}_{\ve_i}(x) + N P_{\ve_i}(x)\\
 - N^{2-d} \sum_{\substack{a_i \geq 1\\ a_{j\neq i} \geq0}} a_i t_{a_1 \dotsb a_d} \bigl[\prod_{c\neq i} \tilde{x}_c^{-a_c-1}\bigr] \Bigl( x^{a_i-1} \overline{W}_{\sum_{c=1}^d \ve_c}(\tilde{x}_1, \tilde{x}_2,\dotsc, \tilde{x}_i=x, \dotsc, \tilde{x}_d) - \overline{U}^{(a_i)}_{\ve_i;\sum_{c \neq i} \ve_c}(x,\{\tilde{x}_c\}_{c\neq i}) \Bigr)
\end{multline}

We re--express the generating functions in terms of connected ones, using \eqref{Connected}. The first term is $\overline{W}_{2\ve_i}(x,x) = W_{\ve_i}(x)^2 + W_{2\ve_i}(x,x)$. For terms with coupling $t_{a_1\dotsb a_d}$ in \eqref{DiscEq}, applying \eqref{Connected} gives for instance
\begin{equation}
\overline{W}_{\sum_{c=1}^d \ve_c}(\tilde{x}_1, \tilde{x}_2,\dotsc, \tilde{x}_i=x, \dotsc, \tilde{x}_d) = \sum_{\Lambda \vdash \sum_{c=1}^d \ve_c} \prod_{j=1}^{|\Lambda|} W_{\vlam_j}(\tilde{x}_{\vlam_j}),
\end{equation}
which is not convenient since it loses track of the variable $x\in\C_i$. Therefore, we need to keep track of the part $\vlam_j$ which contains the vector $\ve_i$ when considering decompositions $\sum_{c=1}^d \ve_c = \sum_{j=1}^{|\Lambda|} \vlam_j$. Since there is no ordering, we can also assume that $\ve_i$ is in $\vlam_1$ and redefine the latter to be $\vlam_1 - \ve_i$. In contrast with other $\vlam_j$ for $j\neq 1$ which cannot vanish, the new $\vlam_1$ can be zero. This way, we introduce a family of ``almost--partitions'', akin to a partition where a specific, labeled part can be empty. We denote
\begin{equation}
\Lambda = (\vlam_1,\{\vlam_2,\dotsc, \vlam_{|\Lambda|}\}) \vdash_1 \vk
\end{equation}
a list such that $\sum \vlam_j = \vk$ and $\vlam_2, \dotsc, \vlam_{|\Lambda|} \neq 0$ while $\vlam_1$ may be zero. Then,
\begin{equation}
\overline{W}_{\sum_{c=1}^d \ve_c}(\tilde{x}_1, \tilde{x}_2,\dotsc,\tilde{x}_i=x, \dotsc, \tilde{x}_d) = \sum_{(\vlam_1,\{\vlam_2,\dotsc, \vlam_{|\Lambda|}\}) \vdash_1 \sum_{c\neq i} \ve_c} W_{\ve_i + \vlam_1}(x,\tilde{x}_{\vlam_1}) \prod_{j=2}^{|\Lambda|} W_{\vlam_j}(\tilde{x}_{\vlam_j}),
\end{equation}
and similarly for $\overline{U}^{(a_i)}_{\ve_i;\sum_{c\neq i }^d \ve_c}(x;\{\tilde{x}_c\}_{c\neq i})$,
\begin{equation}
\overline{U}^{(a_i)}_{\ve_i;\sum_{c\neq i}^d \ve_c}(x;\{\tilde{x}_c\}_{c\neq i}) = \sum_{(\vlam_1,\{\vlam_2,\dotsc, \vlam_{|\Lambda|}\}) \vdash_1 \sum_{c\neq i} \ve_c} U^{(a_i)}_{\ve_i;\vlam_1}(x, \tilde{x}_{\vlam_1}) \prod_{j=2}^{|\Lambda|} W_{\vlam_j}(\tilde{x}_{\vlam_j}).
\end{equation}
We then have to extract some coefficients out of them.

The disc equation now reads in terms of connected generating functions
\begin{multline} \label{ConnectedDiscEq}
W_{\ve_i}(x)^2 + W_{2\ve_i}(x,x) - N V_i'(x) W_{\ve_i}(x) + N P_{\ve_i}(x)\\
 - N^{2-d} \sum_{\Lambda \vdash_1 \sum_{c\neq i} \ve_c} \sum_{\substack{a_i \geq 1\\ a_{j\neq i}\geq0}} a_i t_{a_1 \dotsb a_d} \bigl[\prod_{c\neq i} \tilde{x}_c^{-a_c-1}\bigr] \prod_{j=2}^{|\Lambda|} W_{\vlam_j}(\tilde{x}_{\vlam_j}) \Bigl(x^{a_i-1} W_{\ve_i + \vlam_1}(x,\tilde{x}_{\vlam_1}) - U^{(a_i)}_{\ve_i;\vlam_1}(x, \tilde{x}_{\vlam_1}) \Bigr) = 0,
\end{multline}
for $i\in\{1, \dotsc, d\}$

\subsection{Multicolored loop equations}

We now act on \eqref{ConnectedDiscEq} with derivatives with respect to the potentials $V_1, \dotsc, V_c$. If $V_c(M) = \sum_k t^c_k M_c^k$, then the derivative with respect to $V_c(x)$ is $\delta/\delta V_c(x) = \sum_k x^{-k-1} \partial/\partial t^c_k, \: x\in \C_c$. It is such that
\begin{equation}
\frac{\delta}{\delta V_c(x)} W_{\vk}(x_{\vk}) = W_{\vk + \ve_c}(x_{\vk + \ve_c}),
\end{equation}
where $x_{\vk + \ve_c}$ has the variable $x\in \C_c$ added to the set of variables of color $c$. Skipping the details which are pretty standard, we obtain an equation labeled by the vector $\vn = \sum_{c=1}^d n_c \ve_c$ and the variables $x_0\in \C_i$ and $x_{\vn}\in \C_1^{n_1}\times \ldots \times \C_d^{n_d}$,
\begin{multline} \label{FullLoopEq}
\sum_{\vk} W_{\ve_i+\vk}(x_0,x_{\vk}) W_{\ve_i + \vn - \vk}(x_0,x_{\vn - \vk}) + W_{2\ve_i+\vn}(x_0,x_0,x_{\vn}) - N V_i'(x_0) W_{\ve_i+\vn}(x_0,x_{\vn}) + N P_{\ve_i;\vn}(x_0;x_{\vn})\\
+ \sum_{j=1}^{n_i} \frac{\partial}{\partial x^{(i)}_{j}} \biggl(\frac{W_{\vn}(x_{\vn}) - W_{\vn}(x_{\vn}\setminus \{x^{(i)}_j\}\cup \{x_0\})}{x^{(i)}_j - x_0}\biggr) - N^{2-d} \sum_{\Lambda \vdash_1 \sum_{c\neq i} \ve_c} \sum_{\substack{(\vmu_1,\dotsc,\vmu_{|\Lambda|}) \\ \sum_{j=1}^{|\Lambda|} \vmu_j = \vn}} \sum_{\substack{a_i \geq 1\\ a_{j\neq i}\geq0}} a_i t_{a_1 \dotsb a_d}\\
\bigl[\prod_{c\neq i} \tilde{x}_c^{-a_c-1}\bigr] \prod_{j=2}^{|\Lambda|} W_{\vlam_j + \vmu_j}(x_{\vmu_j},\tilde{x}_{\vlam_j}) \Bigl(x_0^{a_i-1} W_{\ve_i + \vlam_1 + \vmu_1}(x_0, x_{\vmu_1}, \tilde{x}_{\vlam_1}) - U^{(a_i)}_{\ve_i;\vlam_1 + \vmu_1}(x_0, x_{\vmu_1}, \tilde{x}_{\vlam_1}) \Bigr) = 0.
\end{multline}
Notice here that $\vmu_1, \dotsc, \vmu_{|\Lambda|}$ are ordered and may vanish (but not all at the same time since $\sum_{j=1}^{|\Lambda|} \vmu_j = \vn$).

\section{Disc and cylinder functions at genus zero and first correction} \label{sec:LargeN}

\subsection{Colored Loop equations at genus zero}
We now specialize the loop equations \eqref{ConnectedDiscEq} and \eqref{FullLoopEq} to the case of tensor models and in the large $N$ limit. First, from the matrix model \eqref{MatrixModel}, we specify the following quantities.
\begin{itemize}
\item The potentials are $V_c(x) = x^2/2$ for all $c=1, \dotsc, d$ and $x\in\C_c$. This implies $V_c'(x) = x$ and $P_{\ve_i}(x) = N$ and $P_{\ve_i; \vn}(x;x_{\vn}) = 0, \forall x_{\vn}\in \C_1^{n_1}\times \ldots \times \C_d^{n_d}$ at all orders in $N$.
\item The color couplings are
\begin{equation} \label{TensorCouplings}
t_{a_1 \dotsb a_d} = -\frac{1}{\sum_{c=1}^d a_c} \binom{\sum_{c=1}^d a_c}{a_1, \dotsc, a_d}\, \prod_{c=1}^d\alpha_c^{a_c}\, N^{d-2 - \frac{d-2}{2}\sum_{c=1}^d a_c},
\end{equation}
when $\sum_{c=1}^d a_c \geq 2$, as in \eqref{eq:LogExpansion}.
\end{itemize}
Crucially the couplings $t_{a_1 \dotsb a_d}$ do depend on $N$. It is clear from \eqref{TensorCouplings} that the dominant ones at large $N$ are those with $\sum a_c = 2$. With $a_i\geq0$, the only relevant couplings at large $N$ are thus the coupling of the form $t_{0\dotsb 2 \dotsb 0}$ and $t_{0\dotsb 1 \dotsb 0\dotsb 0 1 0 \dotsb 0}$. It means that the potential \eqref{eq:LogExpansion} reduces in the large $N$ limit to its quadratic terms, thus the large $N$ action reads
\begin{equation} \label{LargeNAction}
S_{N\infty} = -\frac{N}{2} \sum_{c=1}^d\bigl(1 - \alpha_c^2\bigr) \tr M_c^2 +  \sum_{c\neq c'}\alpha_c\alpha_{c'} \tr M_c\,\tr M_{c'}.
\end{equation}
One can then redefine $V_c(x), \: x\in \C_c$ as $V_c(x)=\frac{1-\alpha_c^2}{2}x^2$.

The large $N$ limit also provides the following extra information.
\begin{itemize}
\item We assume as usual that
\begin{equation}
W_{\vn}(x_{\vn}) = N^{2-|\vn|} W_{\vn}^0(x_{\vn}) + \text{corrections},
\end{equation}
where $|\vn| = \sum_{c=1}^d n_c$. This already implies that the term $W_{2\ve_i+\vn}(x_0,x_0,x_{\vn})$ in \eqref{FullLoopEq} is suppressed by $1/N^2$ with respect to the dominant ones and therefore does not appear at large $N$ (and similarly for $W_{2\ve_i}(x,x)$ in \eqref{ConnectedDiscEq}).
\item This also implies that Equation \eqref{FullLoopEq} has terms which scale like $N^{2 - |\vn|}$ at most (and $N^2$ for \eqref{ConnectedDiscEq}). To reach $N^{2 - |\vn|}$, $\Lambda$ in \eqref{ConnectedDiscEq} and \eqref{FullLoopEq} needs to have as many parts as possible, i.e. $|\Lambda| = d$ and $\vlam_1 = 0$ and $\{\vlam_2, \dotsc, \vlam_{d}\} = \{\ve_{c}\}_{c\neq i}$.
\end{itemize}
\subsubsection{Colored disc functions at genus zero}

We here compute the leading order of the disc function $W_{\ve_1}^0(x)$ for $x\in \C_1$, the generalization to $W_{\ve_c}^0(x)$ for $x\in \C_c$ and arbitrary $c\in\{1, \dotsc, d\}$ being immediate. Together with the above ingredients, Equation \eqref{ConnectedDiscEq} reduces to
\begin{equation}\label{eq:planconnecdisc}
W_{\ve_1}^{0}(x)^2 - x\,W_{\ve_1}^{0}(x) + 1 - \sum_{\substack{a_1 \geq 1\\ \sum_{j=1}^d a_j = 2}} a_1 t_{a_1 \dotsb a_d} \bigl[\prod_{c=2}^d \tilde{x}_c^{-a_c-1}\bigr] \prod_{c=2}^d W_{\ve_1}^{0}(\tilde{x}_c) \Bigl(x^{a_1 - 1} W_{\ve_1}^{0}(x) - U_{\ve_1}^{0(a_1)}(x)\Bigr) = 0.
\end{equation}
At this stage, one uses the fact that the dominant couplings are those with $\sum a_c = 2$. Since $a_1\geq 1$ in the loop equation, the condition $\sum a_c = 2$ is realized with either $a_1=2$ and the others being zero, or $a_1 = a_c = 1$ for some $c\neq 1$ and the others vanishing. For those vanishing $a_c$, one uses
\begin{equation}
[\tilde{x}^{-1}] W_{\ve_c}(\tilde{x}) = 1
\end{equation}
which holds by definition.  When $a_{c\neq1}=1$, the symmetry $M_c \mapsto -M_c$ of the large $N$ action \eqref{LargeNAction} implies that $[\tilde{x}^{-2}] W_{\ve_c}^{0}(\tilde{x}) = \lim_{N\to\infty} \langle \tr M_1 \rangle/N = 0$. Only the case $a_1 = 2$ remains, for which $U_{\ve_1}^{0(a_1)}(x)=1$. With $t_{20\dotsb 0}= - \alpha_1^2/2$, the disc equation eventually reduces to
\begin{equation} \label{LargeNLoopEqW1}
W_{\ve_1}^{0}(x)^2 - (1 - \alpha_1^2)\, x\,W_{\ve_1}^{0}(x) + (1 - \alpha_1^2) = 0,
\end{equation}
hence
\begin{equation} \label{LargeNResolvent}
W_{\ve_1}^{0}(x) = \frac{1-\alpha_1^2}{2} \biggl( x - \sqrt{x^2 - \frac{4}{1-\alpha_1^2}} \biggr),
\end{equation}
for all $x\in\C_1$.

This is the disc function of a Gaussian Unitary Ensemble. It has a cut on $[a_1,b_1]$ with $a_1 = - b_1 = -2/\sqrt{1-\alpha_1^2}$. The disc function $W_{\ve_c}(x)$ at large $N$ for the color $c\in\{1, \dotsc, d\}$ and $x\in\C_c$ is obtained through the replacement $\alpha_1 \to \alpha_c$,
\begin{equation} \label{eq:LargeNResolventcolor}
W_{\ve_c}^{0}(x) = \frac{1-\alpha_c^2}{2} \biggl( x - \sqrt{x^2 - \frac{4}{1-\alpha_c^2}} \biggr), 
\end{equation}
which has a cut on $[a_c,b_c]$ with $a_c = - b_c = -2/\sqrt{1-\alpha_c^2}$. 

\subsubsection{Colored cylinder functions at genus zero}

We need to compute both the monocolored cylinder function $W_{2\ve_i}(x_1, x_2), \: x_1, x_2\in\C_i $ for all $i\in \{1, \dotsc, d\}$ and the bicolored cylinder function $W_{\ve_i+\ve_j}(x_1, x_2), \: x_1\in \C_i, x_2\in \C_j$ for $i,j\in \{1, \dotsc, d\}$ and $i\neq j$. The system of equations which determines those two types of functions at large $N$ is obtained by setting $\vn = \ve_i, \ve_j$ in \eqref{FullLoopEq} and taking the large $N$ limit.

There are two differences between the equations satisfied in the two different cases. First, in the monocolored case, there is a derivative term while it is not present in the bicolored case. Second, the terms which involve extracting coefficients of generating functions are different.

For $\vn = \ve_i$ one gets
\begin{multline}
2 W_{\ve_i}^{0}(x_1) W_{2\ve_i}^{0}(x_1,x_2) - x_1 W_{2\ve_i}^{0}(x_1,x_2) + \frac{\partial}{\partial x_2} \frac{W_{\ve_i}^{0}(x_1) - W_{\ve_i}^{0}(x_2)}{x_1 - x_2} \\
- \sum_{\substack{(\vmu_1,\dotsc, \vmu_d)\\\sum \vmu_j = \ve_i}} 2 t_{0\dotsb0,a_i=2,0\dotsb 0} [\prod_{c\neq i} \tilde{x}_c^{-1}] \prod_{j\neq i} W_{\ve_j + \vmu_j}^{0}(\tilde{x}_j, x_{\vmu_j}) x_1 W^{0}_{\ve_i + \vmu_i}(x_1, x_{\vmu_i}) \\
- \sum_{\substack{(\vmu_1,\dotsc, \vmu_d)\\\sum \vmu_j = \ve_i}} \sum_{j\neq i}t_{0,a_j=1,0\dotsb0,a_i=1,0\dotsb 0} [\tilde{x}_j^{-2} \prod_{c\neq i,j} \tilde{x}_c^{-1}] \prod_{j\neq i} W_{\ve_j + \vmu_j}^{0}(\tilde{x}_j, x_{\vmu_j}) W^{0}_{\ve_i + \vmu_i}(x_1, x_{\vmu_1}) = 0.
\end{multline}
The second line corresponds to the multi-matrix potential with $a_i =2$ and $a_c = 0$ for $c\neq i$. It involves $[\tilde{x}^{-1}] W_{\ve_j + \vmu_j}^{0}(\tilde{x}, x_{\vmu_j}) = \delta_{\vmu_j,0}$ which vanishes unless $\vmu_j = 0$. The sum over $(\vmu_1, \dotsc, \vmu_d)$ with $\sum_k \vmu_k =\ve_i$ thus reduces to $\vmu_i = \ve_i$ and
\begin{equation}
-\sum_{\substack{(\vmu_1,\dotsc, \vmu_d)\\\sum \vmu_j = \ve_i}} 2 t_{0\dotsb0,a_i=2,0\dotsb 0} [\prod_{c\neq i} \tilde{x}_c^{-1}] \prod_{j\neq i} W_{\ve_j + \vmu_j}^{0}(\tilde{x}_j, x_{\vmu_j}) x_1 W^{0}_{\ve_i + \vmu_i}(x_1, x_{\vmu_1}) = \alpha^2_i x_1 W_{2\ve_i}^{0}(x_1, x_2)
\end{equation}
The third line corresponds to the multi-matrix potential with $a_i = a_j = 1$. The sum over $(\vmu_1, \dotsc, \vmu_d)$ with $\sum_j \vmu_j =\ve_i$ reduces to either $\vmu_i = \ve_i$ or $\vmu_{j\neq i} = \ve_i$. For $\vmu_i = \ve_i$, the quantities $[\tilde{x}_j^{-2}] W^{0}_{\ve_j}(\tilde{x}_j)$ have to be evaluated. It is the expectation value of $\tr M_j$ at leading order which vanishes. Therefore
\begin{multline}
\sum_{\substack{(\vmu_1,\dotsc, \vmu_d)\\\sum \vmu_j = \ve_i}} \sum_{j\neq i}t_{0,a_j=1,0\dotsb0,a_i=1,0\dotsb 0} [\tilde{x}_j^{-2} \prod_{c\neq i,j} \tilde{x}_c^{-1}] \prod_{j\neq i} W_{\ve_j + \vmu_j}^{0}(\tilde{x}_j, x_{\vmu_j}) W^{0}_{\ve_i + \vmu_i}(x_1, x_{\vmu_i}) \\=  \sum_{j\in\{1, \dotsc, d\}, \: j\neq i}\alpha_i\alpha_j [\tilde{x}_j^{-2}] W_{\ve_j + \ve_i}^{0}(\tilde{x}_j, x_2) W_{\ve_i}^{0}(x_1).
\end{multline}
Putting all together one finds
\begin{multline} \label{LeadingMono2Pt}
2 W_{\ve_i}^{0}(x_1) W_{2\ve_i}^{0}(x_1,x_2) - (1-\alpha_i^2) x_1 W_{2\ve_i}^{0}(x_1,x_2) + \frac{\partial}{\partial x_2} \frac{W_{\ve_i}^{0}(x_1) - W_{\ve_i}^{0}(x_2)}{x_1 - x_2} \\ + \sum_{j\in\{1, \dotsc, d\}, \: j\neq i}\alpha_i\alpha_j [\tilde{x}_j^{-2}] W_{\ve_j + \ve_i}^{0}(\tilde{x}_j, x_2) W_{\ve_i}^{0}(x_1) = 0.
\end{multline}

For $\vn = \ve_j$ a similar reasoning leads to
\begin{equation} \label{LeadingBicolored2Pt}
2 W_{\ve_i}^{0}(x_1) W_{\ve_i + \ve_j}^{0}(x_1,x_2) - (1-\alpha^2_i) x_1 W_{\ve_i + \ve_j}^{0}(x_1,x_2) +\sum_{p\neq i} \alpha_i\alpha_p [\tilde{x}_p^{-2}] W^{0}_{\ve_p+\ve_j}(\tilde{x}, x_2)W_{\ve_i}^{0}(x_1) = 0.
\end{equation}

Equations \eqref{LeadingMono2Pt}, \eqref{LeadingBicolored2Pt} hold in powers of $1/x_1$ for $x_1$ close to infinity. At the order $1/x_1$ the following relations are obtained
\begin{equation} \label{LoopEq2Pt1/x}
\begin{gathered}
-(1-\alpha^2) [\tilde{x}^{-2}] W^{0}_{2 \ve_i}(\tilde{x},x_2) +\sum_{j\in\{1, \dotsc, d\}, \: j\neq i}\alpha_i\alpha_j [\tilde{x}_j^{-2}] W_{\ve_j + \ve_i}^{0}(\tilde{x}_j, x_2) = \frac{\partial}{\partial x_2} W^{0}_{\ve_i}(x_2)\\
-(1-\alpha_i^2) [\tilde{x}^{-2}] W^{0}_{ \ve_i+\ve_j}(\tilde{x},x_2) + \sum_{p\neq i} \alpha_i\alpha_p [\tilde{x}_p^{-2}] W^{0}_{\ve_p+\ve_j}(\tilde{x}_p, x_2)= 0.
\end{gathered}
\end{equation}
We define, for convenience, the quantities $Q_i(x_2)$ and $F_j^{(i)}(x_2)$
\begin{equation}
\begin{gathered}
Q_i(x_2)=\sum_{j\in\{1, \dotsc, d\}, \: j\neq i}\alpha_i\alpha_j [\tilde{x}_j^{-2}] W_{\ve_j + \ve_i}^{0}(\tilde{x}_j, x_2) \\
F_j^{(i)}(x_2)=\sum_{p\neq i} \alpha_i\alpha_p [\tilde{x}_p^{-2}] W^{0}_{\ve_p+\ve_j}(\tilde{x}, x_2).
\end{gathered}
\end{equation}
These quantities can be determined with some algebra, 
\begin{equation}\label{eq:Qi}
Q_i(x_2)\frac{1-\sum_{p\neq i}\alpha_p^2}{\sum_{p\neq i}(\alpha_p\alpha_i)^2}=[\tilde{x}^{-2}]W^0_{2\ve_j}(\tilde{x},x_2)
\end{equation}
and first noticing 
\begin{equation}\label{eq:FjiA}
F_j^{(i)}(x_2)=\frac{\alpha_i}{\alpha_j}\left(Q_j(x_2)-\alpha_i\alpha_j[\tilde{x}_i^{-2}]W^0_{\ve_i+\ve_j}(\tilde{x}_i,x_2)\right)+\alpha_i\alpha_j[\tilde{x}_j^{-2}]W^0_{2\ve_j}(\tilde{x}_j,x_2),
\end{equation}
one shows
\begin{equation}\label{eq:FjiB}
F_j^{(i)}(x_2)=(1-\alpha_i^2)\left(\frac{\alpha_i\alpha_j}{\sum_{p\neq j}(\alpha_p\alpha_j)^2}\right)Q_j(x_2).
\end{equation}
Equation \eqref{eq:Qi} holds for all $i\in\{1, \dotsc, d\}$ and $x_2\in \C_i$, while Equation \eqref{eq:FjiB} is true for all $i\neq j$, $x_2\in \C_j$. 

Using \eqref{eq:Qi}, \eqref{eq:FjiB}, \eqref{LoopEq2Pt1/x}, the two large $N$, cylinder equations are
\begin{multline}
\bigl(2 W_{\ve_i}^{0}(x_1)  - (1-\alpha^2) x_1\bigr) W_{2\ve_i}^{0}(x_1,x_2) + \frac{\partial}{\partial x_2} \frac{W_{\ve_i}^{0}(x_1) - W_{\ve_i}^{0}(x_2)}{x_1 - x_2} \\ + \frac{\sum_{p\neq i} (\alpha_i\alpha_p)^2}{\sum_{p=1}^d\alpha_p^2-1}\partial_{x_2}W^0_{\ve_i}(x_2) W_{\ve_i}^{0}(x_1) = 0,\quad \text{for $x_1,x_2\in \C_i$}
\end{multline}
and
\begin{multline}
\bigl(2 W_{\ve_i}^{0}(x_1)  - (1-\alpha^2_i) x_1\bigr)W_{\ve_i + \ve_j}^{0}(x_1,x_2) +(1-\alpha_i^2)\frac{\alpha_i\alpha_j}{\sum_{p=1}^d\alpha_p^2-1}\partial_{x_2}W^0_{\ve_j}(x_2)W_{\ve_i}^{0}(x_1) = 0, \quad \text{for $x_1\in \C_i, \: x_2 \in \C_j$}.
\end{multline}
We denote
\begin{equation}
\sigma_i(x)=\sqrt{x^2-\frac{4}{1-\alpha_i^2}}, \quad \text{for $x \in \C_i\setminus[a_i,b_i]$}.
\end{equation}
Using the fact that
\begin{equation}
\partial_x W^{0}_{\ve_i}(x) = -\frac{W^{0}_{\ve_i}(x)}{\sigma_i(x)},
\end{equation}
we show,
\begin{equation}
\begin{gathered}
W^0_{2\ve_i}(x_1,x_2)=\frac{x_1x_2-\sigma_i(x_1)\sigma_i(x_2)-\frac{4}{1-\alpha_i^2}}{2(x_1-x_2)^2\sigma_i(x_1)\sigma_i(x_2)}-\frac{\sum_{p\neq i}(\alpha_i\alpha_p)^2}{\sum_{p=1}^d\alpha_p^2-1}\frac{W^0_{\ve_i}(x_1)W^0_{\ve_i}(x_2)}{(1-\alpha_i^2)\sigma_i(x_1)\sigma_i(x_2)}, \quad \text{for $(x_1,x_2)\in \C_i^2$},\\
W^0_{\ve_i+\ve_j}(x_1,x_2)=-\frac{\alpha_i\alpha_j}{\sum_{p=1}^d\alpha_p^2-1}\frac{W_{\ve_i}^0(x_1)W^0_{\ve_j}(x_2)}{\sigma_i(x_1)\sigma_j(x_2)}, \quad \text{for $(x_1,x_2)\in \C_i\times \C_j$}.
\end{gathered}
\end{equation}

We notice that the first contribution in the above expression for $W^0_{2\ve_i}(x_1,x_2)$ is the large $N$, cylinder function $W_{2,i}^0(x_1,x_2)$ of a GUE with covariance $(1-\alpha_i^2)$. It satisfies
\begin{equation}
\bigl(2 W_{\ve_i}^{0}(x_1) - (1-\alpha_i^2) x_1\bigr) W^{0}_{2,i}(x_1, x_2) + \frac{\partial}{\partial x_2} \frac{W_{\ve_i}^{0}(x_1) - W_{\ve_i}^{0}(x_2)}{x_1 - x_2} = 0,
\end{equation}
and equals
\begin{equation} \label{W2GUE}
W_{2,i}^{0}(x_1, x_2) = \frac{x_1 x_2 + \sigma_i(x_1) \sigma_i(x_2) - \frac{4}{1-\alpha^2}}{2 \sigma_i(x_1) \sigma_i(x_2) (x_1 - x_2)^2},
\end{equation}
so that
\begin{equation}
W^0_{2\ve_i}(x_1,x_2) = W^0_{2,i}(x_1,x_2)-\frac{\sum_{p\neq i}(\alpha_i\alpha_p)^2}{\sum_{p=1}^d\alpha_p^2-1}\frac{W^0_{\ve_i}(x_1)W^0_{\ve_i}(x_2)}{(1-\alpha_i^2)\sigma_i(x_1)\sigma_i(x_2)}, \quad \text{for $(x_1,x_2)\in \C_i^2$},
\end{equation}

The remarkable property of $W_{2\ve_i}^0(x_1, x_2)$ is that it has a double pole at $x_1 = x_2$. Importantly, this is not the case for $W_{\ve_i + \ve_j}^0(x_1, x_2)$ for $i\neq j$. This means that the function defined on $(\cup_{c=1}^d \C_c)^2$,
\begin{equation} \label{GlobalW2}
W_2^0(x_1, x_2) = \sum_{i, j= 1}^d \mathbf{1}_{\C_i}(x_1) \mathbf{1}_{\C_j}(x_2) W_{\ve_i + \ve_j}^0(x_1, x_2),
\end{equation}
has a double pole at $x_1 = x_2$. This is the property required to use $W_2^0$ to build the kernel for the topological recursion.

\subsection{Joukowsky transformation}\label{subsec:Joukowsky}

In this subsection, we introduce the Joukowsky transformation. The main motivation in this subsection is to simplify expressions. However, it is also important when we introduce the spectral curve later on. 

For each copy $\C_i$ of $\C$ we introduce a parametrization of it using $z^{(i)}$ variables,
\begin{equation}
x^{(i)}(z^{(i)})=\frac{1}{\sqrt{1-\alpha_i^2}}\Bigl(z^{(i)} + \frac1{z^{(i)}}\Bigr) \qquad \text{for $|z^{(i)}|\geq 1$ and $i = 1, \dotsc, d$}.
\end{equation}
The variable $z^{(i)}$ lives on a copy of $\C\setminus\mathbb{U}$, where $\mathbb{U}$ is the unit disc. This transformation opens up the cut $[a_i,b_i]$ of the $i$-colored  disc function in $\C_i$ in the sense that the images of the unit circle in the $z$ variable planes are $[a_i,b_i]$. Notice that $x^{(i)}(z^{(i)})$ is invariant for all $i$ under the involution
\begin{equation}
\iota (z) = 1/z.
\end{equation}
The fixed points of $\iota$ are $\pm 1$ and their images are the extremities of the cuts $a_i, b_i$. The inverse transform writes
\begin{equation}
z^{(i)}(x^{(i)})=\frac2{a_i-b_i}\left( x^{(i)}+\sqrt{(x^{(i)}-a_i)(x^{(i)}-b_i)}\right),
\end{equation}
though as we will see later one has to carefully choose the sign in front of the square root. One choice of sign is related to the other by the involution $\iota$.

To write the topological recursion, we will pullback the correlation functions on the $z^{(i)}$--planes using $x^{(i)}(z^{(i)})$ and turn them to differential forms by defining
\begin{equation}
\omega_{\vk}(z_{\vk}) = W_{\vk}(x(z_{\vk})) \prod dx(z_{\vk}) + \sum_{c=1}^d \delta_{\vk, 2\ve_c} \frac{dx^{(c)}(z_1) dx^{(c)}(z_2)}{(x^{(c)}(z_1) - x^{(c)}(z_2))^2}
\end{equation}
where $z_{\vk}$ is the list of variables $(z^{(c)}_{1}, \dotsc, z^{(c)}_{k_c})_{c=1, \dotsc, d}$ and the product is over all those elements. Those differential forms admit the same expansions as the correlation functions.

Let us write the disc functions. A little bit of caution is required to rewrite the correlation functions though because they are singular along the cuts $[a_i,b_i]$. The factor $\sigma_i(x^{(i)})^2 = (x^{(i)})^2 - 4/(1-\alpha_i^2)$ becomes $\sigma_i(x^{(i)}(z^{(i)}))^2 = (z^{(i)} - (z^{(i)})^{-1})^2/(1-\alpha_i^2)$. To choose the sign of the square--root we notice that for large $z^{(i)}$, $x^{(i)}$ is linear in $z^{(i)}$ and the disc functions should behave as $1/x^{(i)}\sim 1/z^{(i)}$. Therefore we find for the disc functions
\begin{equation} \label{DiffDiscFunction}
W_{\ve_i}^{0}(x^{(i)}(z^{(i)})) = \sqrt{1-\alpha^2} \frac{1}{z^{(i)}} \quad \Rightarrow \quad \omega_{\ve_1}^{0}(z^{(i)}) = \frac{1}{z^{(i)}}\Bigl(1 - \frac1{(z^{(i)})^2}\Bigr) dz^{(i)},
\end{equation}
which has simple roots at $z^{(i)}=\pm 1$. They are those of $dx^{(i)}(z^{(i)})$.

As for the cylinder function of the GUE in \eqref{W2GUE}, one finds
\begin{equation}
W_2^{0}(x(z_1), x(z_2)) dx(z_1) dx(z_2) = \frac{dz_1\, dz_2}{(z_1 z_2 - 1)^2},
\end{equation}
and thus
\begin{equation}
\omega_2^{0}(z_1, z_2) = W_2^{0}(x(z_1), x(z_2)) dx(z_1) dx(z_2) + \frac{dx(z_1) dx(z_2)}{(x(z_1) - x(z_2))^2} = \frac{dz_1\,dz_2}{(z_1 - z_2)^2},
\end{equation}
which has a double pole on the diagonal $z_1=z_2$ and is the Bergmann kernel on the sphere. One further finds
\begin{equation}
\frac{W_{\ve_c}^{0}(x^{(c)}(z^{(c)})) dx^{(c)}(z^{(c)})}{\sigma_c(x^{(c)}(z^{(c)}))} = \sqrt{1-\alpha_c^2} \frac{dz^{(c)}}{{(z^{(c)})}^2}, \: \forall c\in \{1, \dotsc, d\}
\end{equation}
so that
\begin{equation}\label{eq:monocylform}
\omega_{2\ve_i}^{0}(z_1, z_2) = \frac{dz_1 dz_2}{(z_1 - z_2)^2} -\frac{\sum_{p\neq i}(\alpha_i\alpha_p)^2}{\sum_{p=1}^d\alpha_p^2-1}\frac{dz_1 dz_2}{z_1^2 z_2^2},  \quad \text{and} \quad\omega_{\ve_i+\ve_j}(z_1,z_2)=-\frac{\alpha_i\alpha_j\sqrt{(1-\alpha_i)(1-\alpha_j)}}{\sum_{p=1}^d\alpha_p^2-1}\frac{dz_1dz_2}{z_1^2z_2^2}.
\end{equation}

\subsection{First correction at $d=3$}

In this subsection, we specialize to the symmetric case, this implies that $\alpha_c=\alpha, \: \forall c\in \{1, \dotsc, d\}$. This has several other consequences. The main one is that the functions $W_{\ve_c}^0$ all have a cut in the same position in their copy of the complex plane, i.e. $[a_c,b_c]=[a,b]$, then we can use the same Joukowsky transformation for all the $x$ variables. Moreover we can identify all the copies of $\C$ one to another using the identity map, and pull back the generating function $W_{\vn}^p(x_{\vn})$ accordingly. Using these facts one can perform the following computations. 

\smallskip    

The couplings \eqref{TensorCouplings} have an expansion in powers of $1/N^{\frac{d-2}{2}} = 1/\sqrt{N}$ at $d=3$. This suggests that the first correction to the correlation functions comes with a $1/\sqrt{N}$ suppression with respect to their leading orders. In particular,
\begin{equation}
W_{\ve_c}(x) = N\Bigl(W_{\ve_c}^0(x) + \frac{1}{\sqrt{N}} W_{\ve_c}^{1/2}(x) + \mathcal{O}(1/N)\Bigr).
\end{equation}
To find $W_{\ve_c}^{1/2}(x)$, one plugs the above expansion into the loop equation on the disc function \eqref{ConnectedDiscEq}. The equation one obtains this way is the same as the loop equation for the matrix model with potential given by the action \eqref{eq:LogExpansion} truncated at cubic order (all 3 types of cubic terms have the same scaling, found by evaluating all traces to $N$). This fact can be recovered by analyzing the generic disc equation \eqref{ConnectedDiscEq} at next--to--leading order.

By plugging the above $1/\sqrt{N}$ expansion into \eqref{ConnectedDiscEq}, it can be seen that a term of the summand in the bottom line,
\begin{equation}
N^{2-d} a_1 t_{a_1 \dotsb a_d} \bigl[\prod_{c=2}^d \tilde{x}_c^{-a_c-1}\bigr] \prod_{j=1}^{|\Lambda|} N^{2-|\vlam_j|}N^{h_j}  \prod_{j=2}^{|\Lambda|} W_{\vlam_j}^{h_j}(\tilde{x}_{\vlam_j}) \Bigl(x^{a_1-1} W_{\ve_1 + \vlam_1}^{h_1}(x,\tilde{x}_{\vlam_1}) - U^{(a_1)h_1}_{\ve_1;\vlam_1}(x, \tilde{x}_{\vlam_1}) \Bigr)
\end{equation}
with $h_1, h_2, h_3 \in \mathbbm{N}/2$, scales like $N^\beta$ with
\begin{equation}
\beta = -\sum_{c=1}^3 \frac{a_c}{2} + 2|\Lambda| - 3 -\sum_{j=1}^{|\Lambda|} h_j.
\end{equation}
Here we have used $\sum_{j=1}^{|\Lambda|} |\vlam_j| = 2$ (since $\sum_{j=1}^{|\Lambda|} \vlam_j = \ve_2 + \ve_3$) and the explicit $N$--dependence of the couplings given in \eqref{TensorCouplings}. It is bounded by $2$ (the leading order of the loop equation, with $|\Lambda|=3, \sum a_c = 2$ and $h_j=0$), so the first correction is for $\beta = 3/2$. If $|\Lambda|\leq 2$, then $\beta = 3/2$ becomes $-\sum a_c/2 - \sum h_j = 9/2 - 2|\Lambda| \geq 1/2$, which is impossible since $a_c, h_j \geq 0$. It enforces $|\Lambda|=3$ and thus $\vlam_1 =0$ and $\vlam_2 + \vlam_3 = \ve_2 + \ve_3$. The equation $\beta = 3/2$ thus becomes
\begin{equation}
\sum_{c=1}^3 \Bigl(\frac{a_c}{2} + h_c\Bigr) = \frac{3}{2}
\end{equation}
with $a_1 \geq 1$. The cases $\sum h_c \geq 3/2$ are thus impossible. Similarly, $\sum_c h_c = 1$ does not contribute as it enforces $a_1 = 1, a_2 = a_3 = 0$ for which the coupling $t_{a_1 a_2 a_3}$ vanishes. One is left with the terms 
\begin{itemize}
\item $\sum h_c = 0$ and $\sum a_c = 3$, which are the cubic terms of the action coupled to leading--order disc functions,
\item $\sum h_c = 1/2$ and $\sum a_c =2$ which are the Gaussian terms coupled to one next--to--leading--order disc function.
\end{itemize}
Some of those contributions vanish due to $[x^{-2}] W_{\ve_c}^{0}(x) = [x^{-1}] W_{\ve_c}^{1/2}(x) = 0$ as well as from $\langle \tr \frac{x^{a_1-1} - M_1^{a_1-1}}{x - M_1}\rangle^{(h_1)} = 0$ for $a_1 = 1$ and for $a_1=2, h_1=1/2$. The surviving terms are $(a_1=2, h_1 = 1/2), (a_1 = a_2 = 1, h_2 = 1/2), (a_1 = 2, a_2 = 2), (a_1 = 3)$ and their symmetrical ones exchanging the colors 2 and 3. Eventually, one gets
\begin{equation} \label{Omega1/2Eq}
\Bigl(2 W_{\ve_1}^{0}(x) - (1-\alpha^2)x \Bigr) W_{\ve_1}^{1/2}(x) + \Bigl(2\alpha^2 [\tilde{x}^{-2}] W_{\ve_1}^{1/2}(\tilde{x}) + 2\alpha^3 [\tilde{x}^{-3}] W_{\ve_1}^{0}(\tilde{x}) + \alpha^3 x^2\Bigr) W_{\ve_1}^{0}(x) - \alpha^3 x = 0.
\end{equation}

From the leading--order disc function one obtains
\begin{equation}
[\tilde{x}^{-3}] W_{\ve_1}^{0}(\tilde{x}) = \frac1N \langle \tr M_1^2\rangle^{(0)} = \frac{1}{1-\alpha^2}.
\end{equation}
Equation \eqref{Omega1/2Eq} holds order by order in the large $x$ expansion. At order $x$, it reduces to $[\tilde{x}^{-1}]W_{\ve_1}^{0}(\tilde{x}) = 1$ and at order $x^0$, $[\tilde{x}^{-2}]W_{\ve_1}^{0}(\tilde{x}) = 0$, as already known. The first non--trivial order is $1/x$ which leads to
\begin{equation}
[\tilde{x}^{-2}] W_{\ve_1}^{1/2}(\tilde{x}) = \frac{3\alpha^2}{(1-\alpha^2)(1-3\alpha^2)}.
\end{equation}
Plugging it back into \eqref{Omega1/2Eq}, one finds
\begin{equation}
W_{\ve_1}^{1/2}(x) = \frac{\alpha^3}{(1-\alpha^2) \sigma(x)} \Bigl( x (xW_{\ve_1}^{0}(x) - 1) + \frac{2}{(1-\alpha^2)(1-3\alpha^2)} W_{\ve_1}^{0}(x) \Bigr),
\end{equation}
or written in differential form,
\begin{equation}
\omega_{\ve_1}^{1/2}(z) = W_{\ve_1}^{1/2}(x(z)) dx(z) = \frac{\alpha^3}{(1-\alpha^2)^{3/2}} \Bigl(\frac1{z^2} + \frac{3(1-\alpha^2)}{1-3\alpha^2} \Bigr) \frac{dz}{z^2}.
\end{equation}
Unlike the usual generating functions of maps in the topological $1/N^2$-expansion of matrix models, that one does not have poles at $z=\pm 1$. This is because the correction at order $1/\sqrt{N}$ is non--topological, in the sense that it does not come at the order $1/N^{2g}$ for an integral genus $g$. It is absent from regular matrix models whose potentials have a $1/N^2$ expansion (as opposed to a $1/\sqrt{N}$ expansion here for $d=3$).

At $d=4$ and $d=5$, the correlation function have expansions in powers of $1/N$ and $1/N^{3/2}$ respectively and the first corrections are regular at $z=\pm1$. Only at $d=6$, and more generally for $d = 4 \delta + 2$, one recovers the traditional form of the $1/N^2$ expansion. In those cases, one expects the correlation functions to have poles at $z=\pm1$ as in regular matrix models (this can be shown by an extension of the argument used in \cite{BlobbedTR}, Lemma 4.1).

\section{Putting the pieces together: colored loop equations as local loop equations on the spectral curve} \label{sec:SpectralCurve}

\subsection{Geometry of the curve}\label{sec:geometrycurve}

We start by introducing a few notations. We define $\hat{\C}$ as the compactification of $\C$ obtained by adding a point at infinity, $\hat{\C}=\C\cup \{\infty\}$, this space is often called the Riemann sphere.  

As we have seen the colored planar disc functions satisfy algebraic equations of order $2$. For each color $i\in\{1, \dotsc, d\}$ we have a polynomial relationship between $W_{\ve_i}^0$ and $x^{(i)}$. Temporarily changing notations, we have to solve a family of polynomials in two variables $(x^{(i)}, y_i)\in \hat{\C}_i^2$,
\begin{equation}\label{eq:family}
\begin{matrix}
f_1(x^{(1)},y_1):=y_1^2-(1-\alpha_1^2)\ x^{(1)}y_1+(1-\alpha_1^2)=0 \\
f_2(x^{(2)},y_2):=y_2^2-(1-\alpha_2^2)\ x^{(2)}y_2+(1-\alpha_2^2)=0 \\
\vdots \\
f_d(x^{(d)},y_d):=y_d^2-(1-\alpha_d^2)\ x^{(d)}y_d+(1-\alpha_d^2)=0,
\end{matrix}
\end{equation}  
where we denote $y_i=W_{\ve_i}^0$. These polynomials vanish along a curve $\mathcal{C}$ in $\bigcup_{i=1}^d\hat{\C}_i^2$. 
Indeed, it is possible to write a single polynomial equation for variables $(x,y)$ living in $\bigcup_{i=1}^d\hat{\C}_i^2$. To this aim, define
\begin{equation}
\tilde{V}'(x)=\sum_{i=1}^d\mathbf{1}_{\hat{\C}_i}(x)(1-\alpha_i^2)x,
\end{equation}
and 
\begin{equation}
P_1^0(x)=\sum_{i=1}^d\mathbf{1}_{\hat{\C}_i}(x)P_{\ve_i}(x)=\sum_{i=1}^d\mathbf{1}_{\hat{\C}_i}(x)(1-\alpha_i^2),
\end{equation}
where $\mathbf{1}_A$ is the indicator function of the set $A$, so to say, $\mathbf{1}_A(x) = 1$ if $x\in A$ and $0$ else.
Then, the unique polynomial equation recasting \eqref{eq:family} writes
\begin{equation} \label{PolynomialSpectralCurve}
f(x,y):=y^2-\tilde{V}'(x)y+P_1^0(x)=0, \quad \text{for $(x,y)\in \bigcup_{i=1}^d\hat{\C}_i^2$}.
\end{equation}
The colored variables $(x^{(i)},y_i)$ now appear as local coordinates on each component $\hat{\C}_i^2$ of the union $\bigcup_{i=1}^d\hat{\C}_i^2$. The global planar resolvent can be written as well
\begin{equation}
W_1^0(x):=\sum_{i=1}^d\mathbf{1}_{\hat{\C}_i}(x)W^0_{\ve_i}(x).
\end{equation}

\smallskip

$\mathcal{C}$ comes with an explicit complex structure defined component-wise. We explain how a natural choice of complex charts leads to the Joukowsky coordinates. Consider the component $\mathcal{C}_i=\mathcal{C}\cap \hat{\C}_i^2$. On each $\hat{\mathcal{C}}_i$, one can define,
\begin{eqnarray}
&g_i^+(x)=\frac{1-\alpha_i^2}{2}\left(x+\sqrt{(x-a_i)(x-b_i)} \right),\\
&g_i^-(x)=W_{\ve_i}^0(x)=\frac{1-\alpha_i^2}{2}\left(x-\sqrt{(x-a_i)(x-b_i)} \right),
\end{eqnarray}
for $x\in \hat{\C}_i\setminus [a_i,b_i]$. Then one defines the sets
\begin{equation}
\mathcal{C}_i^+=\{(x,g_i^+(x))|x\in \hat{\C_i}\setminus[a_i,b_i]\}, \quad \text{and} \quad
\mathcal{C}_i^-=\{(x,g_i^-(x))|x\in \hat{\C_i}\setminus[a_i,b_i]\}.
\end{equation}
The projections $\pi_i^{\pm}:\mathcal{C}_i^{\pm}\rightarrow \C_i\setminus[a_i,b_i]$ sending $(x,g_i^{\pm}(x))\mapsto x$ are homeomorphisms whose inverses are given by $(\pi_i^{\pm})^{-1}(x)=(x,g_i^{\pm}(x))$. Moreover as long as $\frac{\partial f_i}{\partial y}\neq 0$, $\pi_i^{\pm}$ are holomorphic\footnote{It is an easy consequence of the implicit function theorem.}, consequently they are complex charts away from the cut $[a_i,b_i]$.  These charts are sufficient to parametrize the components $\mathcal{C}_i$ of $\mathcal{C}$ away from the cuts. 

\smallskip

Another parametrization valid along the cuts leading to the Joukowsky variables introduced earlier can be constructed as follows. We again localize on one chosen component $\mathcal{C}_i$ of $\mathcal{C}$. As we have seen, $\frac{\partial f_i}{\partial y}(a_i,\sqrt{1-\alpha_i^2})=0$ and $\frac{\partial f_i}{\partial y}(b_i,-\sqrt{1-\alpha_i^2})=0$. That is why we cannot define a complex structure with the former parametrization around the cut. However $\frac{\partial f_i}{\partial x}(a_i,\sqrt{1-\alpha_i^2})$ and $\frac{\partial f_i}{\partial x}(b_i,-\sqrt{1-\alpha_i^2})$ are non-zero. Also notice that $ \frac{\partial f_i}{\partial x}(x,0)=0, \: \forall x\in \C_i$.
If one sets\footnote{We 'solve' the polynomial $f_i(x,y)$ in the $x$ variable.} for $y\neq 0$
\begin{equation}
s(y)=\left(\frac1{y}+\frac{y}{(1-\alpha_i^2)}\right),
\end{equation}
then $f(s(y),y)=0$. We define a holomorphic map $p_i:\C\rightarrow\hat{\C}^*_i$ by $\zeta\mapsto\frac{\sqrt{1-\alpha_i^2}}{\zeta}$. So the open set $O=\{(s(y),y)|y\in \C_i^*\cup\{\infty\}\}$ can be sent via the projection on the second factor $\tilde{\pi}_i:O\rightarrow \C_i^*\cup\{\infty\}$, turning $(O,p^*_i\tilde{\pi}_i)$ into a complex chart valid away from the point $y=0$. Now notice that,
\begin{equation}
s\circ(p^*_i\tilde{\pi}_i)(\zeta)=\frac{1}{\sqrt{1-\alpha_i^2}}\left(\zeta+\frac1{\zeta}\right), \: \zeta\in \C,
\end{equation}
which looks like the expression for the Joukowsky coordinates. However there is one difference, in section \ref{subsec:Joukowsky}, the $z$ variables live on $\hat{\C}\setminus \mathbb{U}$, while the $\zeta$ variables live on $\C$.  This can be solved by introducing a parametrization of $\C$ by variables in $\hat{\C}$, just set $\zeta(z)=\frac1{z},\: z\in \hat{\C}^*$. One ends up with the Joukowsky variables,
\begin{equation}
x^{(i)}(z):=(p^*_i\tilde{\pi}_i)(\zeta(z))=\frac{1}{\sqrt{1-\alpha_i^2}}\left(z+\frac1{z}\right), \: z\in  \hat{\C}^*.
\end{equation} 
Then for each component $\mathcal{C}_i$ we can construct a set of Joukowsky coordinates. This ultimately justifies their introduction in section \ref{subsec:Joukowsky}.
\begin{remark}
We often call just $x(z)$ the collection of local coordinates $x^{(i)}(z)$. So if $x(z)$ describes the position of a point in the connected component $\mathcal{C}_i$, then $x(z)=x^{(i)}(z)$.
\end{remark}

\begin{remark}
In the next sections we need to integrate around cuts $\Gamma_c$ living in each copies of $\hat{\C}$. We denote these cuts $\Gamma_c$ whatever variables we choose to use ($x$ variables or $z$ variables), but the reader has to be aware that when using $z$ variables, $\Gamma_c$ means the pre-image $x^{(-1)}(\Gamma_c)$.  
\end{remark}

\subsection{Global loop equations} 

To define a spectral curve we also need a bi-differential. A globally defined bi-differential can be set as follows,
\begin{equation}\label{eq:bi-diff}
\omega_{2}^0(z_1,z_2):=\sum_{i,j =1}^d \mathbf{1}_{\C_i}(z_1)\mathbf{1}_{\C_j}(z_2)\omega_{\ve_i+\ve_j}^0(z_1,z_2).
\end{equation} 
The spectral curve is then given by the triplet $\mathcal{S}=(\mathcal{C}, \omega_1^0(z), \omega_{2}^0(z_1,z_2))$, where
\begin{itemize}
\item $\mathcal{C}$ is the curve defined in \eqref{PolynomialSpectralCurve} as the locus of zeroes of the polynomial $f(x,y)$ in $\bigcup_{i=1}^d\hat{\C}_i^2$,
\item $\omega_1^0(z) = W_1^0(x(z))dx(z) = \sum_{c=1}^d \mathbf{1}_{\C_c}(z) \omega_{\ve_c}(z)$, with $\omega_{\ve_c}(z)$ defined in \eqref{DiffDiscFunction},
\item $\omega_2^0(z_1,z_2)$ is defined in Equation \eqref{eq:bi-diff}.
\end{itemize}
Moreover the local involutions $\iota_i$ are identified with $\iota$ sending $z\mapsto\frac1{z}$, \textit{i.e.} we generally forget the index $i$ since all the $\iota_i$ have the same expression in local $z$ coordinates.

\smallskip

It is possible to rewrite the colored loop equations \eqref{FullLoopEq} as the local expressions for global loop equations on the spectral curve. Indeed, instead of specifying on which copy of $\C$ a variable $x$ or $z$  lives on, one can define the correlation functions on the spectral curve, as envisioned in Section \ref{sec:globalobservablesnotations},
\begin{equation}\label{eq:globalWandw}
\begin{gathered}
W_n(x_1,\ldots,x_n)=\sum_{i_1,\ldots, i_n =1}^d \prod_{j=1}^n\mathbf{1}_{\hat{\C}_{i_j}\setminus[a_{i_j},b_{i_j}]}(x_j) \ W_{\sum_{k=1}^n\ve_{i_k}}(x_1,\ldots,x_n),\\
\omega_n(z_1,\ldots,z_n)=\sum_{i_1,\ldots, i_n =1}^d \prod_{j=1}^n\mathbf{1}_{\C_{i_j}}(z_j)\ \omega_{\sum_{k=1}^n\ve_{i_k}}(z_1,\ldots,z_n),
\end{gathered}
\end{equation}
as well as the other globally defined objects such as
\begin{equation}
V'(x)=\sum_{i=1}^d\mathbf{1}_{\hat{\C}_i}(x)V'_i(x) \qquad \text{and} \qquad
P_n(x;x_1,\ldots,x_n)=\sum_{k,i_1,\ldots,i_n=1}^d\mathbf{1}_{\hat{\C}_k}(x)\prod_{j=1}^n\mathbf{1}_{\hat{\C}_{i_j}}(x_j) \ P_{\ve_k;\vn}(x;x_1,\ldots,x_n).
\end{equation}
We also need to define the multi-matrix potential in a global fashion. We first define the function
\begin{equation}
\mathbf{1}_{(x,y)} = \sum_{i=1, \dotsc, d} \mathbf{1}_{\C_i}(x) \mathbf{1}_{\C_i}(y) = \begin{cases}  1 & \: \text{if there exists $i\in\{1, \dotsc, d\}$ such that $x,y \in \hat{\C}_i$} \\
 0 & \text{otherwise.} \end{cases}
\end{equation}
Let $x_1,\ldots, x_d\in \bigcup_{i=1}^d \hat{\C}_i$, then  
\begin{equation}
T(x_1,\ldots, x_d)=\prod_{i=1}^d\mathbf{1}_{\hat{\C}_i}(x_i)\sum_{\substack{a_1,\ldots,a_d \ge 0 \\ \sum_i a_i \ge 1}}t_{a_1\ldots a_d} \prod_{j=1}^dx_j^{a_{j}}.
\end{equation}
Notice that $T(x_1,x_2,\ldots,x_d)$ is a formal series in $1/N$. We need to define several families of polynomials. We construct from all bijective maps $\pi_i:\{x_2,x_3,\ldots,x_d\}\rightarrow \{x_1,\ldots,x_{\hat{i}},\ldots ,x_d\}$ for each $i \in \{1, \dotsc, d\}$
\begin{equation}\label{eq:localmultitracepolynom}
\mathcal{U}_{i,d}(x;\zeta;x_2,\ldots,x_d)= 
\frac{1}{(d-1)!} \sum_{\pi_i} \frac{\partial_{x_i} T(\pi_i^{-1}(x_1),\dotsc,x,\dotsc,\pi_i^{-1}(x_d))-\partial_{x_i} T(\pi_i^{-1}(x_1),\dotsc,\zeta,\dotsc,\pi_i^{-1}(x_d))}{x-\zeta},
\end{equation}
where $x$ and $\zeta$ appear in the above numerator as the $i$-th arguments of $T$. We also introduce
\begin{equation}\label{eq:globalmultitracepolynom}
\mathcal{U}_d(x;\zeta;x_2,\ldots,x_d)=\sum_{i=1}^d\mathbf{1}_{\hat{\C}_i}(x) \mathcal{U}_{i,d}(x;\zeta;x_2,\ldots,x_d).
\end{equation}

We denote $\Gamma_i=[a_i,b_i]$ and $\Gamma=\bigcup_{i=1}^d \Gamma_i$. We use the notation $I=\{1, \dotsc, n\}$. We claim that, 
\begin{multline}\label{eq:GlobalLoopEq}
\sum_{J\subseteq I}W_{|J|+1}(x_0,x_J)W_{n+1-|J|}(x_0,x_{I\setminus J}) + W_{n+2}(x_0,x_0,x_I) - NV'(x_0)W_{n+1}(x_0,x_1,\ldots,x_n)\\ 
+ NP_n(x_0; x_1,\ldots,x_n) + \sum_{i\in I}\partial_{x_i}\left( \mathbf{1}_{(x_0,x_i)}\frac{W_{n}(x_0,x_{I\setminus \{i\}})-W_{n}(x_I)}{x-x_i}\right) \\
+N^{2-d}\sum_{\substack{K \vdash \{1, \dotsc, d\} \\ J_1\sqcup J_2 \sqcup \ldots \sqcup J_{[K]} = I}}\oint_{\Gamma^d}\left[ \prod_{j=1}^d \frac{d\zeta_j}{2i\pi}\right]\left(\sum_{q=1}^d\partial_{\zeta_q}T(\zeta_1,\ldots,\zeta_d)\frac{\mathbf{1}_{(x_0,\zeta_q)}}{x_0-\zeta_q}\right)\prod_{p=1}^{[K]}W_{|K_p|+|J_p|}(\zeta_{K_p},x_{J_p})\\
-N^{2-d}\sum_{\substack{K \vdash \{1, \dotsc, d\} \\ J_1\sqcup J_2 \sqcup \ldots \sqcup J_{[K]} = I}}\oint_{\Gamma^d}\left[\prod_{j=1}^d\frac{d\zeta_j}{2i\pi}\right]\mathcal{U}_d(x_0;\zeta_1;\zeta_2,\ldots,\zeta_d)\prod_{p=1}^{[K]}W_{|K_p|+|J_p|}(\zeta_{K_p},x_{J_p})=0, 
\end{multline} 
is the global expression of the colored loop equations of \eqref{FullLoopEq}, so to say they are valid for all $(x_0,\ldots,x_n)\in \left(\bigcup_{i=1}^d\hat{\C}_i\setminus\Gamma_i\right)^{n+1}$ and one recovers the colored equations by assigning one component of $\bigcup_{i=1}^d\hat{\C}_i\setminus\Gamma_i$ to each variable $x_0,x_1,\ldots,x_n$. 

\section{Topological recursion for the six--dimensional case} \label{sec:TR}

In this section we focus on the case $d=6$ as it is the first topological case beyond dimension $d=2$ (which corresponds to regular matrix models). The main reasons are to improve readability and simplify notations and computations. However we stress that the technical results extend to all topological cases $d=4\delta+2$, and demonstration for those cases can be obtained by using the same techniques.

\subsection{$1/N$ expansion of the loop equations}

At $d=6$, the potential has an expansion in $1/N^2$ which induces the following expansions of the colored correlation functions and differential forms
\begin{equation}
W_{\vn}(x_{\vn}) = \sum_{g\geq 0} N^{2-|\vn|-2g}\,W_{\vn}^{g}(x_{\vn})\qquad \omega_{\vn}(z_{\vn}) = \sum_{g\geq 0} N^{2-|\vn|-2g}\,\omega_{\vn}^{g}(z_{\vn}),
\end{equation}
consequently inducing a topological expansion for the globally defined correlation functions and differential forms
\begin{equation}
W_{n}(x_1,\ldots, x_n) = \sum_{g\geq 0} N^{2-n-2g}\,W_{n}^{g}(x_1,\ldots,x_n)\qquad \omega_{n}(z_1,\ldots, z_n) = \sum_{g\geq 0} N^{2-n-2g}\,\omega_{n}^{g}(z_{1},\ldots,z_n).
\end{equation}
In order to write the loop equations at fixed genus $g$, we make the $N$--dependence of the couplings explicit,
\begin{equation}
t_{a_1 \dotsb a_6} = N^{4 - 2\sum_{c=1}^6 a_c} \ \tilde{t}_{a_1 \dotsb a_6} \qquad \text{with $\tilde{t}_{a_1 \dotsb a_6} = -\frac{\prod_{c=1}^{d=6}\alpha_c^{ a_c}}{\sum_{c=1}^6 a_c} \binom{\sum_{c=1}^6 a_c}{a_1, \dotsc, a_6}$}.
\end{equation}
It is natural to define the corresponding family of multi-matrix potential functions. Again for all map $f:\{1,\ldots,6\}\rightarrow \{1,\ldots,6\}$ consider the family of variables $\{x_i\}_{i=1}^d$ such that $x_i\in \C_{f(i)}$, then
\begin{equation}
T^p(x_1,\ldots,x_6)=\begin{cases}\sum_{\substack{a_1,\ldots,a_d \ge 0 \\\sum_i a_i=p}}\tilde{t}_{a_1 \dotsb a_6}x_1^{f(1)}\ldots x_6^{f(6)} & \: \text{if and only if} \: f=id \\
0 & \: \text{Otherwise.} 
\end{cases}
\end{equation}
One also needs to write the $\mathcal{U}$ polynomial at fixed genus, define similarly to \eqref{eq:localmultitracepolynom},
\begin{multline}\label{eq:localfixedgenus}
\mathcal{U}_{i,d=6}^g(x;\zeta;x_2,\ldots,x_6)=\\ \sum_{\pi_i}
 \frac{\partial_{x_i}T^g(x_1=\pi_i^{-1}(x_1),\ldots,x_i=x,\ldots,x_6=\pi_i^{-1}(x_6))-\partial_{x_i}T^g(x_1=\pi_i^{-1}(x_1),\ldots,x_i=\zeta,\ldots,x_6=\pi_i^{-1}(x_6))}{x-\zeta},
\end{multline}
as well as,
\begin{equation}
\mathcal{U}_6^g(x;\zeta;x_2,\ldots,x_6)=\sum_{i=1}^d\mathbf{1}_{\hat{\C}_i}(x) \mathcal{U}_{i,6}^g(x;\zeta;x_2,\ldots,x_6).
\end{equation}
From these potentials one can write the $1/N$ expansion of the loop equations \eqref{eq:GlobalLoopEq}. One obtains
\begin{multline}\label{eq:GlobalLoopEqExpansion}
\sum_{\substack{J\subseteq I \\ 0\le h \le g}}W_{|J|+1}^h(x_0,x_J)W_{n+1-|J|}^{g-h}(x_0,x_{I\setminus J})+ W_{n+2}^{g-1}(x_0,x_0,x_I) - V'(x_0)W_{n+1}^g(x_0,x_1,\ldots,x_n)  +P_n^g(x_0; x_1,\ldots,x_n) \\
+ \sum_{i\in I}\partial_{x_i}\left( \mathbf{1}_{(x_0,x_i)}\frac{W_{n}^g(x_0,x_{I\setminus \{i\}})-W_{n}^g(x_I)}{x-x_i}\right) \\
+\sum_{\substack{K \vdash \{1, \dotsc, d\} \\ J_1\sqcup J_2 \sqcup \ldots \sqcup J_{[K]} = I}}\sum_{\substack{\{f_p\}_{p=1}^{[K]}\\g=h_0+\sum_{p=1}^{[K]}h_p+\frac{d+2}{2}-[K]}}\oint_{\Gamma^d}\left[ \prod_{j=1}^d \frac{d\zeta_j}{2i\pi}\right]\left(\sum_{q=1}^{d=6}\partial_{\zeta_q}T^{h_0}(\zeta_1,\ldots,\zeta_6)\frac{\mathbf{1}_{(x_0,\zeta_q)}}{x_0-\zeta_q}\right)\prod_{p=1}^{[K]}W_{|K_p|+|J_p|}^{h_p}(\zeta_{K_p},x_{J_p})\\
-\sum_{\substack{K \vdash \{1, \dotsc, d\} \\ J_1\sqcup J_2 \sqcup \ldots \sqcup J_{[K]} = I}}\sum_{\substack{\{f_p\}_{p=1}^{[K]}\\g=h_0+\sum_{p=1}^{[K]}h_p+\frac{d+2}{2}-[K]}}\oint_{\Gamma^d}\left[ \prod_{j=1}^d \frac{d\zeta_j}{2i\pi}\right]\mathcal{U}^{h_0}_6(x_0;\zeta_1;\zeta_2,\ldots,\zeta_6)\prod_{p=1}^{[K]}W_{|K_p|+|J_p|}^{h_p}(\zeta_{K_p},x_{J_p})=0.
\end{multline}
In this equation we did not replace $d$ by its value $6$ in the constraints of the sums appearing in the two last lines. This should allow the reader to track more easily the origin of the constraint. If we replace $d$ by its value $6$ then the term $(d+2)/2=4$, which the reason why $4$ appear in the constraint of the equation below.
The $1/N$ expansion of the global loop equations can be localized for any assignments of connected components to variables $\{x_0,x_1,\ldots,x_n\}$, leading to the following $1/N$ expanded loop equations 
\begin{multline} \label{LoopEqGenus}
\sum_{0\le h\le g, \vk} W_{\ve_i+\vk}^{h}(x_0,x_{\vk}) W_{\ve_i + \vn - \vk}^{g-h}(x_0,x_{\vn - \vk}) + W_{2\ve_i+\vn}^{g-1}(x_0,x_0,x_{\vn}) - V_i'(x_0) W_{\ve_i+\vn}^{g}(x_0,x_{\vn}) + P_{\ve_i;\vn}^{g}(x_0;x_{\vn})\\
+ \sum_{j=1}^{n_i} \frac{\partial}{\partial x^{(i)}_{j}} \biggl(\frac{W_{\vn}^{g}(x_{\vn}) - W_{\vn}^{g}(x_{\vn}\setminus \{x^{(i)}_j\}\cup \{x_0\})}{x^{(i)}_j - x_0}\biggr) + \Sigma_{\ve_i;\vn}^{g}(x_0; x_{\vn}) = 0, \: x_0\in \hat{\C}_{i}\setminus \Gamma_i
\end{multline}
with  
\begin{multline}\label{Sigma}
\Sigma_{\ve_i;\vn}^{g}(x_0; x_{\vn}) = - \sum_{\Lambda \vdash_1\! \! \mathop{\sum}\limits_{\substack{c\neq i \\ 1\le c\le d}} \! \! \ve_c} \ \sum_{\substack{(\vmu_1,\dotsc,\vmu_{|\Lambda|}) \\ \sum_{j=1}^{|\Lambda|} \vmu_j = \vn}} \ \sum_{\substack{(h_1, \dotsc, h_{|\Lambda|}) \\ \sum_{j=1}^{|\Lambda|} h_j = g + |\Lambda| - 4 - \sum_{c=1}^6 a_c}} \sum_{\substack{a_i \geq 1\\ a_{j\neq i}\geq0}}
a_i \tilde{t}_{a_1 \dotsb a_d}\\
 \bigl[\prod_{c\neq i} \tilde{x}_c^{-a_c-1}\bigr] \prod_{j=2}^{|\Lambda|} W_{\vlam_j + \vmu_j}^{h_j}(x_{\vmu_j},\tilde{x}_{\vlam_j}) \Bigl(x_0^{a_i-1} W_{\ve_i + \vlam_1 + \vmu_1}^{h_1}(x_0, x_{\vmu_1}, \tilde{x}_{\vlam_1}) - U^{(a_i)h_1}_{\ve_i;\vlam_1 + \vmu_1}(x_0, x_{\vmu_1}, \tilde{x}_{\vlam_1}) \Bigr).
\end{multline}
$\Sigma_{\ve_i;\vn}^{g}(x_0; x_{\vn})$ is the contribution coming the multi--trace potential. The constraints on the sums which define it play a crucial role. For each term of the potential characterized by $a_1, \dotsc, a_6$, each part of $\Lambda = (\vlam_1, \{\vlam_2, \dotsc, \vlam_{|\Lambda|}\})$ with $\vlam_j \neq0$ for $j=2, \dotsc, |\Lambda|$ and $|\Lambda| \leq 6$ receives a vector $\vmu_j$ and an integer $h_j$, both possibly vanishing, and such that $\sum_{j=1}^{|\Lambda|} \vmu_j = \vn$ and $\sum_{j=1}^{|\Lambda|} h_j = g + |\Lambda| - 4 - \sum_{c=1}^6 a_c$. The correspondence between the terms of equations \eqref{LoopEqGenus} and \eqref{eq:GlobalLoopEqExpansion} is quite straightforward unless maybe for the $\Sigma_{\ve_i;\vn}^{g}(x_0; x_{\vn})$ term. This term of equation \eqref{LoopEqGenus} corresponds to the two last lines of \eqref{eq:GlobalLoopEqExpansion}. \\

This form of the expansion is particularly useful in the study of the linear loop equations that are local by nature. 

\subsection{Linear loop equations} \label{sec:LLE}

We define the following operators, $\Delta$ and $S$, \textit{via} their local expressions for $x\in U_{\Gamma_i}$, a neighborhood of the cut $\Gamma_i$ for some $i\in\{1, \dotsc, d\}$,
\begin{equation} \label{Sx}
\Delta f(x)=f(x+i0) - f(x-i0), \qquad S f(x)=f(x+i0)+f(x-i0). 
\end{equation}
They satisfy the relations
\begin{align} \label{Polarization}
&\Delta (f g)(x) = \frac12 \bigl(\Delta f(x) S g(x) + S f(x) \Delta g(x)\bigr)\\
&S (f g)(x) = \frac12 \bigl(S f(x) S g(x) + \Delta f(x) \Delta g(x)\bigr).
\end{align}
We also define similar operators to be used on differential forms. Let $\lambda$ be a $1$-form on the spectral curve then
\begin{equation} \label{Sz}
\Delta \lambda(z)=(\lambda-\iota^*\lambda)(z), \qquad S \lambda(z)= (\lambda+\iota^*\lambda)(z).
\end{equation}
Note that the operators of \eqref{Sx} and \eqref{Sz} only match (via the Joukowsky transformation) on the cuts.

\subsubsection{Linear loop equations for planar disc functions}

Let us start with the simplest case of linear loop equation, the one for the planar resolvent. It is obtained by applying the operator $\Delta$ on the equation \eqref{LargeNLoopEqW1}, or by applying the operator $S$ directly on the solution \eqref{LargeNResolvent}. One gets
\begin{equation} \label{eq:localLLEdisc}
S W_{\ve_c}^0 (x) - (1-\alpha_c^2) x = 0,
\end{equation}
for $x$ on the (interior of the) cut $\Gamma_c$, as usual for a GUE.

It is easily turned into an equation on the global resolvent $W_1^0(x)$ for $x\in \Gamma = \cup_{c=1}^d \Gamma_c$, by defining
\begin{equation}
\tilde{V}(x) = \sum_{c=1}^d \mathbf{1}_{\C_c}(x)\ (1-\alpha_c^2) x
\end{equation}
so that
\begin{equation}
S W_1^0(x) - \tilde{V}'(x) = 0.
\end{equation}

As explained in \cite{BlobbedTR} by Borot, one can continue analytically this equation by going to the variable $z$ through the Joukowsky transformation and using the operator $S$ of \eqref{Sz}. From the invariance of the maps $x^{(c)}(z)$ under $\iota(z)=1/z$ one deduces that 
\begin{equation}
S \omega_{\ve_c}^0(z) = SW_{\ve_c}^0(x^{(c)}(z)) dx^{(c)}(z) = (1-\alpha_c^2) x^{(c)}(z) dx^{(c)}(z)
\end{equation} 
which now holds everywhere on the component $\cC_c$ of the spectral curve of color $c$. It shows that $S\omega_{\ve_c}^0(z)$ is holomorphic and has a simple zero at $\pm 1$.


\subsubsection{Linear loop equations for planar cylinder functions}

One can similarly compute the linear loop equation for the cylinder functions, by applying $\Delta$ to \eqref{LeadingMono2Pt} and \eqref{LeadingBicolored2Pt}. For \eqref{LeadingMono2Pt} one gets
\begin{multline}
\Delta_{x_1} W_{\ve_i}^{0}(x_1) S_{x_1} W_{2\ve_i}^{0}(x_1,x_2)+ S_{x_1} W_{\ve_i}^{0}(x_1)\Delta_{x_1} W_{2\ve_i}^{0}(x_1,x_2) - (1-\alpha_i^2) x_1\Delta_{x_1} W_{2\ve_i}^{0}(x_1,x_2) +  \frac{\Delta_{x_1} W_{\ve_i}^{0}(x_1)}{(x_1 - x_2)^2} \\ + \sum_{j\in\{1, \dotsc, d\}, \: j\neq i}\alpha_i\alpha_j [\tilde{x}_j^{-2}] W_{\ve_j + \ve_i}^{0}(\tilde{x}_j, x_2) \Delta_{x_1} W_{\ve_i}^{0}(x_1) = 0.
\end{multline}
Gathering the factors of both $\Delta_{x_1} W_{\ve_i}^{0}(x_1)$ and $\Delta_{x_1} W_{2\ve_i}^{0}(x_1,x_2)$ we get
\begin{multline}
\Delta_{x_1} W_{\ve_i}^{0}(x_1)\Bigl(S_{x_1} W_{2\ve_i}^{0}(x_1,x_2) + \frac{1}{(x_1 - x_2)^2}+ \sum_{j\in\{1, \dotsc, d\}, \: j\neq i}\alpha_i\alpha_j [\tilde{x}_j^{-2}] W_{\ve_j + \ve_i}^{0}(\tilde{x}_j, x_2)\Bigr)\\
+\Delta_{x_1} W_{2\ve_i}^{0}(x_1,x_2)\left(S_{x_1} W_{\ve_i}^{0}(x_1)- (1-\alpha_i^2) x_1\right)=0.
\end{multline}
The factor of $\Delta_{x_1} W_{2\ve_i}^{0}(x_1,x_2)$ is actually the linear loop equation \eqref{eq:localLLEdisc} on $W_{\ve_i}^0(x_1)$. This implies that 
\begin{equation}\label{eq:localLLEmonocyl}
S_{x_1} W_{2\ve_i}^{0}(x_1,x_2)+\frac{1}{(x_1 - x_2)^2}+ \sum_{j\in\{1, \dotsc, d\}, \: j\neq i}\alpha_i\alpha_j [\tilde{x}_j^{-2}] W_{\ve_j + \ve_i}^{0}(\tilde{x}_j, x_2)=0, 
\end{equation}
for $x_1$ on the interior of the cut $\Gamma_i$ of color $i$ and $x_2 \in \C\setminus \Gamma_i$.

Similarly applying $\Delta$ to \eqref{LeadingBicolored2Pt} gives
\begin{equation}\label{eq:localLLEbicyl}
S_{x_1} W^0_{\ve_i+\ve_j}(x_1,x_2) + \sum_{k\in\{1, \dotsc, d\}, k\neq i}\alpha_i\alpha_k[\tilde{x}_k^{-2}]W^0_{\ve_k+\ve_j}(\tilde{x}_k^{-2},x_2)=0, 
\end{equation} 
for $x_1 \in \Gamma_i$ and $x_2 \in \C_j\setminus \Gamma_j$ with $i\neq j$.

We can rewrite both \eqref{eq:localLLEmonocyl} and \eqref{eq:localLLEbicyl} as a single equation on the function $W_2^0(x_1, x_2)$, \eqref{GlobalW2}. To this aim, we introduce
\begin{equation}
q(x_1,x_2) = x_1 x_2 \sum_{\substack{i, j = 1, \dotsc, d\\ i\neq j}} \mathbf{1}_{\C_i}(x_1) \mathbf{1}_{\C_j}(x_2),
\end{equation}
which vanishes when $x_1$ and $x_2$ are both on a copy of $\C$ of the same color, so that
\begin{equation}
S_{x_1} W_2^0(x_1, x_2) + \frac{\mathbf{1}_{(x_1, x_2)}}{(x_1 - x_2)^2} + \oint_\Gamma \frac{dy}{2i\pi}\ \partial_{x_1} q(x_1, y)\,W_2^0(y, x_2) = 0,
\end{equation}
for $x_1$ in the interior of $\Gamma$ and $x_2 \in \cup_{c=1}^d \C_c\setminus \Gamma_c$.

As in the case of $W_1^0$, the above equations can be analytically continued using Joukowsky's transformation. This way, the differential form
\begin{equation}
\omega_2^0(z_1, z_2) = \biggl(W_2^0(x(z_1), x(z_2)) + \frac{1}{(x(z_1) - x(z_2))^2}\biggr)\, dx(z_1) dx(z_2)
\end{equation}
satisfies the linear loop equation
\begin{equation} \label{LinearLoopOmega2}
S_{z_1}\omega_2^0(z_1, z_2) + \cO_{z_1} \omega_2^0(z_1, z_2) = \frac{dx(z_1) dx(z_2)}{(x(z_1) - x(z_2))^2},
\end{equation}
with the operator $\cO$ acts on any form on the spectral curve like
\begin{equation}
\cO_{z_1} \omega(z_1, \dotsc, z_n) = \oint_\Gamma d_{z_1} q(x(z_1), x(z))\,\omega(z, \dotsc, z_n),
\end{equation}
the integral being perform with respect to $z$. Therefore, $\omega_2^0$ satisfies an equation exactly of the same form as in \cite{BlobbedTR}. This shows that it is a Cauchy kernel. If a form $\omega$ satisfies
\begin{equation}
(S_{z} + \cO_{z})\, \omega(z) + \chi(z) = 0
\end{equation}
everywhere expect possibly around the origin on each component of the spectral curve, for some holomorphic $\chi$, then a solution for $\omega(z)$ can be given in term of $\omega_2^0, \chi(z)$ and the knowledge of $\Delta \omega(z)$ close to the points $z=\pm1$. Moreover, this solution is unique if $\omega$ has a perturbative expansion in the parameters $\alpha_1, \dotsc, \alpha_d$. We refer to \cite{BlobbedTR} for the full set of details.

\subsubsection{Generic case}\label{sec:LLEGenericcase}

We now show that indeed the correlation functions satisfy
\begin{equation} \label{eq:GlobalLLEforms}
(S_{z_1} + \cO_{z_1})\, \omega_n^g(z_1, \dotsc, z_n) + l_n^{'g}(z_1; z_2, \dotsc, z_n) = 0
\end{equation}
where $l_n^{'g}(z_1; z_2, \dotsc, z_n)$ is holomorphic on the spectral curve except around the origin of each component.

Proving that equation can be done by proving it component-wise, which can be done by applying the operator $\Delta$ on the generic loop equation for $W_{\vn}^g(x_{\vn})$. Let us prove by induction on $(g,|\vn|)$ that
\begin{multline} \label{ExplicitLinearLoop}
S_{x_0} W_{\ve_i + \vn}^{g}(x_0, x_{\vn}) - \sum_{\substack{ \{\vlam_2, \dotsc, \vlam_{|\Lambda|}\} \\ \sum_{j=2}^{|\Lambda|} \vlam_j = \sum_{c=2}^6 \ve_c }} \sum_{\substack{ (\vmu_2, \dotsc, \vmu_{|\Lambda|}) \\ \sum_{j=2}^{|\Lambda|} \vmu_j = \vn }} \sum_{\substack{ a_i \geq 1, a_{j\neq i} \geq 0 \\ \sum_{c=1}^6 a_c \geq 2}} \sum_{\substack{ (h_2, \dotsc, h_{|\Lambda|})\\ \sum_{j=2}^{|\Lambda|} h_j = g + |\Lambda| - 4 - \sum_{c=1}^6 a_c}}\\
a_i \tilde{t}_{a_1 \dotsb a_6} \bigl[\prod_{c\in\{1, \dotsc, 6\}\setminus\{i\}} \tilde{x}_c^{-a_c-1} \bigr] \prod_{j=2}^{|\Lambda|} W_{\vmu_j + \vlam_j}^{h_j}(x_{\vmu_j}, \tilde{x}_{\vlam_j}) x_0^{a_i - 1} = 0, \quad x_0\in U_{\Gamma_i}.
\end{multline}
To do so, we apply $\Delta$ to the loop equation \eqref{LoopEqGenus} along the cut and find
\begin{multline}
\Delta_{x_0}  W_{\ve_i + \vn}^{g}(x_0, x_{\vn}) \Bigl(S_{x_0} W_{\ve_i}^{0}(x_0) - (1-\alpha_i^2) x_0\Bigr) + \Delta_{x_0} W_{2\ve_i + \vn}^{g-1}(x_0, x_0, x_{\vn}) + \Delta_{x_0} \Sigma_{\ve_i ; \vn}^{g}(x_0 ; x_{\vn}) \\
+ \sum_{\substack{h, \vk\\ (h, \vk) \neq (g, \vn)}} \Delta_{x_0} W_{\ve_i + \vk}^{h}(x_0, x_{\vk}) \Bigl( S_{x_0} W_{\ve_i + \vn - \vk}^{g-h}(x_0, x_{\vn - \vk}) + \sum_{j=1}^{n_i} \frac{\delta_{h,0} \delta_{\vk, \vn - \ve_i}}{(x_j^{(i)} - x_0)^2} \Bigr) = 0.
\end{multline}
One recognizes on the first line the linear loop equation for the disc function at genus zero, $S_{x_0} W_{\ve_i}^{0}(x_0) - (1-\alpha_i^2) x_0 = 0$, provided the integral term vanishes for the specific chosen form of $V_i(x_0)$. We also rewrite $\Delta_{x_0} W_{2\ve_i + \vn}^{g-1}(x_0, x_0, x_{\vn}) = \lim_{x_0' \to x_0} \Delta_{x_0} S_{x_0'}  W_{2\ve_i + \vn}^{g-1}(x_0, x_0', x_{\vn})$ and get
\begin{multline} \label{LinearIntermediate}
\lim_{x_0' \to x_0} \Delta_{x_0} S_{x_0'}  W_{2\ve_i + \vn}^{g-1}(x_0, x_0', x_{\vn}) + \Delta_{x_0} \Sigma_{\ve_i ; \vn}^{g}(x_0 ; x_{\vn}) \\
+ \sum_{\substack{h, \vk\\ (h, \vk) \neq (g, \vn)}} \Delta_{x_0} W_{\ve_i + \vk}^{h}(x_0, x_{\vk}) \Bigl( S_{x_0} W_{\ve_i + \vn - \vk}^{g-h}(x_0, x_{\vn - \vk}) + \sum_{j=1}^{n_i} \frac{\delta_{h,0} \delta_{\vk, \vn - \ve_i}}{(x_j^{(i)} - x_0)^2} \Bigr) = 0.
\end{multline}
Then one has to investigate the form of the term $\Delta_{x_0} \Sigma_{\ve_i ; \vn}^{g}(x_0 ; x_{\vn})$. In the sums which define $\Sigma_{\ve_i ; \vn}^{g}(x_0 ; x_{\vn})$ in \eqref{Sigma}, all $\vlam_j$ are non--zero except possibly $\vlam_1$. We separate the terms according to $\vlam_1=0$ or $\vlam_1 \neq 0$ in the following way,
\begin{equation}\label{eq:splitting}
\Delta_{x_0} \Sigma_{\ve_i ; \vn}^{g}(x_0 ; x_{\vn}) = \sum_{h_1, \vmu_1} L_{\ve_i ;\vn - \vmu_1}^{g - h_1}(x_0 ; x_{\vn-\vmu_1}) \Delta_{x_0} W_{\ve_i + \vmu_1}^{h_1}(x_0, x_{\vmu_1}) + \lim_{x'_0 \to x_0} \Delta_{x_0} M_{\ve_i ; \vn}^{g}(x'_0; x_0, x_{\vn})
\end{equation}
with $L_{\ve_i ; \vn - \vmu}^{g - h}(x_0 ; x_{\vn-\vmu})$ taking into account the terms for which $\vlam_1 = 0$, and for fixed $\vmu_1, h_1$,
\begin{multline}\label{eq:Lgn}
L_{\ve_i ; \vn - \vmu_1}^{g - h_1}(x_0 ; x_{\vn-\vmu}) = \sum_{\substack{ \{\vlam_2, \dotsc, \vlam_{|\Lambda|}\} \\ \sum_{j=2}^{|\Lambda|} \vlam_j = \kern-3mm \mathop{\sum}\limits_{c\in\{1, \dotsc, 6\}\setminus \{i\}} \kern-3mm \ve_c }} \ \sum_{\substack{ (\vmu_2, \dotsc, \vmu_{|\Lambda|}) \\ \sum_{j=2}^{|\Lambda|} \vmu_j = \vn - \vmu_1}} \ \sum_{\substack{ (h_2, \dotsc, h_{|\Lambda|}) \\ \sum_{j=2}^{|\Lambda|} h_j = g - h_1 + |\Lambda| - 4 - \sum_{c=1}^6 a_c}} \ \sum_{ \substack{ a_1 \geq 1, a_2, \dotsc, a_6 \geq 0 \\ \sum_{c=1}^6 a_c \geq 2}}\\
a_i \tilde{t}_{a_1 \dotsb a_6} x_0^{a_i - 1} [\prod_{c\neq i} \tilde{x}_c^{-a_c - 1}] \prod_{j=2}^{|\Lambda|} W_{\vmu_j + \vlam_j}^{h_j}(x_{\vmu_j}, \tilde{x}_{\vlam_j})
\end{multline}
and $M_{\ve_i ; \vn}^{g}(x'_0 ; x_0, x_{\vn})$ being the part of $\Sigma_{\ve_i ; \vn}^{g}(x_0 ; x_{\vn})$ with $\vlam_1 \neq 0$ (the terms polynomial in $x_0$ are dropped), with an additional distinction between the variables $x'_0$ and $x_0$, 
\begin{multline}
M_{\ve_i ; \vn}^{g}(x'_0 ; x_0, x_{\vn}) = \sum_{\substack{ \vlam_1 \neq 0, \{\vlam_2, \dotsc, \vlam_{|\Lambda|}\} \\ \sum_{j=1}^{|\Lambda|} \vlam_j = \kern-3mm \mathop{\sum}\limits_{c\in\{1, \dotsc, 6\}\setminus \{i\}} \kern-3mm \ve_c }} \ \sum_{\substack{ (\vmu_1, \dotsc, \vmu_{|\Lambda|}) \\ \sum_{j=1}^{|\Lambda|} \vmu_j = \vn}} \ \sum_{\substack{ (h_1,h_2, \dotsc, h_{|\Lambda|}) \\ \sum_{j=1}^{|\Lambda|} h_j = g + |\Lambda| - 4 - \sum_{c=1}^6 a_c}} \ \sum_{ \substack{ a_i \geq 1, a_{j\neq i} \geq 0 \\ \sum_{c=1}^6 a_c \geq 2}}\\
a_i \tilde{t}_{a_1 \dotsb a_6} x_0'^{a_i - 1} [\prod_{c\neq i} \tilde{x}_c^{-a_c - 1}] \prod_{j=2}^{|\Lambda|} W_{\vmu_j + \vlam_j}^{h_j}(x_{\vmu_j}, \tilde{x}_{\vlam_j}) W_{\ve_i + \vmu_1 + \vlam_1}^{h_1}(x_0, x_{\vmu_1}, \tilde{x}_{\vlam_1}).
\end{multline}
Let us look at 
\begin{multline}
L_{\ve_i;\ve_i + \vn}^{g-1}(x_0' ; x_0, x_{\vn}) = \sum_{\substack{ \{\vlam_2, \dotsc, \vlam_{|\Lambda|}\} \\ \sum_{j=2}^{|\Lambda|} \vlam_j = \kern-3mm \mathop{\sum}\limits_{c\in\{1, \dotsc, 6\}\setminus \{i\}} \kern-3mm \ve_c }} \ \sum_{\substack{ (\vmu_2, \dotsc, \vmu_{|\Lambda|}) \\ \sum_{j=2}^{|\Lambda|} \vmu_j = \vn + \ve_i}} \ \sum_{\substack{ (h_2, \dotsc, h_{|\Lambda|}) \\ \sum_{j=2}^{|\Lambda|} h_j = g + |\Lambda| - 1 - 4 - \sum_{c=1}^6 a_c}} \ \sum_{ \substack{ a_i \geq 1, a_{j\neq i}\geq 0 \\ \sum_{c=1}^6 a_c \geq 2}}\\
a_i \tilde{t}_{a_1 \dotsb a_6} x_0'^{a_i - 1} [\prod_{c\neq i} \tilde{x}_c^{-a_c - 1}] \prod_{j=2}^{|\Lambda|} W_{\vmu_j + \vlam_j}^{h_j}(x_{\vmu_j}, \tilde{x}_{\vlam_j})
\end{multline}
The variable $x_0$ is in one of the sets, say $x_{\vmu_{j^*}}$ for some $j^* \in \{2, \dotsc, |\Lambda|\}$ with $\vmu_{j^*} \neq 0$ and $\vlam_{j^*} \neq 0$. We now distinguish this $j^*$ and decide to rename it $j^*=1$. The set $\{\vlam_2, \dotsc, \vlam_{|\Lambda|}\}$ becomes $(\vlam_1, \{\vlam_2, \dotsc, \vlam_{|\Lambda|-1}\})$ where all vectors are non--zero, and the list $(\vmu_2, \dotsc, \vmu_{|\Lambda|})$ becomes $(\vmu'_1, \dotsc, \vmu_{|\Lambda|-1})$ with $\vmu'_1 + \ve_i = \vmu_{j^*}$, so that $\vmu'_1 + \vmu_2 + \dotsb + \vmu_{|\Lambda|-1} = \vn$. Similarly $(h_2, \dotsc, h_{|\Lambda|})$ becomes $(h_1, \dotsc, h_{|\Lambda|-1})$. We can then replace $|\Lambda|$ with $|\Lambda| + 1$ and rename $\vmu'_1$ as $\vmu_1$ and get 
\begin{multline}
L_{\ve_i;\ve_i + \vn}^{g-1}(x_0' ; x_0, x_{\vn}) = \sum_{\substack{ \vlam_1 \neq 0, \{\vlam_2, \dotsc, \vlam_{|\Lambda|}\} \\ \sum_{j=1}^{|\Lambda|} \vlam_j = \kern-3mm \mathop{\sum}\limits_{c\in\{1, \dotsc, 6\}\setminus \{i\}} \kern-3mm \ve_c }} \ \sum_{\substack{ (\vmu_1, \dotsc, \vmu_{|\Lambda|}) \\ \sum_{j=1}^{|\Lambda|} \vmu_j = \vn}} \ \sum_{\substack{ (h_1, \dotsc, h_{|\Lambda|}) \\ \sum_{j=1}^{|\Lambda|} h_j = g + |\Lambda| - 4 - \sum_{c=1}^6 a_c}} \ \sum_{ \substack{ a_1 \geq 1, a_2, \dotsc, a_6 \geq 0 \\ \sum_{c=1}^6 a_c \geq 2}}\\
a_i \tilde{t}_{a_1 \dotsb a_6} x_0'^{a_i - 1} [\prod_{c\neq i} \tilde{x}_c^{-a_c - 1}] \prod_{j=2}^{|\Lambda|} W_{\vmu_j + \vlam_j}^{h_j}(x_{\vmu_j}, \tilde{x}_{\vlam_j})\ W_{\ve_i + \vmu_1 + \vlam_1}^{h_1}(x_0, x_{\vmu_1}, \tilde{x}_{\vlam_1})
\end{multline}
i.e.
\begin{equation}
M_{\ve_i ; \vn}^{g}(x'_0 ; x_0, x_{\vn}) = L_{\ve_i; \ve_i + \vn}^{g-1}(x_0' ; x_0, x_{\vn})
\end{equation}
and thus
\begin{equation}
\Delta_{x_0} \Sigma_{\ve_i ; \vn}^{g}(x_0 ; x_{\vn}) = \sum_{h_1, \vmu_1} L_{\ve_i ;\vn - \vmu_1}^{g - h_1}(x_0 ; x_{\vn-\vmu_1}) \Delta_{x_0} W_{\ve_i + \vmu_1}^{h_1}(x_0, x_{\vmu_1}) + \lim_{x'_0 \to x_0}\Delta_{x_0} L_{\ve_i; \ve_i + \vn}^{g-1}(x_0' ; x_0, x_{\vn}).
\end{equation}
Equation \eqref{LinearIntermediate} therefore rewrites as
\begin{multline}
\sum_{\substack{h, \vk\\ (h, \vk) \neq (g, \vn)}} \Delta_{x_0} W_{\ve_i + \vk}^{h}(x_0, x_{\vk}) \Bigl( S_{x_0} W_{\ve_i + \vn - \vk}^{g-h}(x_0, x_{\vn - \vk}) + \sum_{j=1}^{n_i} \frac{\delta_{h,0} \delta_{\vk, \vn - \ve_i}}{(x_j^{(i)} - x_0)^2} + L_{\ve_i;\vn - \vk}^{g - h}(x_0 ; x_{\vn-\vk}) \Bigr)\\
+ \lim_{x_0' \to x_0} \Delta_{x_0} \Bigl(S_{x_0'}  W_{2\ve_i + \vn}^{g-1}(x_0, x_0', x_{\vn}) + L_{\ve_i;\ve_i + \vn}^{g-1}(x_0' ; x_0, x_{\vn}) \Bigr)= 0.
\end{multline}
An induction thus shows that if $(\vn, g) \neq (\ve_i, 0)$ then
\begin{equation}
S_{x_0} W_{\ve_i + \vn}^{g}(x_0, x_{\vn}) + L_{\ve_i;\vn}^{g}(x_0; x_{\vn}) = 0
\end{equation}
which is a compact form for \eqref{ExplicitLinearLoop}.

The term $L_{\ve_i;\vn}^{g}(x_0; x_{\vn})$ contains correlation functions $W_{\vn'}^{g'}$ such that $2g' + |\vn'| \leq 2g + |\vn| + 1$, since this is a property of the loop equations themselves. Let us make fully explicit the terms for which $2g' + |\vn'| = 2g + |\vn| + 1$. Using the notations of \eqref{ExplicitLinearLoop}, we look for $k \in \{2, \dotsc, |\Lambda|\}$ such that
\begin{equation} \label{EqualityConstraint}
2 h_k + |\vmu_k| + |\vlam_k| = 2g + 1 + |\vn|.
\end{equation}
Taking a simple linear combination of the above equation and the constraint $\sum_{j=2}^{|\Lambda|} h_j = g + |\Lambda| - 4 - \sum_{c=1}^6 a_c$ from \eqref{ExplicitLinearLoop} leads to 
\begin{equation}
9 + 2\sum_{c=1}^6 a_c + 2 \sum_{\substack{j=2 \\ j \neq k}}^{|\Lambda|} h_j + |\vn| = 2|\Lambda| + |\vlam_k| + |\vmu_k|.
\end{equation}
Since $\sum_{c=1}^6 a_c \geq 2$ and $h_j \geq 0$, the left hand side is found to be greater than or equal to $13 + |\vn|$. As for the right hand side, we show that it is less than or equal to $13 + |\vn|$. Indeed, from \eqref{ExplicitLinearLoop} one has $\vlam_k + \sum_{j=2, j\neq k}^{|\Lambda|} \vlam_j = \sum_{c\neq i}\ve_c$. Recalling that $|\vlam_j| \geq 1$ one gets
\begin{equation}
5 = |\sum_{c\neq i}\ve_c| = |\vlam_k| + \sum_{j=2, j\neq k}^{|\Lambda|} |\vlam_j| \geq |\vlam_k| + (|\Lambda| - 2)
\end{equation}
and therefore $2|\Lambda| + |\vlam_k| \leq |\Lambda| + 7 \leq 13$ as $|\Lambda| \leq 6$. One concludes with the fact that $|\vmu_k| \leq |\vn|$.

The equality is then satisfied when $\sum_{c=1}^6 a_c = 2$, $h_j = 0$ for $j\neq k$ and $h_k = g$, $|\Lambda| = 6$ meaning $\vlam_j = \ve_j$ for $j = 2, \dotsc, 6$, and $\vmu_k = \vn$ and $\vmu_j = 0$ for $j\neq k$. The terms from $L_{\ve_i;\vn}^{g}(x_0; x_{\vn})$ satisfying those restrictions are
\begin{equation}
- \sum_{k\neq i} \sum_{\substack{a_i \geq 1, a_{j\neq i} \geq 0 \\ \sum_{c=1}^6 a_c = 2}} a_i \tilde{t}_{a_1 \dotsb a_6} [ \prod_{c\neq i} \tilde{x}_c^{-a_c-1}] \prod_{ c \neq i,k}^6 W_{\ve_c}^{0}(\tilde{x}_c)\ W_{\ve_k + \vn}^{g}(\tilde{x}_k, x_{\vn}) x_0^{a_1-1}.
\end{equation}
This is non--vanishing if and only if $a_i = a_k = 1$ and $a_c = 0$ for $c \neq  i,k$. We thus find
\begin{equation}\label{eq:Otermextract}
L_{\ve_i;\vn}^{g}(x_0; x_{\vn}) = \sum_{\substack{1\le c \le 6 \\ c\neq i}} \alpha_i\alpha_c [\tilde{x}_c^{-2}]W_{\ve_c + \vn}^{g}(\tilde{x}_c, x_{\vn}) + L_{\ve_i;\vn}^{'g}(x_0; x_{\vn})
\end{equation}
where $L_{\ve_i;\vn}^{'g}(x_0; x_{\vn})$ is defined as $L_{\ve_i;\vn}^{g}(x_0; x_{\vn})$ excluding the terms where the equality \eqref{EqualityConstraint} is satisfied. Moreover, we notice that the first term of the r.h.s of \eqref{eq:Otermextract} writes in term of contour integral as
\begin{equation}
\sum_{\substack{1\le c \le 6 \\ c\neq i}} \alpha_i\alpha_c [\tilde{x}_c^{-2}]W_{\ve_c + \vn}^{g}(\tilde{x}_c, x_{\vn})=\sum_{\substack{1\le c \le 6 \\ c\neq i}} \alpha_i\alpha_c\oint_{\Gamma_c}\frac{d\tilde{x}_c}{2i\pi}\tilde{x}_c W_{\ve_c + \vn}^{g}(\tilde{x}_c, x_{\vn}).
\end{equation}
The linear loop equation thus far takes the form
\begin{equation}
S_{x_0} W_{\ve_i + \vn}^{g}(x_0, x_{\vn}) +  \sum_{\substack{1\le c \le 6 \\ c\neq i}} \alpha_i\alpha_c [\tilde{x}^{-2}] W_{\ve_{c} + \vn}^{g}(\tilde{x}, x_{\vn}) + L_{\ve_i;\vn}^{'g}(x_0; x_{\vn}) = 0.
\end{equation}

This is the local expression of linear loop equations satisfied by the functions $W^g_n$ defined globally on the spectral curve. To this aim, we define 
\begin{equation}
L^{'g}_{n+1}(x_0; x_{1},\ldots,x_n)=\sum_{i_0,i_1,\ldots, i_n =1}^d \mathbf{1}_{\hat{\C}_{i_0}}(x_0) \prod_{j=1}^n\mathbf{1}_{\hat{\C}_{i_j}\setminus[a_{i_j},b_{i_j}]}(x_j) \ L_{\ve_i;\vn}^{'g}(x_0; x_{1},\ldots,x_n)
\end{equation}  
this allows us to recast the linear loop equations using only globally defined functions,
\begin{equation}
S_{x_0} W^g_{n+1}(x_0,x_1,\dotsc,x_n) + \oint_{\Gamma}\frac{dy}{2i\pi} \partial_{x_0} q(x_0, y) W^g_{n+1}(y,x_1, \dotsc, x_n) + L^{'g}_{n+1}(x_0; x_{1},\dotsc,x_n) = 0,
\end{equation}
with $x_0 \in \Gamma$ and $x_1, \dotsc, x_n \in \prod_{i=1}^{d=6}\hat{\C}_i\setminus \Gamma_i$.

Just like for $W_1^0$ and $W_2^0$, the above equation can be analytically continued to the spectral curve using the map $x(z)$. This leads directly to \eqref{eq:GlobalLLEforms}, showing that $S_{z_1} \omega_n^g(z_1, \dotsc, z_n)$ is holomorphic except in a neighborhood of the origin for each component of the spectral curve and has a simple zero at $z_1 = \pm 1$.


\subsection{Quadratic loop equations}\label{sec:QLE}

The aim of this subsection is to show that the quadratic forms $Q^g_n$ defined below have a double zero at the vanishing points of $dx$. When this statement is true, we say that the quadratic loop equations are satisfied. This is necessary for the blobbed topological recursion to be satisfied. 

\subsubsection{Quadratic forms: definition and first computation}

The quadratic forms are defined for all $(g,n)\neq (0,1),(0,2)$ as follows \cite{BlobbedTR2}   
\begin{equation}
Q^g_n(z,\iota(z),z_2,\ldots, z_n)=\omega^{g-1}_{n+1}(z,\iota(z),z_2,\ldots,z_n)+\sum_{\substack{0\le h \le g \\ I\sqcup J=\{2,\ldots, n\}}} \omega_{1+|I|}^h(z,z_I)\omega^{g-h}_{1+|J|}(\iota(z),z_J),
\end{equation}
where $z,z_2,\ldots,z_n \in \mathcal{C}$. One notices that this definition closely mimics the form of the first two terms of equation \eqref{eq:GlobalLoopEq} with the notable differences that we used $\omega_n^g$ differential forms for the definition of $Q^g_n$ instead of $W^g_n$ functions and that the forms are evaluated at $z$ and $\iota(z)$. Let us prove that for all $(g,n)\neq (0,1),(0,2),(1,1)$, the quadratic forms $Q^g_n$ have double zeroes at the zeroes of $dx$.


To study $Q_n^g$, it is sufficient to consider its local expression obtained by assigning a color to each variable,
\begin{equation}
Q_{\ve_i+\vn}^g(z,\iota(z);z_{\vn})=\omega_{2\ve_i+\vn}^{g-1}(z,\iota(z),z_{\vn})+ \sum_{0\le h\le g, \vk} \omega_{\ve_i+\vk}^{h}(z,z_{\vk}) \omega_{\ve_i + \vn - \vk}^{g-h}(\iota(z),z_{\vn - \vk}). 
\end{equation} 
The function featuring $\iota$ on the right hand side can be replaced with the same functions without $\iota$ plus additional terms, using the linear loop equations. It gives
\begin{multline}\label{eq:QgnstepA}
Q_{\ve_i+\vn}^g(z,\iota(z);z_{\vn})=-\omega_{2\ve_i+\vn}^{g-1}(z,z,z_{\vn}) -\sum_{0\le h\le g, \vk} \omega_{\ve_i+\vk}^{h}(z,z_{\vk}) \omega_{\ve_i + \vn - \vk}^{g-h}(\iota(z),z_{\vn - \vk}) \\
-\sum_{\substack{1\le c \le 6 \\ c\neq i}} dx^{(i)}(z)\alpha_i\alpha_c [\tilde{x}^{(c)}(\tilde{z})^{-2}] \omega_{\ve_i+\ve_{c} + \vn}^{g}(z,\tilde{x}^{(c)}(\tilde{z}), z_{\vn}) - l_{\ve_i;\ve_i+\vn}^{'g}(z;z, z_{\vn})\\
-\sum_{0\le h\le g, \vk} \omega_{\ve_i+\vk}^{h}(z,z_{\vk}) (\sum_{\substack{1\le c \le 6 \\ c\neq i}} dx^{(i)}(z)\alpha_i\alpha_c [\tilde{x}^{(c)}(\tilde{z})^{-2}] \omega_{\ve_{c} + \vn}^{g}(\tilde{x}^{(c)}(\tilde{z}), z_{\vn}) + l_{\ve_i;\vn}^{'g}(z; z_{\vn})).
\end{multline}
We call
\begin{equation}\label{eq:rgndef}
r^g_{\ve_i;\vn}(z;z_{\vn})=-\omega_{2\ve_i+\vn}^{g-1}(z,z,z_{\vn}) -\sum_{0\le h\le g, \vk} \omega_{\ve_i+\vk}^{h}(z,z_{\vk}) \omega_{\ve_i + \vn - \vk}^{g-h}(z,z_{\vn - \vk}),
\end{equation}
whose expression can be obtained from the local loop equations \eqref{LoopEqGenus}
\begin{multline}\label{eq:rgnresult}
r^g_{\ve_i;\vn}(z;z_{\vn})=- d_zV_i(x(z)) \omega_{\ve_i+\vn}^{g}(z,z_{\vn}) + p_{\ve_i;\vn}^{g}(z;z_{\vn})\\
+ dx(z)dx(z) \sum_{j=1}^{n_i} d_{z^{(i)}_j} \biggl(\frac{\omega_{\vn}^{g}(x_{\vn})}{dx^{(i)}_j(z^{(i)}_j)(x^{(i)}_j(z^{(i)}_j) - x(z))}\biggr) 
+ \sigma_{\ve_i;\vn}^{g}(z; z_{\vn}),
\end{multline}
where $\sigma_{\ve_i;\vn}^{g}(z; z_{\vn})$ is the differential form version of $\Sigma_{\ve_i;\vn}^{g}(x; x_{\vn})$
\begin{equation}
\sigma_{\ve_i;\vn}^{g}(z; z_{\vn})=\Sigma_{\ve_i;\vn}^{g}(x(z); x_{\vn}(z_{\vn})) dx(z)dx(z)\prod dx(z_{\vn}).
\end{equation}
We rewrite $\sigma_{\ve_i;\vn}^{g}(z; z_{\vn})$ using a splitting very similar to the one used in \eqref{eq:splitting}
\begin{equation}\label{eq:splitting2}
\sigma_{\ve_i;\vn}^{g}(z; z_{\vn})=\sum_{0\le h_1\le g, \vk}l^{g-h_1}_{\ve_i; \vn-\vk}(z;z_{\vn-\vk})\omega_{\ve_i+\vk}^{h_1}(z,z_{\vk})+l^{g-1}_{\ve_i;\ve_i+\vn}(z;z,z_{\vn})+u_{\ve_i;\vn}^g(z;z_{\vn})
\end{equation}
where $u_{\ve_i;\vn}^g(z;z_{\vn})$ comes from the last term of equation \eqref{Sigma} and writes
\begin{multline}\label{eq:ugn}
u_{\ve_i;\vn}^g(z;z_{\vn})=dx(z)dx(z)\sum_{\Lambda \vdash_1\! \! \mathop{\sum}\limits_{\substack{c\neq i \\ 1\le c\le d}} \! \! \ve_c} \ \sum_{\substack{(\vmu_1,\dotsc,\vmu_{|\Lambda|}) \\ \sum_{j=1}^{|\Lambda|} \vmu_j = \vn}} \ \sum_{\substack{(h_1, \dotsc, h_{|\Lambda|}) \\ \sum_{j=1}^{|\Lambda|} h_j = g + |\Lambda| - 4 - \sum_{c=1}^6 a_c}} \sum_{\substack{a_i \geq 1\\ a_{j\neq i}\geq0}}\\
a_i \tilde{t}_{a_1 \dotsb a_d} \bigl[\prod_{c\neq i} \tilde{x}_c^{-a_c-1}(\tilde{z}_c)\bigr] \prod_{j=2}^{|\Lambda|} \omega_{\vlam_j + \vmu_j}^{h_j}(z_{\vmu_j},\tilde{z}_{\vlam_j}) U^{(a_i)h_1}_{\ve_i;\vlam_1 + \vmu_1}(x(z), x_{\vmu_1}(z_{\vmu_1}), \tilde{x}_{\vlam_1}(\tilde{z}_{\vlam_1}))\prod dx(z_{\vn}).
\end{multline}
Combining equations \eqref{eq:QgnstepA}, \eqref{eq:rgnresult}, \eqref{eq:splitting2} and \eqref{eq:ugn} one obtains that 
\begin{equation}\label{eq:localQgnfinal}
Q_{\ve_i+\vn}^g(z,\iota(z);z_{\vn})= p_{\ve_i;\vn}^{g}(z;z_{\vn})\\
+ dx(z)dx(z) \sum_{j=1}^{n_i} d_{z^{(i)}_j} \biggl(\frac{\omega_{\vn}^{g}(z_{\vn})}{dx^{(i)}_j(z^{(i)}_j)(x^{(i)}_j(z^{(i)}_j) - x(z))}\biggr) + u_{\ve_i;\vn}^g(z;z_{\vn}).
\end{equation}
From equation \eqref{eq:localQgnfinal} we deduce that $Q_{\ve_i+\vn}^g(z,\iota(z);z_{\vn})$ has double zeroes on the zeroes of $dx$ on the connected component of color $i$ of the spectral curve. Indeed, $Q_{\ve_i+\vn}^g(z,\iota(z);z_{\vn})$ writes as $dx^{(i)}(z)dx^{(i)}(z)$ times a polynomial in $x^{(i)}(z)$ which is holomorphic around the zeroes of $dx$ at $z=\pm 1$. Since $dx$ has simple zeroes at these points, $Q_{\ve_i+\vn}^g(z,\iota(z);z_{\vn})$ has double zeroes at $z=\pm 1$.

\subsubsection{Quadratic forms: (1,1) case}

In the last computations we left aside the case $(g,n)=(1,1)$. This is because the expression for $Q^1_1$ involves $\omega_2^0(z,\iota(z))$ and $\omega_2^0(z_1,z_2)$ has a double pole on the diagonal. The calculation is the same although one needs to take care of the diagonal poles, which is done by taking a limit. The local expression is,
\begin{equation}
Q^1_{\ve_i}(z,\iota(z))= \omega_{2\ve_i}^0(z,\iota(z))+\omega_{\ve_i}^0(z)\omega_{\ve_i}^1(\iota(z))+\omega^0_{\ve_i}(\iota(z))\omega^1_{\ve_i}(z),
\end{equation}  
which by definition of $\tilde{\omega}_{2 \ve_i}^0(z_1, z_2) = \omega_{2 \ve_i}^0(z_1, z_2) - dx(z_1) dx(z_2)/( x(z_1) - x(z_2) )^2 = W_{2 \ve_i}^0(x(z_1), x(z_2)) dx(z_1) dx(z_2)$ rewrites
\begin{equation}
Q^1_{\ve_i}(z,\iota(z))=\mathop{\lim}\limits_{z^{'}\rightarrow z}\tilde{\omega}_{2\ve_i}^0(z,\iota(z^{'}))+\frac{dx(z)dx(z^ {'})}{(x(z)-x(z^{'}))^2}+ \omega_{\ve_i}^0(z)\omega_{\ve_i}^1(\iota(z))+\omega^0_{\ve_i}(\iota(z))\omega^1_{\ve_i}(z).
\end{equation}
Using the linear loop equations we get
\begin{multline}
Q^1_{\ve_i}(z,\iota(z))=\mathop{\lim}\limits_{z^{'}\rightarrow z}-\tilde{\omega}_{2\ve_i}^0(z,z^{'})-\mathcal{O}\tilde{\omega}_{2\ve_i}^0(z,z^{'})-\frac{dx(z)dx(z^{'})}{(x(z)-x(z^{'}))^2}+\frac{dx(z)dx(z^{'})}{(x(z)-x(z^{'}))^2} \\
- \omega_{\ve_i}^0(z)\omega_{\ve_i}^1(z)-\omega^0_{\ve_i}(z)\omega^1_{\ve_i}(z) -\omega_{\ve_i}^0(z)l^1_{\ve_i}(z)-\omega^1_{\ve_i}(z)l^0_{\ve_i}(z).
\end{multline}
The key point is now that $\tilde{\omega}_{2 \ve_i}^0(z_1, z_2)$ has a double pole along $z_1 = 1/z_2$ (and not along $z_1 = z_2$ like $\omega_{2 \ve_i}^0(z_1, z_2)$), so that the limit $z'\to z$ can be taken. Elementary algebra leads to
\begin{equation}
Q^1_{\ve_i}(z,\iota(z))=p_{\ve_i}^{1}(z)+ u_{\ve_i}^1(z),
\end{equation}
which has a double zero at $\pm 1$ on the connected component of color $i$ of the spectral curve.

\subsection{The blobbed topological recursion} \label{sec:BlobbedTR}

In this section, we need to use the results derived in sections \ref{sec:LLE}, \ref{sec:QLE}. In \cite{BlobbedTR2}, Borot and Shadrin introduce a method to solve general linear and quadratic loop equations recursively which generalizes the Eynard-Orantin topological recursion method. They call this new method the \emph{blobbed} topological recursion. As it turns out after having obtained linear and quadratic loop equations in the previous sections, we can now apply their method, with the novelty that we have a disconnected spectral curve. This method is sketched in the following subsections.

\subsubsection{First topological recursion formula}\label{sec:firstformula}

Assume $\{\omega^g_n\}$ are a solution of linear and quadratic loop equations. One defines their polar $P\omega^g_n$ and holomorphic part $H\omega^g_n$ by,
\begin{equation} \label{SplitPolarHol}
P\omega^g_n(z_1,\ldots,z_n)=\sum_{p\in \mathcal{C}}\underset{z\rightarrow p}{\textrm{Res}} \ \omega^g_n(z,z_2,\ldots,z_n)G(z,z_1), \qquad \omega_n^g(z_1,\ldots,z_n)=P\omega^g_n(z_1,\ldots,z_n)+H\omega^g_n(z_1,\ldots,z_n),
\end{equation}
where $p$ are the zeroes of $dx$ and where $G(z,z_1)$ is defined from $\omega^0_2$ by
\begin{equation}
G(z,z_1)=\int^z\omega^0_2(\cdot,z_1).
\end{equation}
The reason for doing so is that, in simple enough cases\footnote{Those are cases in which linear loop equations can be put into the $(S+\mathcal{O})\omega$ form with $\mathcal{O}$ a linear operator.}, $P\omega^g_n(z_1,\ldots,z_n)$ is in the kernel of the linear operator $(S+\mathcal{O})$ (see section \ref{sec:LLEGenericcase} equation \eqref{eq:GlobalLLEforms}).

Moreover there is a simple expression for $P\omega^g_n(z_1,\ldots,z_n)$. Indeed as was shown in\footnote{We repeat carefully the step of their proof here as we think it strongly justifies the previous computations. However we will not repeat the proof of all the results from their paper which we use.} \cite{BlobbedTR2}, one has
\begin{align}
P\omega^g_n(z_1,\ldots,z_n)&=\sum_{p\in \mathcal{C}}\underset{z\rightarrow p}{\textrm{Res}} \ \omega^g_n(z,z_2,\ldots,z_n)G(z,z_1) \\
&=\frac12 \sum_{p\in \mathcal{C}} \underset{z\rightarrow p}{\textrm{Res}} \ [G(z,z_1)\omega^g_n(z,z_2,\ldots,z_n)+G(\iota(z),z_1)\omega^g_n(\iota(z),z_2,\ldots,z_n)],
\end{align}
using the fact that the vanishing points of $dx$ are left invariant by $\iota$. Then formulae \eqref{Polarization} shows that
\begin{align}
P\omega^g_n(z_1,\ldots,z_n)&=\frac14 \sum_{p\in \mathcal{C}} \underset{z\rightarrow p}{\textrm{Res}}[\Delta G(z,z_1)\ \Delta \omega^g_n(z,z_2,\ldots,z_n)+SG(z,z_1)\ S\omega^g_n(z,z_2,\ldots,z_n)].
\end{align}
We showed in section \ref{sec:LLE} that $S\omega^g_n(z,z_2,\ldots,z_n)$ was holomorphic around the zeroes of $dx$. Moreover, both $G(z,z_1)$ and $SG(z,z_1)$ are holomorphic away from $z_1$, then the last term of the above expression vanishes. One ends with
\begin{equation}\label{eq:PwstepA}
P\omega^g_n(z_1,\ldots,z_n)=\frac14 \sum_{p\in \mathcal{C}} \underset{z\rightarrow p}{\textrm{Res}}\Delta G(z,z_1)\ \Delta\omega^g_n(z,z_2,\ldots,z_n),
\end{equation} 
as a result one needs to express $\Delta\omega^g_n(z,z_2,\ldots,z_n)$. This is done using the $Q^g_n$ quadratic forms. Indeed, one has
\begin{equation}
Q^g_n(z,\iota(z),z_2,\ldots z_n)=\frac12(S\omega_n^g(z,z_2,\ldots,z_n)S\omega_1^0(z)-\Delta \omega_n^g(z,z_2,\ldots,z_n)\Delta\omega^0_1(z))+ \tilde{Q}^g_n(z,\iota(z),z_2,\ldots z_n),
\end{equation}
where $\tilde{Q}^g_n(z,\iota(z),z_2,\ldots z_n)$ is
\begin{equation}
\tilde{Q}^g_n(z,\iota(z),z_2,\ldots z_n)=\omega^{g-1}_{n+1}(z,\iota(z),z_2,\ldots z_n)+\sum_{\substack{0\le h \le g \\ I\sqcup J=\{2,\ldots, n\}}} ^{'}\omega_{1+|I|}^h(z,z_I)\omega^{g-h}_{1+|J|}(\iota(z),z_J),
\end{equation}
where $\sum^{'}$ is the traditional notation in topological recursion literature meaning that the terms containing $\omega^0_1$ are excluded from the sum. The key point to notice here is that $\tilde{Q}^g_n(z,\iota(z),z_2,\ldots z_n)$ only depends on $\omega^{g'}_{n'}$ such that $2g'-2+n'<2g-2+n$. We express\footnote{The term $\frac{2}{\Delta \omega^0_1(z)}$ in the expression of $\Delta\omega^g_n(z,z_2,\ldots,z_n)$ is the ``inverse'' of a differential form. It actually means that ``$dz^{-1}$'' performs the contraction with $\partial_z$. } $\Delta\omega^g_n(z,z_2,\ldots,z_n)$ using $Q^g_n$, $\tilde{Q}^g_n$ and $S\omega_n^g(z,z_2,\ldots,z_n)S\omega_1^0(z)$,
\begin{equation}
\Delta\omega^g_n(z,z_2,\ldots,z_n)=\frac{2}{\Delta \omega^0_1(z)}\Bigl(\tilde{Q}^g_n(z,\iota(z),z_2,\ldots z_n)-Q^g_n(z,\iota(z),z_2,\ldots z_n)+\frac12 S\omega_n^g(z,z_2,\ldots,z_n)S\omega_1^0(z)\Bigr).
\end{equation}
Then making the replacement in equation \eqref{eq:PwstepA},
\begin{equation}\label{eq:PwstepB}
P\omega^g_n(z_1,\ldots,z_n)=\sum_{p\in \mathcal{C}} \underset{z\rightarrow p}{\textrm{Res}}\frac{\Delta G(z,z_1)}{2\Delta\omega^0_1(z)}\Bigl(\tilde{Q}^g_n(z,\iota(z),z_2,\ldots,z_n)-Q^g_n(z,\iota(z),z_2,\ldots,z_n)+\frac12 S\omega_n^g(z,z_2,\ldots,z_n)S\omega_1^0(z)\Bigr).
\end{equation}
Then notice that $\omega_1^0$ has a double zero when $z\rightarrow p$, while $\Delta G(z,z_1)$ has a simple zero for $z\rightarrow p$. This implies that $\frac{\Delta G(z,z_1)}{2\Delta\omega^0_1(z)}$ has a simple pole when $z\rightarrow p$. This is why we need the quadratic loop equations to be satisfied. Indeed, thanks to these equations, the combination $\frac{\Delta G(z,z_1)}{2\Delta\omega^0_1(z)}Q^g_n(z,\iota(z),z_2,\ldots,z_n)$ has a simple zero when $z\rightarrow p$ meaning that it has no residue at $p$ and thus does not contribute in equation \eqref{eq:PwstepB}. Moreover both $S\omega_n^g(z,z_2,\ldots,z_n)$ and $S\omega_1^0(z)$ have a simple zero when $z\rightarrow p$, then the combination $\frac{\Delta G(z,z_1)}{2\Delta\omega^0_1(z)}S\omega_n^g(z,z_2,\ldots,z_n)S\omega_1^0(z)$ does not contribute neither to the sum \eqref{eq:PwstepB}.
Then we have a recursive formula for $P\omega^g_n(z_1,\ldots,z_n)$
\begin{equation}\label{eq:Polarpart}
P\omega^g_n(z_1,\ldots,z_n)=\sum_{p\in \mathcal{C}} \underset{z\rightarrow p}{\textrm{Res}}\ K(z,z_1)\Bigl[\omega^{g-1}_{n+1}(z,\iota(z),z_2,\ldots z_n)+\sum_{\substack{0\le h \le g \\ I\sqcup J=\{2,\ldots, n\}}} ^{'}\omega_{1+|I|}^h(z,z_I)\omega^{g-h}_{1+|J|}(\iota(z),z_J) \Bigr],
\end{equation}
where we define the topological recursion kernel $K(z,z_1)=\frac{\Delta G(z,z_1)}{2\Delta\omega^0_1(z)}$.
The problem is to find a formula for $H\omega^g_n(z_1,\ldots,z_n)$. It is a key result of \cite{BlobbedTR} that it can be constructed as 
\begin{equation}
H\omega^g_n(z_1,\ldots,z_n)=-\frac{1}{2i\pi}\oint G(z,z_1)l^{'g}_n(z;z_2,\ldots,z_n),
\end{equation}
and if we denote $v^g_n(z;z_2,\ldots,z_n)$ the primitive of $l^{'g}_n(z;z_2,\ldots,z_n)$,
\begin{equation}\label{eq:Holomorphicpart}
H\omega^g_n(z_1,\ldots,z_n)=\frac{1}{2i\pi}\oint_{\Gamma} \omega^0_2(z,z_1)v^g_n(z;z_2,\ldots,z_n).
\end{equation}
We then have an expression for the correlation function viewed as a form $\omega^g_n$ on the spectral curve. This mimics the results presented in \cite{BlobbedTR} in the case of the one-cut Hermitian multi-trace one-matrix model. Our results generalize \cite{BlobbedTR} in two directions, as we have several Hermitian matrices and several cuts.

Another key result of \cite{BlobbedTR} ensures the uniqueness of the solution, as we are interested in perturbative solutions (which have an expansion in $\alpha_c$ around $\alpha_c=0$). In the next parts we are going to explain how to find a graphical expansion for both the polar and the holomorphic part.

\subsubsection{Graphical expansion of the $\omega^g_n$}\label{sec:graphexpansionw}

Thanks to Corollary 2.5 in \cite{BlobbedTR2}, it is possible to find a combinatorial representation of the $\omega^g_n$. This combinatorial expansion supplements the one of the Eynard-Orantin topological recursion. Indeed, it allows us to write the $\omega^g_n$ satisfying the blobbed topological recursion in terms of \emph{normalized}\footnote{The added $0$ superscript indicates that this is a normalized form.} forms $\omega^{g,0}_n$  satisfying the usual Eynard-Orantin topological recursion, and additional functions called \emph{blobs} which act as initial conditions for the recursion. 

\medskip

\begin{figure}
 \begin{center}
  \includegraphics[scale=0.8]{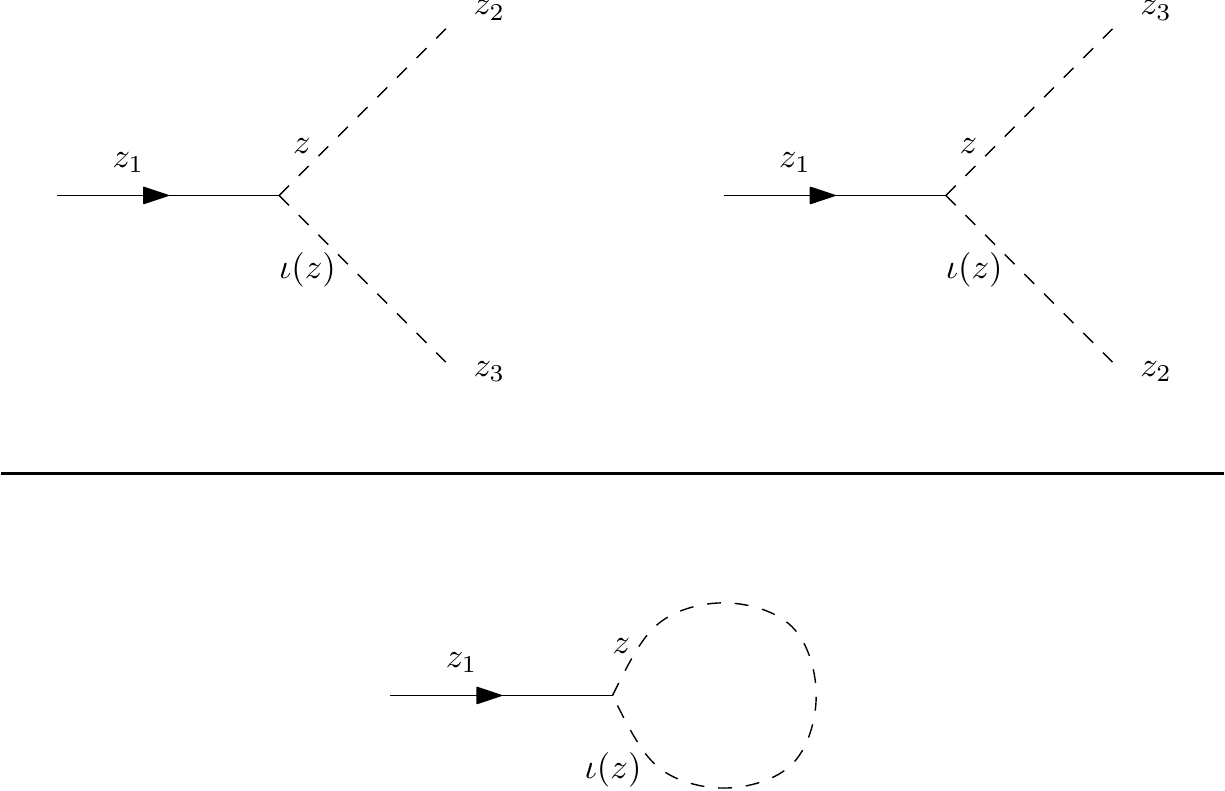}
  \caption{First half of the figure: skeleton graphs appearing in the computation of $\omega^{0,0}_3(z_1,z_2,z_3)$. Second half of the figure: skeleton graphs appearing in the computation of $\omega^{1,0}_1(z_1)$. Dashed edges represent $\omega^0_2$, trivalent vertices represent the topological recursion kernel $K$.\label{fig:skeleton}}
 \end{center}
\end{figure}

We present the objects and the results needed to apply the blobbed topological recursion in this case. The normalized forms $\omega^{g,0}_n$ are defined as the forms obtained for $2g-2+n>0$ by successively applying the Eynard-Orantin recurrence to a choice of $\omega^0_1$ and $\omega_2^0$. In our case, $\omega_1^0=W^1_0(x(z))dx(z)$ and $\omega^0_2$ is the one of equation \eqref{eq:bi-diff}. For instance\footnote{For the interested and brave reader we give here an explicit result for $\omega^{1,0}_1(z_1)=\sum_{i=1}^6\mathbf{1}_{\mathcal{C}_i}(z_1)4z_1\frac{z_1^2-\eta_i(z_1^2-1)^2}{(z_1^2-1)^4}$ where $\eta_i=\frac{\sum_{p\neq i }(\alpha_i\alpha_p)^2}{\sum_{p=1}^6\alpha_p^2-1}$. The expression for $\omega^{0,0}_3(z_1,z_2,z_3)$ can also be computed but the expression is involved and long so that we do not display it here.}, by definition we have,
\begin{align}
&\omega^{0,0}_3(z_1,z_2,z_3)=\sum_{p\in \mathcal{C}}\underset{z\rightarrow p}{\textrm{Res}}\ K(z,z_1)[\omega_2^0(z,z_2)\omega_2^0(\iota(z),z_3)+\omega_2^0(\iota(z),z_2)\omega_2^0(z,z_3)]\\
&\omega^{1,0}_1(z_1)=\sum_{p\in \mathcal{C}}\underset{z\rightarrow p}{\textrm{Res}} \ K(z,z_1)\omega^0_2(z,\iota(z)),
\end{align}  
while more generally we have
\begin{equation}
\omega^{g,0}_n(z_1,\ldots,z_n)=\sum_{p\in \mathcal{C}}\underset{z\rightarrow p}{\textrm{Res}}\ K(z,z_1)[\omega^{g-1}_{n+1}(z,\iota(z),z_2,\ldots,z_n)+\sum_{\substack{0\le h \le g \\ I\sqcup J=\{2,\ldots, n\}}} ^{'}\omega_{1+|I|}^{h,0}(z,z_I)\omega^{g-h,0}_{1+|J|}(\iota(z),z_J) ].
\end{equation}
These normalized forms appear as weights in the blobbed topological recursion graphical expansion with the \emph{blobs} denoted $\phi^g_n$, such that $\phi^g_n=H_1\ldots H_n \omega^g_n$ by definition. There is a graphical expansion for the normalized forms as well, this is the one discovered in \cite{TR} and described in terms of skeleton graphs in \cite{BlobbedTR2}. Skeleton graphs indexing the expansion of the forms $\omega^{0,0}_3(z_1,z_2,z_3)$ and $\omega^{1,0}_1(z_1)$ are displayed on Fig.\ref{fig:skeleton}.  

\medskip

In order to show how these normalized forms appear in the computation of  the $\omega^g_n$, we compute what we think is an illuminating example. Let us look at $P\omega^0_4(z_1,z_2,z_3,z_4)$, we have from equation \eqref{eq:Polarpart}
\begin{multline}
P\omega^0_4(z_1,z_2,z_3,z_4)=\sum_{p\in \mathcal{C}}\underset{z\rightarrow p}{\textrm{Res}} \ K(z,z_1)[\omega^0_2(z,z_2)\omega^0_3(\iota(z),z_3,z_4)\\
+\omega_2^0(z,z_3)\omega^0_3(\iota(z),z_2,z_4)+\omega^0_2(z,z_4)\omega^0_3(\iota(z),z_2,z_3) + z\leftrightarrow \iota(z)].
\end{multline}
In order to further compute this last quantity, we split $\omega^0_3=P\omega^0_3+H\omega^0_3$. Focusing on the expression for $P\omega^0_3$, we notice that $P\omega^0_3=\omega^{0,0}_3$ (by just comparing with the definition of $\omega^{0,0}_3$). Consequently we obtain
\begin{multline}
P\omega^0_4(z_1,z_2,z_3,z_4)=\sum_{p\in \mathcal{C}}\underset{z\rightarrow p}{\textrm{Res}} \ K(z,z_1)[\omega^0_2(z,z_2)\omega^{0,0}_3(\iota(z),z_3,z_4)
+\omega_2^0(z,z_3)\omega^{0,0}_3(\iota(z),z_2,z_4) \\
+\omega^0_2(z,z_4)\omega^{0,0}_3(\iota(z),z_2,z_3) 
+\omega^0_2(z,z_2)H\omega^0_3(\iota(z),z_3,z_4)
+\omega_2^0(z,z_3)H\omega^0_3(\iota(z),z_2,z_4)+\omega^0_2(z,z_4)H\omega^0_3(\iota(z),z_2,z_3)\\
 + z\leftrightarrow \iota(z)].
\end{multline}
In the first line we recognize the expression for $\omega^{0,0}_4(z_1,z_2,z_3,z_4)$, then
\begin{multline}
P\omega^0_4(z_1,z_2,z_3,z_4)=\omega_4^{0,0}(z_1,z_2,z_3,z_4)+\sum_{p\in \mathcal{C}}\underset{z\rightarrow p}{\textrm{Res}} \ K(z,z_1) \Bigl[\omega^0_2(z,z_2)H\omega^0_3(\iota(z),z_3,z_4) \\
+\omega_2^0(z,z_3)H\omega^0_3(\iota(z),z_2,z_4)+\omega^0_2(z,z_4)H\omega^0_3(\iota(z),z_2,z_3) + z\leftrightarrow \iota(z)\Bigr].
\end{multline}
Using Cauchy formula, 
\begin{multline}
P\omega^0_4(z_1,z_2,z_3,z_4)=\omega_4^{0,0}(z_1,z_2,z_3,z_4)+\sum_{p\in \mathcal{C}}\underset{z\rightarrow p}{\textrm{Res}} \ K(z,z_1)[\omega^0_2(z,z_2)\underset{z'\rightarrow \iota(z)}{\textrm{Res}}\omega^0_2(\iota(z),z')\int^{z'}\kern-2mm H\omega^0_3(\cdot,z_3,z_4) \\
+\omega^0_2(z,z_3)\underset{z'\rightarrow \iota(z)}{\textrm{Res}}\omega^0_2(\iota(z),z')\int^{z'}\kern-2mm H\omega^0_3(\cdot,z_2,z_4) +\omega^0_2(z,z_4)\underset{z'\rightarrow \iota(z)}{\textrm{Res}}\omega^0_2(\iota(z),z')\int^{z'}\kern-2mm H\omega^0_3(\cdot,z_2,z_3) \\
+z\leftrightarrow \iota(z).
\end{multline}
We can commute the residues $\underset{z\rightarrow p}{\textrm{Res}}$ and $\underset{z'\rightarrow \iota(z)}{\textrm{Res}}$ using the following formula
\begin{equation}
\sum_{p\in \mathcal{C}}\underset{z\rightarrow p}{\textrm{Res}} \ \underset{z'\rightarrow \iota(z)}{\textrm{Res}}=\sum_{p,p'\in \mathcal{C}} \underset{z'\rightarrow p'}{\textrm{Res}}\ \underset{z\rightarrow p}{\textrm{Res}}-\underset{z\rightarrow p}{\textrm{Res}} \ \underset{z'\rightarrow p'}{\textrm{Res}}
\end{equation}
and using the fact that the integrand $\omega^0_2(\iota(z),z')\int^{z'}\kern-2mm H\omega^0_3(\cdot,\ldots)$ has no poles when $z'\rightarrow p'$ we obtain,
\begin{multline}
P\omega^0_4(z_1,z_2,z_3,z_4)=\omega_4^{0,0}(z_1,z_2,z_3,z_4)+\sum_{p'\in \mathcal{C}}\underset{z'\rightarrow p'}{\textrm{Res}}\Bigl[\int^{z'}\kern-2mm H\omega^0_3(\cdot,z_3,z_4) \sum_{p\in \mathcal{C}}\underset{z\rightarrow p}{\textrm{Res}} \  K(z,z_1)\omega^0_2(\iota(z),z')\omega^0_2(z,z_2) \\
+ \int^{z'}\kern-2mm H\omega^0_3(\cdot,z_2,z_4) \sum_{p\in \mathcal{C}}\underset{z\rightarrow p}{\textrm{Res}} \  K(z,z_1)\omega^0_2(\iota(z),z')\omega^0_2(z,z_3) \\
+\int^{z'}\kern-2mm H\omega^0_3(\cdot,z_2,z_3) \sum_{p\in \mathcal{C}}\underset{z\rightarrow p}{\textrm{Res}} \  K(z,z_1)\omega^0_2(\iota(z),z')\omega^0_2(z,z_4)+z\leftrightarrow \iota(z) \Bigl].
\end{multline}
In expressions such as $ \sum_{p\in \mathcal{C}}\underset{z\rightarrow p}{\textrm{Res}} \  K(z,z_1)[\omega^0_2(\iota(z),z')\omega^0_2(z,z_2)+ z\leftrightarrow \iota(z)]$, we recognize $\omega_3^{0,0}$, then 
\begin{equation}
P\omega^0_4(z_1,z_2,z_3,z_4)=\omega_4^{0,0}(z_1,z_2,z_3,z_4)+\sum_{i\in N=\{1, \dotsc, 4\}\setminus \{1\}}\Bigl\langle \omega^{0,0}_3(z_1,z_i,z')H\omega^0_3(z',z_{N\setminus \{i\}}) \Bigr\rangle_{z'},
\end{equation} 
where we introduced the pairing notation used in \cite{BlobbedTR2} whose definition is, when specialized to our case,
\begin{equation}\label{eq:Pairingdef}
\langle\omega(z_e,\ldots)\varphi(z_e,\ldots) \rangle_{z_e}=\sum_{p\in \mathcal{C}}\underset{z_e\rightarrow p}{\textrm{Res}}  \ \omega(z_e, \ldots)\int^{z_e}\varphi(\cdot,\ldots). 
\end{equation}
One property of normalized forms is that they are invariant (see \cite{BlobbedTR2}, Theorem 2.1 and above) under the action of the polar operator $P$. In particular, if we denote $P_i$ the action of $P$ on the i$^{\textrm{th}}$ variable, then 
\begin{equation}
P_1\ldots P_n\omega^{g,0}_n(z_1,\ldots,z_n)=\omega^{g,0}_n(z_1,\ldots,z_n)
\end{equation}
is true and it tells us that $P_i\omega^{0}_3\in \textrm{Im}(P_1P_2P_3), \ \forall i \in \{1,2,3\}$, this implies that $H_i\omega^0_3\in \textrm{Im}(H_1H_2H_3), \ \forall i \in \{1,2,3\}$ and thus $H_i\omega^0_3=\phi^0_3, \ \forall i\in \{1,2,3\}$. These considerations lead us to write
\begin{equation}\label{eq:Pw04}
P\omega^0_4(z_1,z_2,z_3,z_4)=\omega_4^{0,0}(z_1,z_2,z_3,z_4)+\sum_{i\in N=\{1, \dotsc, 4\}\setminus \{1\}}\Bigl\langle \omega^{0,0}_3(z_1,z_i,z')\phi^0_3(z',z_{N\setminus \{i\}}) \Bigr\rangle_{z'}.
\end{equation}
Using equation \eqref{eq:Pw04}, we can obtain the following sequence of result
\begin{align}\label{eq:decompw04-1}
P_1P_2P_3P_4\omega^0_4(z_1,z_2,z_3,z_4)=\omega_4^{0,0}(z_1,z_2,z_3,z_4), \quad P_1P_iH_jH_k\omega^0_4(z_1,z_2,z_3,z_4)=\Bigl\langle \omega^{0,0}_3(z_1,z_i,z')\phi^0_3(z',z_j,z_k) \Bigr\rangle_{z'}, 
\end{align}
and more generally, using the same types of calculation
\begin{align}\label{eq:decompw04-2}
P_iP_jH_kH_l\omega^0_4(z_1,z_2,z_3,z_4)=\Bigl\langle \omega^{0,0}_3(z_i,z_j,z')\phi^0_3(z',z_k,z_l) \Bigr\rangle_{z'}, \quad H_1H_2H_3H_4 \omega^0_4(z_1,z_2,z_3,z_4)=\phi^0_4(z_1,z_2,z_3,z_4), 
\end{align}
while all other cases vanish.

\medskip

These equalities are special cases of results expressed by both Formula 7 and Corollary 2.5 of \cite{BlobbedTR2}. We do not give a proof here (see \cite{BlobbedTR2}), but we use our above example to compare with the statement of the formula and Corollary in \cite{BlobbedTR2}. We recall that the statement of Formula 7 in \cite{BlobbedTR2} is the following 
\begin{equation}\label{eq:formula7}
H_AP_B \omega^g_n=\sum_{G\in \textrm{Bip}^0_{g,n}(A,B)}\frac{\varpi^0_G(z_1,\ldots,z_n)}{|\textrm{Aut}(G)|},
\end{equation}
where $A\sqcup B=\{1, \dotsc, n\}$, and $\textrm{Bip}^0_{g,n}(A,B)$ is the set of bipartite graphs $G$ that satisfy the following properties (described in \cite{BlobbedTR2}):
\begin{itemize}
\item Vertices $v$ are either of type $\phi$ or $\omega^0$, and carry an integer label $h(v)$ (their genus) such that the valency $d(v)$ satisfies $2h(v)-2+d(v)>0$.
\item Edges connect only $\phi$ to $\omega^0$ vertices.
\item There are $n$ unbounded edges called leaves labelled from $1$ to $n$. The leaves with label $a\in A$ must be incident to $\phi$ vertices while the leaves with label $b\in B$ must be incident to $\omega^0$ vertices.  
\item Each $\omega^0$ vertices must be incident to at least a leaf.
\item $G$ is connected and satisfies $b_1(G)+\sum_{v\in G}h(v)=g$ ($b_1(G)$ being its first Betti number).
\end{itemize}
Each $G$ comes with a weight $\varpi^0(z_1,\ldots,z_n)$ computed as follows. Assign variables $z_1,\ldots,z_n\in \mathcal{C}$ to the leaves and integration variables $z_e$ to edges $e$. To each vertex $v$ whose set of variables attached to its incident edges is $Z(v)$, assign a weight $\omega^{h(v),0}_{d(v)}(Z(v))$ (respectively $\phi^{h(v),0}_{d(v)}(Z(v))$) if it is of type $\omega^0$ (resp. $\phi$). Then multiply all weights, and integrate all edge variables $z_e$ with the pairing defined above in equation \eqref{eq:Pairingdef}.

\begin{remark}\label{rem:Bipcases}
The case with no internal edges is allowed and corresponds to graphs in either $\textrm{Bip}^0_{g,n}(\{1, \dotsc, n\},\emptyset)$ or $\textrm{Bip}^0_{g,n}(\emptyset,\{1, \dotsc, n\})$. In both cases there is only one graphs which is in the first case a $\phi$ vertex with $n$ leaves and $g$ as integer label and in the second case a $\omega^0$ vertex with $n$ leaves and $g$ as integer label. 
\end{remark}

\begin{figure}
 \begin{center}
  \includegraphics[scale=1.2]{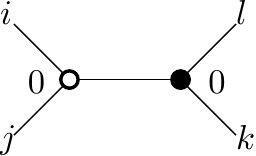}
  \caption{Graph in $\textrm{Bip}^0_{0,4}(\{k,l\},\{i,j\})$. The integers next to the vertices are the genus labels of the vertex. The white vertex represents a $\omega^0$ type vertex,  while the black vertex represents a $\phi$ type vertex.\label{fig:Gijkl}}
 \end{center}
\end{figure}
Let us match this formula \eqref{eq:formula7} with our previous computations by looking at our $\omega^0_4$ example. For $P_1\ldots P_4\omega^0_4$ we found that 
\begin{equation}
P_1\ldots P_4\omega^0_4=\omega^{0,0}_4,
\end{equation} 
which is coherent with our remark \ref{rem:Bipcases} for $n=4$ and $g=0$, \textit{i.e.} there exists only one graph in $\textrm{Bip}^0_{0,4}(\emptyset,\{1, \dotsc, 4\})$ which is the $4$-valent vertex of $\omega^0$ type. For $H_1\ldots H_4\omega^0_4$  remark \ref{rem:Bipcases} still applies and is coherent with our findings in equation \eqref{eq:decompw04-2}. Then comes the cases of $P_iP_jH_kH_l \omega^0_4$. There is one graph $G_{i,j,k,l}$ in each $\textrm{Bip}^0_{0,4}(\{k,l\},\{i,j\})$ pictured one Fig. \ref{fig:Gijkl}. When computing the weight associated to each $G_{i,j,k,l}$ we end up with 
\begin{equation}
\varpi^0_{G_{i,j,k,l}}(z_i,z_j,z_k,z_l)=\Bigl\langle \omega^{0,0}_3(z_i,z_j,z')\phi^0_3(z',z_k,z_l) \Bigr\rangle_{z'},
\end{equation}   
which is what we found in equation \eqref{eq:decompw04-2}. Then we find that any left sets of graphs $\textrm{Bip}^0_{0,4}(A,B)$ is empty. This is coherent with our claim below equation \eqref{eq:decompw04-2} that only $P_1\ldots P_4\omega^0_4$, $H_1\ldots H_4\omega^0_4$ and $P_iP_jH_kH_l \omega^0_4$ for $\{i,j,k,l\}=\{1,2,3,4\}$ are non-vanishing.

We come to Corollary 2.5 of \cite{BlobbedTR2}. We reproduce it here for convenience
\begin{corollary}[Borot-Shadrin,\cite{BlobbedTR2}]
For any $2g-2+n>0$, we have:
\begin{equation}
\omega^g_n(z_1,\ldots,z_n)=\sum_{G\in \textrm{Bip}^0_g,n}\frac{\varpi_G^0(z_1,\ldots,z_n)}{|\textrm{Aut}(G)|}
\end{equation}
where $\textrm{Bip}^0_{g,n}=\bigsqcup_{A\sqcup B =\{1, \dotsc, n\}}\textrm{Bip}^0_{g,n}(A,B)$.
\end{corollary}
This corollary relies on the fact that for any $2g-2+n>0$
\begin{equation}\label{eq:identitydecompwgn}
\omega^g_n(z_1,\ldots,z_n)=\left(\bigotimes_{i=1}^n (P_i+H_i)\right)\omega^g_n(z_1,\ldots,z_n).
\end{equation} 
This equality can be expanded in a sum
\begin{equation}
\left(\bigotimes_{i=1}^n (P_i+H_i)\right)\omega^g_n(z_1,\ldots,z_n)=\sum_{w_n(H,P)}w_n(H,P)\omega_n^g(z_1,\ldots,z_n)
\end{equation}
where $w_n(H,P)$ are words of length $n$ made from an alphabet $\{H,P\}$ whose letters represent the operator $H$ and $P$. The ordered position from left to right of a letter in a word $w_n$ indicates on which variable it acts. Then using the formula \eqref{eq:formula7}, we can expand in graphs any quantity of the form $w_n(H,P)\omega^g_n(z_1,\ldots,z_n)$. Thanks to this corollary, and to the fact that weights $\omega^0$ are recursively computable, we can compute any $\omega^g_n$ recursively as long as we know the expression of the blobs $\phi^g_n$. This is the goal of the next section.

\medskip

We come back to our $\omega^0_4$ example to match our previous computations with the Corollary of Borot-Shadrin. We use \eqref{eq:identitydecompwgn} as well as equations \eqref{eq:decompw04-1}, \eqref{eq:decompw04-2} and the claim below it to obtain 
\begin{equation}\label{eq:examplegraphexpansion}
\begin{aligned}
\omega^0_4(z_1,z_2,z_3,z_3)&= (P_1P_2P_3P_4 + H_1H_2H_3H_4)\omega^{0}_4(z_1,z_2,z_3,z_4)+\sum_{\substack{i,j,k,l\\ \{i,j,k,l\}=\{1,2,3,4\}}}P_iP_jH_kH_l\omega^0_4(z_1,z_2,z_3,z_4)\\
&= \ \raisebox{-4ex}{\includegraphics{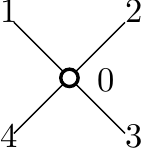}}  \ + \ \raisebox{-4ex}{\includegraphics{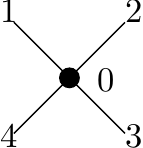}} \ + \ \sum_{\substack{i,j,k,l\\ \{i,j,k,l\}=\{1,2,3,4\}}}\raisebox{-4.5ex}{\includegraphics{PPHHw04ter.pdf}}
\end{aligned}
\end{equation}
where the white vertices represent $\omega^0$ weights while black vertices represent $\phi$ weights. The integers next to the vertices indicate their associated genus $h(v)$. Equation \eqref{eq:examplegraphexpansion} corresponds to what we expect from \cite{BlobbedTR2}.

\subsection{Graphical expansion of the blob}

In section \ref{sec:firstformula}, we give the formula \eqref{eq:Holomorphicpart} for the holomorphic part $H\omega^g_n$ of $\omega^g_n$, $(g,n)\neq (0,1),(0,2)$. This formula is a generalization of the formula described in \cite{BlobbedTR}. Moreover in \ref{sec:graphexpansionw}, the graphical expansion of $\omega^g_n$ is recalled. The weights of the graphs appearing in the expansion \eqref{eq:identitydecompwgn} depend on $\phi^g_n(z_1,\ldots,z_n)=H_1\ldots H_n \omega^g_n(z_1,\ldots,z_n)$. These quantities are representable as sums of weighted graphs. The goal of this section is to describe how to perform this sum in our case. Most complications of this part come from the disconnectedness of the spectral curve. However once the right global quantities are used to express $v^{g}_{n}$ this generalization can be obtained by following the lines of the construction described in \cite{BlobbedTR2} section 7.5. 

We recall the formula  \eqref{eq:Holomorphicpart} expressing the holomorphic part of $\omega^g_n$
\begin{equation}
H_1\omega^g_n(z_1,\ldots,z_n)=\frac{1}{2i\pi}\oint_{\Gamma} \omega^0_2(z,z_1)v^g_n(z;z_2,\ldots,z_n),
\end{equation}
we have 
\begin{equation}
H_1\ldots H_n \omega^g_n(z_1,\ldots,z_n)=\frac{1}{2i\pi}\oint_{\Gamma} \omega^0_2(z,z_1)H_2\ldots H_n v^g_n(z;z_2,\ldots,z_n).
\end{equation} 
We need to find an expression for $H_2\ldots H_n v^g_n(z;z_2,\ldots,z_n)$. $v^g_n$ is a primitive for $l^{'g}_n$. In order to express the graphical expansion of the blobs we need an explicit representation of $v^g_n$ and its function counterpart hereafter denoted $V^g_n$. The local expressions of $v_n^g$ can be obtained recalling the fact that in the expression of $L^g_{\ve_i;\vn}$ coming from \eqref{eq:Lgn}, the factor $a_i$ in front of the coefficient $\tilde{t}_{a_1\ldots a_6}$ comes from a derivation with respect to $x_0$. Getting rid of the terms containing $W^{h_i}_{\vmu_i+\vlam_i}$ with $2h_i-2+|\vmu_i+\vlam_j|=2g-2+|\vn|+1$ in the expression of $L^g_{\ve_i;\vn}$ one can construct a local function $V^g_{\ve_i;\vn}(x_0;x_{\vn})$ whose derivative with respect to the first variable is $L^{'g}_{\ve_i;\vn}$
\begin{multline}
V^g_{\ve_i;\vn}(x_0;x_{\vn})=\sum_{ \substack{\{\vlam_2,\ldots,\vlam_{|\Lambda|}\}\\\sum_j \vlam_j=\sum_{c\neq i}\ve_c}}\sum_{\substack{(\vmu_2,\ldots,\vmu_{|\Lambda|})\\ \sum_{j=2}^{|\Lambda|}\vmu_j=\vn}}\sum^{'}_{\substack{h_0;h_2,\ldots, h_{|\Lambda|}\\ h_0+\sum_{j=2}^{|\Lambda|}h_j=g+|\Lambda|-4}}\oint_{\Gamma^5}\left(\prod_{c\neq i} \frac{d\tilde{x}_c}{2i\pi} \right) \\
T^{h_0}(\tilde{x}_1,\ldots,\tilde{x}_i=x_0,\ldots,\tilde{x}_6)\prod_{j=2}^{|\Lambda|}W_{\vmu_j+\vlam_j}^{h_j}(x_{\vmu_j},\tilde{x}_{\vlam_j}), 
\\ x_0\in \hat{\C}_i, \, x_{\vn} \in (\hat{\C}_1\setminus [a_1,b_1])^{n_1}\times \ldots \times (\hat{\C}_d\setminus [a_d,b_d])^{n_d},
\end{multline}  
where the symbol $\sum^{'}$ here excludes terms containing $W^{h_i}_{\vmu_i+\vlam_i}$ with $2h_i-2+|\vmu_i+\vlam_j|=2g-2+|\vn|+1$. Thanks to the properties of $T^{h_0}$, $V^g_{\ve_i;\vn}$ can be extended as a global function $V^g_{\ve_i;n}$ in its variables $x_{\vn}$ using global $W^g_n$. This writes
\begin{multline}
V^g_{\ve_i;n}(x_0;x_{n})=\sum_{ \substack{K\vdash \{1,\ldots,6\}\setminus \{i\}\\ J_1\sqcup \dotsb \sqcup J_{[K]}=\{1,\ldots,n\}}}\sum^{'}_{\substack{h_0,h_1\ldots h_{[K]}\\ h_0+\sum_{j=1}^{[K]}h_j=g+[K]-4}}\oint_{\Gamma^5}\left(\prod_{c\neq i} \frac{d\tilde{x}_c}{2i\pi} \right) \\
T^{h_0}(\tilde{x}_1,\ldots,\tilde{x}_i=x_0,\ldots,\tilde{x}_6)\prod_{j=1}^{[K]}W_{|J_j|+|K_j|}^{h_j}(x_{J_j},\tilde{x}_{K_j}), 
\end{multline}
for $x_0 \in \hat{\C}_i$ and $x_{k\neq 0} \in \bigcup_{p=1}^d \hat{\C}_p\setminus [a_p,b_p]$. The corresponding global function in $x_0$ can be defined as well
\begin{equation}
V^g_{n+1}(x_0;x_n)=\sum_{i=1}^d \mathbf{1}_{\hat{\C}_{i}}(x_0)\ V_{\ve_i;n}^{g}(x_0; x_{1},\ldots,x_n).
\end{equation}
We further rewrite $V^g_{n+1}(x_0;x_n)$ in a form that is better suited to the graphical expansion. To this aim we define
\begin{equation}
\tilde{T}^p(x_1,\ldots,x_6)=\frac{1}{5!}\sum_{\pi \in \mathfrak{S}_{d=6}}T^p(y_1=x_{\pi(1)},\ldots,y_6=x_{\pi(6)}).
\end{equation}
The summands in the above sum over permutations are non zero if and only if $x_{\pi(i)}\in \hat{\C}_i$. The $1/5!$ comes from the integration of five variables out of six around the disconnected cut $\Gamma=\bigcup_{i=1}^6 \Gamma_i$ in equation below. Then a moment of reflection reveals 
\begin{equation}
V^g_{n+1}(x_0;x_n)=\sum_{\substack{K\vdash \{1, \dotsc, 5\}\\ J_1\sqcup \dotsb \sqcup J_{[K]}=\{1,\ldots,n\} }}\sum^{'}_{\substack{h_0,h_1\ldots h_{[K]}\\ h_0+\sum_{j=1}^{[K]}h_j=g+[K]-4}}\oint_{\Gamma^5}\left(\prod_{k=1}^5 \frac{d\tilde{x}_k}{2i\pi} \right) \\
\tilde{T}^{h_0}(x_0,\tilde{x}_1,\ldots,\tilde{x}_5)\prod_{j=1}^{[K]}W_{|J_j|+|K_j|}^{h_j}(x_{J_j},\tilde{x}_{K_j}).
\end{equation}
Since $v^g_{n+1}$ is the differential form counterpart of $V^g_{n+1}(x_0;x_n)$, we have
\begin{equation}
v^g_{n+1}(z_0;z_1,\ldots,z_n)=V^g_{n+1}(x_0;x_n)\prod_{j=1}^n dx(z_j).
\end{equation}
Notice that $v^g_{n+1}(z_0;z_1,\ldots,z_n)$ is a function and not a differential form in its first variable. Indeed it is a primitive of $l^{'g}_{n+1}$ in its first variable. 

\bigskip

In the following paragraphs we describe the graphical expansion of $\phi^g_n=H_1H_2\ldots H_n\omega^g_n$. We define the family of dressed potential $\mathbb{T}^{h_0}_k(x_1,\ldots,x_k)$ for any $h_0$ and $k\in \{1, \dotsc, 6\}$ 
\begin{align}
&\mathbb{T}^{h_0}_k(x_1,\ldots,x_k) \nonumber \\
&=\sum_{\substack{h\ge 0,r\in \{1, \dotsc, 5\} \\ 2h-2+k>0}}\frac{1}{r!}\sum_{\substack{h_1,\ldots,h_r \\ l_1,\ldots,l_r \\ \sum_jl_j= r \\ h+\sum_j(h_j+l_j-1)=h_0}}\prod_{\substack{1\le j \le r \\ 1\le m \le l_j}}\oint_{\Gamma}\frac{dx^{'}_{j,m}}{2i\pi}\frac{5!}{(k-1)!}\tilde{T}^{h}(x_1,\ldots,x_k,x^{'}_{1,1},\ldots,x^{'}_{r,l_r}) \prod_{1\le j\le r}\frac{W^{h_j}_{l_j}(X^{'}_j)}{l_j !},
\end{align}
where $X^{'}_j=\{x^{'}_{j,m}\}_{1\le m\le l_j}$. This formula uses extensively the fact that $\tilde{T}^h$ is symmetric under permutations of its variables. Then $V^g_{n+1}$ rewrites using $\mathbb{T}^{h_0}_{k+1}$ as
\begin{multline}\label{eq:potentialwithdress}
V^g_{n+1}(x_0;x_n)=\sum_{k\ge 0} \sum_{\substack{K\vdash \{1, \dotsc, k\}\\ J_1\sqcup \dotsb \sqcup J_{[K]}=\{1,\ldots,n\}, J_i\neq \emptyset }} \sum^{'}_{\substack{h_0,h_1\ldots h_{[K]}\\ h_0+\sum_{j=1}^{[K]}h_j+k-[K]=g}} \\ \oint_{\Gamma^k}\left(\prod_{p=1}^k \frac{d\tilde{x}_p}{2i\pi} \right) \mathbb{T}^{h_0}_{k+1}(x_0,\tilde{x}_1,\ldots,\tilde{x}_k)\prod_{j=1}^{[K]}W_{|J_j|+|K_j|}^{h_j}(x_{J_j},\tilde{x}_{K_j}).
\end{multline}
From this last expression one can get the corresponding expression for $v^g_{n+1}(z_0; z_n)$.  
Using \eqref{eq:potentialwithdress} one infers that there is a combinatorial expansion for $n\ge 0$ of $H_0\omega^g_{n+1}$ in term of planted graphs introduced in \cite{BlobbedTR2} with the slight differences that the local weights are integrated on a disconnected cut $\Gamma$ and that the $\omega$ vertices are all incident to a leaf. We define the relevant set of graph below. Mimicking notations adopted in \cite{BlobbedTR2}, we call it $\textrm{Plant}^{\mathbb{T}}_{g,n+1}(0)$. It is a set of bipartite graph $G$ with the following properties:
\begin{itemize}
 \item the set of vertices is made of one root vertex and a set of $\omega$ vertices. Each vertex $v$ comes with a genus index $h(v)$ and a valence $d(v)$. The root vertex must have indices satisfying $2h(v)-2+d(v)>0$, with $d(v)\le 6$. $\omega$ vertices must have indices $(h,d)\neq (0,1)$.
 \item Edges are dashed and can only connect the root vertex to $\omega$ vertices.
 \item There are $n+1$ leaves labeled from $0$ to $n$. They must be incident to $\omega$ vertices and all $\omega$ vertices must be incident to at least a leaf. Moreover, the leaf labeled $0$ is incident to a $\omega^0_2$ vertex, which is incident to the root vertex.
 \item $G$ is connected and $b_1(G)+\sum_v h(v)=g$. 
\end{itemize}
To produce the associated weight $\varpi^{\mathbb{T}}_{G}(z_0,z_1,\ldots,z_n)$ we attach leaf variables $z_0,z_1,\ldots,z_n$ and integration variables $\{z_e\}$ to the edges. To a vertex $v$ with incident variables set $Z(v)$ we associate a weight $\mathbb{T}^{h(v)}_{d(v)}(Z(v))$ if it is the root vertex, while we associate the weight $\omega^{h(v)}_{d(v)}(Z(v))$ otherwise. The weight of $G$ is obtained by multiplying the local weights and integrating $\frac{1}{2i\pi}\oint$ the variables $z_e$ along the disconnected cut $\Gamma$. The leaf variables need to be outside the contour.  

\medskip

In order to deepen this expansion we introduce the purely polar part $\omega^{g,\mathcal{P}}_{n+1}=P_0\ldots P_n \omega^{g}_{n+1}$. It allows to write an expansion of any $H_AP_B\omega^g_{n+1}$ in terms of graphs in $\textrm{Bip}^{\mathcal{P}}_{g,n+1}(A,B)$, \text{i.e.} graphs in $\textrm{Bip}^0_{g,n}(A,B)$ that do not contain internal $\phi$ vertices (that are incident to no leaves). This comes from \cite{BlobbedTR2} proposition 2.3. We have that 
\begin{equation}\label{eq:formulaP}
H_AP_B\omega^g_{n+1}(z_0,z_1,\ldots,z_n)=\sum_{G\in \textrm{Bip}^{\mathcal{P}}_{g,n+1}(A,B)}\frac{\varpi_G^{\mathcal{P}}(z_0,z_1,\ldots,z_n)}{|\textrm{Aut}(G)|}.
\end{equation} 
In this construction the weight $\varpi_G^{\mathcal{P}}(z_0,z_1,\ldots,z_n)$ of a graph $G\in \textrm{Bip}^{\mathcal{P}}_{g,n+1}(A,B)$ is obtained in the same way than in the expansion \eqref{eq:formula7}, but associating a weight $\omega^{h(v),\mathcal{P}}_{d(v)}$ to $\omega$ vertices of genus $h(v)$ and valence $d(v)$ of the graph $G$. Indeed formula \eqref{eq:formulaP} is obtained by performing a resummation of internal $\phi$ vertices (see \cite{BlobbedTR2}).

\medskip

Coming back to our graphical expression for $H_0\omega^g_{n+1}$, in order to obtain the graphical expansion for $\phi^g_{n+1}$ we need to project the variables $z_1,\ldots,z_n$ to the holomorphic part. These variables correspond to leaves of graphs in $\textrm{Plant}^{\mathbb{T}}_{g,n+1}(0)$ and thus it amounts to replace the weight of $\omega$ vertices by weights $H_{A(v)}\omega^{h(v)}_{d(v)}$, where $A(v)$ is its set of leaves. Then we have that
\begin{equation}
H_A\omega^h_d=\sum_{A'\sqcup B =\{1, \dotsc, d\}\setminus A}H_{A\sqcup A'}P_B\omega^h_d,
\end{equation} 
which can be evaluated from formula \eqref{eq:formulaP}. It involves only $\phi$ vertices and $\omega^{\mathcal{P}}$ vertices of lower order. Using the definition of $\omega^{\mathcal P}$ we can write 
\begin{equation}
\omega^{h',\mathcal{P}}_{d'}(z_1,\ldots,z_{d'})=\sum_{p_1,\ldots,p_{d'} \in \mathcal{C}}\underset{z'_1\rightarrow p_1}{\textrm{Res}}\underset{z'_2\rightarrow p_2}{\textrm{Res}}\ldots \underset{z'_{d'}\rightarrow p_{d'}}{\textrm{Res}}\omega^{h'}_{d'}(z_1',\ldots,z'_{d'})\prod_{j=1}^{d'}\int^{z'_j}\omega^{0}_2(\cdot,z_j).
\end{equation}
Still following the lines of \cite{BlobbedTR2}, this results in graphical expansion for the blob.
\begin{equation}
\phi_{n+1}^g(z_0,z_1,\ldots,z_n)=\sum_{G \in \textrm{PreBip}^{\mathbb{T}}_{g,n+1}}\varpi^{\mathbb{T}}_G(z_0,z_1,\ldots,z_n)
\end{equation}
The sets of graphs PreBip has the same properties than in \cite{BlobbedTR2} section 7.5. The only change being the weight that involves disconnected cut $\Gamma$. We recall here the definition of $\textrm{PreBip}^{\mathbb{T}}_{g,n+1}$ here. It is the set of graphs $G$ with the properties
\begin{itemize}
\item vertices $v$ are of type $\omega$ or $\mathbb{T}$, carry a genus $h(v)$ and a valency $d(v)$ such that $2h(v)-2+d(v)>0$ for type $\mathbb{T}$ vertices and $2h(v)-2+d(v)\ge 0$ for type $\omega$ vertices. Moreover $d(v)\le 6$ for $\mathbb{T}$ vertices.
\item Edges are plain or dashed. Dashed edges connect $\omega$ and $\mathbb{T}$ vertices, while plain edges can only connect $\omega^0_2$-vertices to $\omega^h_k$-vertices with $2h-2+k>0$.
\item No plain edge is a bridge
\item There are $n+1$ labeled leaves: they must be incident to a $\omega^0_2$-vertex which is itself incident to a $\mathbb{T}$ vertex.
\item $G$ is connected and $b_1(G)+\sum_vh(v)=g$. 
\end{itemize} 
The weight $\varpi^{\mathbb{T}}_G$ is computed by multiplying all local weights and integrating edge variables $z_e$ with $\frac{1}{2i\pi}\oint_{\Gamma}$ for dashed edges while for plain edges we integrate with the pairing $\sum_{p\in \mathcal{C}}\underset{z_e\rightarrow p}{\textrm{Res}}\left(\int^{z_e}\omega^0_2(z^{'},\ldots) \right)\omega^h_k(z_e,\ldots)$. This last integration is actually the previously introduced pairing \eqref{eq:Pairingdef}.  

\medskip

This combinatorial expansion can be simplified further. This is done by getting rid of the $\omega^0_2$ vertices. Without repeating the arguments of \cite{BlobbedTR2} we define
\begin{equation}
\tau^g_{k}(z_1,\ldots,z_k)=\sum_{\mathcal{G}\in \textrm{Graph}_{g,k}}\Biggl\langle\prod_{j=0}^k\omega^0_2(z_j,z'_j) \prod_{v=\textrm{vertex}}\mathbb{T}^{h(v)}_{d(v)}(Z^{'}(v))\prod_{\substack{e=\{v,w\}\\\textrm{edge}}}\omega^{0}_2(z'_v,z'_w) \Biggr\rangle_{\Gamma},
\end{equation}
where here the bracket $\langle \rangle_{\Gamma}$ means that internal variables $z'$ must be integrated $\frac{1}{2i\pi}\oint_{\Gamma}$ around $\Gamma$. 
$\textrm{Graph}_{g,k}$ contains graphs such that
\begin{itemize}
 \item $\mathcal{G}$ is made of $\mathbb{T}$ vertices with a genus $h(v)$ and a valency $d(v)\le 6$ satisfying $2h(v)-2+d(v)>0$ and of $k$ labeled univalent vertices whose genus is zero by convention.
 \item $\sum_v h(v)=g$.
 \item We stress these graphs can be disconnected.
\end{itemize}
The next representation is then obtained by resumming the $\mathbb{T}$ vertices of PreBip that are surrounded by $\omega^0_2$-vertices in $\tau$ vertices. This corresponds to a slight generalization of \cite{BlobbedTR2} proposition 7.2. For $2g-2+n>0$
\begin{equation}
\phi^g_{n+1}(z_0,z_1,\ldots,z_n)=\sum_{G\in \textrm{Bip}^{\tau}_{g,n+1}}\varpi_G^{\tau}(z_0,z_1,\ldots,z_n),
\end{equation}
where $\textrm{Bip}^{\tau}_{g,n+1}$ is the set of bipartite graphs $G$ such that:
\begin{itemize}
 \item vertices $v$ are of type $\omega$ or $\tau$, and carry a genus $h(v)$, a valency $d(v)$ satisfying $2h(v)-2+d(v)>0$.
 \item edges connect $\omega$ to $\tau$ vertices.
 \item There are $n+1$ leaves, labeled $\{0,\ldots,n\}$, incident to $\tau$ vertices.
 \item $G$ is connected and $b_1(G)+\sum_{v}h(v)=g$. 
\end{itemize}
The weight $\varpi_G^{\tau}(z_0,z_1,\ldots,z_n)$ is obtained by attaching edge variables $z_e$ to all internal edge of $G$. We assign a weight $\tau_{d(v)}^{h(v)}(Z(v))$ to a $\tau$ vertex $v$ with set of incident edges $Z(v)$. We assign a weight $\omega_{d(v)}^{h(v)}(Z(v))$ to a $\omega$ vertex $v$ with set of incident edges $Z(v)$. Then we multiply all local weights and integrate the edge variables with the pairing \eqref{eq:Pairingdef}.

\begin{remark}
We end this section with a small remark. It is easy to show that in dimensions $d=4\delta+2$, the $\phi^h_n$ vanish for $h\le \delta-1$ in the quartic melonic tensor model. In particular, for $d=6$ (the case we chose to investigate in details), all $\phi^0_n$ are zero. This comes from the peculiar $1/N$ expansion of the potential of the model, which is Gaussian at large $N$, and whose first non-trivial correction in $N$ appears at order $h=\delta$. 
\end{remark}

\section{Exact expression for tensor models observables} \label{sec:TensorObs}

\subsection{Formal series of an observable}

In this subsection we propose an exact expression of the formal series associated to the mean value of any observable. This expression depends on the Gaussian Wick contractions of this observables and need as an input the $W$ computed using the recursive process presented in the preceding sections of this paper. 

\medskip

Pick an observable $\cB(T,\overline{T})$ of the tensor model. Its mean value writes 
\begin{equation}
\langle \cB(T,\overline{T})\rangle=\frac{1}{Z_{\textrm{tensor}}}\int dTd\overline{T} \ \cB(T,\overline{T})\exp -N^{d-1}\left(T\cdot \overline{T}+\frac12\sum_{c=1}^dg_c^2B_c(T,\overline{T})\right),
\end{equation}
and using equation \eqref{eq:HubbardSplit}, it rewrites
\begin{equation}
\langle \cB(T,\overline{T})\rangle=\frac{1}{Z_{\textrm{tensor}}}\int dTd\overline{T}\int \prod_{c=1}^d dX_c \ \cB(T,\overline{T})e^{-N^{d-1}T(\mathbbm{1}^{\otimes d}+i\sum_{c=1}^dg_c Y_c)\overline{T}}e^{-\frac{1}{2}\sum_{c=1}^d\tr Y_c^2}.
\end{equation}
This integral is a Gaussian mean value for the variable $T,\overline{T}$ with covariance $\mathbbm{1}^{\otimes d}+i\sum_{c=1}^dg_c Y_c$. The integral over tensor variables can thus be performed by summing over Wick pairing of the $T,\overline{T}$ variables knowing that $\langle T_{i_1\ldots i_d} \overline{T}_{i_1\ldots i_d}\rangle = (\mathbbm{1}^{\otimes d}+i\sum_{c=1}^dg_c Y_c)^{-1}_{i_1\ldots i_d, j_1\ldots j_d}$. We are left with an integral over the $X_c$ of the form
\begin{equation}
\langle \cB(T,\overline{T})\rangle=\frac{1}{Z_{\textrm{tensor}}}\int \prod_{c=1}^d dX_c \ \det{}^{-1}\left(\mathbbm{1}^{\otimes d}+i\sum_{c=1}^dg_c Y_c\right) \left(\sum_{\textrm{pairings } w} w(\cB(T,\overline{T}))\right) \ e^{-\frac{1}{2}\sum_{c=1}^d\tr Y_c^2}.
\end{equation}
We stress that the weight $w(\cB(T,\overline{T}))$ depends on the $Y_c$ as indeed the covariance of the tensor variables depends on them. However, the only pairings that appear are the Gaussian ones. Thus there is a finite number of terms in the sum, $p!$ for an observable of degree $p$. As is well known from tensor models literature \cite{SDLargeN}, $d$-dimensional tensor observables are represented by bipartite colored graphs with $d$ colors whose white (resp. black) vertices represents $T$ (resp. $\overline{T}$) variables and whose edges of colors $1$ to $d$ represent contraction of indices in position $1$ to $d$ of $T$ variables with $\overline{T}$ variables. Each Wick pairing $w$ can be represented by a graph with $d+1$ colors, ranging from $0$ to $d$. The additional $0$ colors represent the Wick pairing $w$ in such a way that by withdrawing the edges of color $0$ we end up with the graph representing the observable $\cB(T,\overline{T})$. The weight associated with a pairing $w$ is to be described below.

Let us first notice that an observable $\cB(T,\overline{T})$ with $2p$ vertices can be encoded by a $d$-uplet of permutations $(\sigma_1,\ldots, \sigma_d)$,
\begin{equation}
\cB(T,\overline{T})=\sum_{\substack{i_1^1,\ldots,i^1_d \\ \ldots \\ i^p_1,\ldots, i_p^d =1}}^N\sum_{\substack{j_1^1,\ldots,j^1_d \\ \ldots \\ j^p_1,\ldots, j_p^d =1}}^N \prod_{a,b=1}^p T_{i^a_1\ldots i^a_d}\overline{T}_{j^b_1\ldots j^b_d}\prod_{k=1}^d\delta_{i^a_kj^{\sigma_k(a)}_k}.
\end{equation}
In order to give the above explicit expression we have chosen an arbitrary labeling of the white vertices (resp. black vertices) of the colored graph representing $\mathcal{B}(T,\overline{T})$ by integers $a\in \{1, \dotsc, p\}$ (resp. $b\in \{1, \dotsc, p\}$). 
A Wick pairing $w$ can be seen here as an involution $s_w$ sending each vertex white vertex $a$ representing a $T$ to a black vertex $b=s_w(a)$. The weight associated to this pairing is 
\begin{equation}
w(\cB(T,\overline{T}))=\frac{1}{N^{p(d-1)}} \sum_{\substack{i_1^1,\ldots,i^1_d \\ \ldots \\ i^p_1,\ldots, i_p^d =1}}^N\sum_{\substack{j_1^1,\ldots,j^1_d \\ \ldots \\ j^p_1,\ldots, j_p^d =1}}^N \prod_{a=1}^p\left[ \bigl(\mathbbm{1}^{\otimes d}+i\sum_{c=1}^dg_c Y_c\bigr)^{-1}_{i_1^a\ldots i_d^a, j_1^{s_w(a)}\ldots j_d^{s_w(a)}}\prod_{k=1}^d\delta_{i^a_kj^{\sigma_k(a)}_k}\right].
\end{equation}
We need to evaluate quantities such as
\begin{multline}
\int \frac{ \prod_{c=1}^d dX_c }{Z_{\textrm{tensor}}}\ w(\cB(T,\overline{T})) e^{-\frac{1}{2}\sum_{c=1}^d\tr Y_c^2 - \tr \log\left(\mathbbm{1}^{\otimes d}+i\sum_{c=1}^dg_c Y_c\right)}  = 
\int \frac{\prod_{c=1}^d dX_c}{Z_{\textrm{tensor}}} \ 
N^{-p(d-1)} \sum_{\substack{i_1^1,\ldots,i^1_d \\ \ldots \\ i^p_1,\ldots, i_p^d =1}}^N\sum_{\substack{j_1^1,\ldots,j^1_d \\ \ldots \\ j^p_1,\ldots, j_p^d =1}}^N \\
\prod_{a=1}^p\left[ \bigl(\mathbbm{1}^{\otimes d}+i\sum_{c=1}^dg_c Y_c\bigr)^{-1}_{i_1^a\ldots i_d^a, j_1^{s_w(a)}\ldots j_d^{s_w(a)}}\prod_{k=1}^d\delta_{i^a_kj^{\sigma_k(a)}_k}\right] e^{-\frac{1}{2}\sum_{c=1}^d\tr Y_c^2 - \tr \log\left(\mathbbm{1}^{\otimes d}+i\sum_{c=1}^dg_c Y_c\right)}. 
\end{multline}
We now perform the shift on the $X_c$ variables. The above integral rewrites
\begin{multline}\label{eq:complicatedintegral}
 \frac{1}{Z(\{\alpha_c\},N)}\int \prod_{c=1}^d dM_c \ [N^{d-1}(1+iG)]^{-p}
 \sum_{\substack{i_1^1,\ldots,i^1_d \\ \ldots \\ i^p_1,\ldots, i_p^d =1}}^N\sum_{\substack{j_1^1,\ldots,j^1_d \\ \ldots \\ j^p_1,\ldots, j_p^d =1}}^N  \\ \prod_{a=1}^p\left[ \bigl(\mathbbm{1}^{\otimes d}-\sum_{c=1}^d\alpha_c N^{(2-d)/2}\mathcal{M}_c\bigr)^{-1}_{i_1^a\ldots i_d^a, j_1^{s_w(a)}\ldots j_d^{s_w(a)}}\prod_{k=1}^d\delta_{i^a_kj^{\sigma_k(a)}_k}\right]  e^{-\frac{N}{2}\sum_{c=1}^d\tr M_c^2 +\tr\sum_{p\ge 2}\frac{N^{\frac{2-d}{2}p}}{p}\left(\sum_c\alpha_c\mathcal{M}_c\right)^p}.
\end{multline}
Then we can expand the terms $\bigl(\mathbbm{1}^{\otimes d}-\sum_{c=1}^d\alpha_c N^{(2-d)/2}\mathcal{M}_c\bigr)^{-1}$ in the product. We have 
\begin{multline}
 \sum_{\substack{i_1^1,\ldots,i^1_d \\ \ldots \\ i^p_1,\ldots, i_p^d =1}}^N\sum_{\substack{j_1^1,\ldots,j^1_d \\ \ldots \\ j^p_1,\ldots, j_p^d =1}}^N \prod_{a=1}^p\left[ \bigl(\mathbbm{1}^{\otimes d}-\sum_{c=1}^d\alpha_c N^{(2-d)/2}\mathcal{M}_c\bigr)^{-1}_{i_1^a\ldots i_d^a, j_1^{s_w(a)}\ldots j_d^{s_w(a)}}\prod_{k=1}^d\delta_{i^a_kj^{\sigma_k(a)}_k}\right] = \\
\sum_{\substack{i_1^1,\ldots,i^1_d \\ \ldots \\ i^p_1,\ldots, i_p^d =1}}^N\sum_{\substack{j_1^1,\ldots,j^1_d \\ \ldots \\ j^p_1,\ldots, j_p^d =1}}^N\sum_{q_1,\ldots,q_p\ge 0 }\sum_{\substack{n^{(1)}_1,\ldots , n^{(1)}_d \ge 0\\ \ldots \\ n^{(p)}_1,\ldots, n^{(p)}_d\ge 0 \\\sum_i n^{(m)}_i =q_m, \ 1\le m \le p}} \prod_{a=1}^p N^{\frac{2-d}{2}q_a}\left(\prod_{c=1}^d\alpha_c^{n_c^{(a)}}\mathcal{M}_c^{n_c^{(a)}} \right)_{i_1^a\ldots i^a_d, j_1^{s_w(a)}\ldots j_d^{s_w(a)}}\prod_{k=1}^d\delta_{i^a_kj^{\sigma_k(a)}_k}.
\end{multline}  
If we consider the cycles of each permutation $s_w\circ \sigma_k^{-1}$, then when summing over $i_k^1,i_k^2,\ldots,i_k^p$, each given cycle closes a trace of a matrix $M_k$ at some power which writes as a sum  over $a$ of all $n^{(a)}_k$ such that $a$ belongs to the given cycle. We denote the cycles of $s_w\circ \sigma_k^{-1}$ by $C_{i,w}^{(k)}$, where $i$ runs from $1$ to the number of cycles $C_w^{(k)}$ of $s_w\circ \sigma_k^{-1}$. It follows that the above equation rewrites
\begin{multline}
\sum_{\substack{i_1^1,\ldots,i^1_d \\ \ldots \\ i^p_1,\ldots, i_p^d =1}}^N\sum_{\substack{j_1^1,\ldots,j^1_d \\ \ldots \\ j^p_1,\ldots, j_p^d =1}}^N\sum_{q_1,\ldots,q_p\ge 0 }\sum_{\substack{n^{(1)}_1,\ldots , n^{(1)}_d \ge 0\\ \ldots \\ n^{(p)}_1,\ldots, n^{(p)}_d\ge 0 \\\sum_i n^{(m)}_i =q_m, \ 1\le m \le p}} \prod_{a=1}^p N^{\frac{2-d}{2}q_a}\left(\prod_{c=1}^d\alpha_c^{n_c^{(a)}}\mathcal{M}_c^{n_c^{(a)}} \right)_{i_1^a\ldots i^a_d, j_1^{s_w(a)}\ldots j_d^{s_w(a)}}\prod_{k=1}^d\delta_{i^a_kj^{\sigma_k(a)}_k} = \\
\sum_{q_1,\ldots,q_p\ge 0 }\sum_{\substack{n^{(1)}_1,\ldots , n^{(1)}_d \ge 0\\ \ldots \\ n^{(p)}_1,\ldots, n^{(p)}_d\ge 0 \\\sum_i n^{(m)}_i =q_m, \ 1\le m \le p}}N^{\frac{2-d}{2}\sum_{a=1}^p q_a}\prod_{c=1}^d\alpha_c^{\sum_{a=1}^pn_c^{(a)}}\left[\prod_{i=1}^{C_w^{(c)}}\tr\left(M_c^{\sum_{a\in C_{i,w}^{(c)}}n_c^{(a)}}\right) \right].
\end{multline}
Then \eqref{eq:complicatedintegral} re-expresses as
\begin{multline}\label{eq:complicatedintegral2}
\frac{1}{Z_{\textrm{tensor}}}\int \prod_{c=1}^d dX_c \ w(\cB(T,\overline{T})) e^{-\frac{1}{2}\sum_{c=1}^d\tr Y_c^2 - \tr \log\left(\mathbbm{1}^{\otimes d}+i\sum_{c=1}^dg_c Y_c\right)}  = 
[N^{d-1}(1+iG)]^{-p}\sum_{q_1,\ldots,q_p\ge 0 } \\
 \sum_{\substack{n^{(1)}_1,\ldots , n^{(1)}_d \ge 0\\ \ldots \\ n^{(p)}_1,\ldots, n^{(p)}_d\ge 0 \\\sum_i n^{(m)}_i =q_m, \ 1\le m \le p}} N^{\frac{2-d}{2}\sum_{a=1}^p q_a}\left( \prod_{c=1}^d\alpha_c^{\sum_{a=1}^pn_c^{(a)}}\right) \left\langle \prod_{c=1}^d\left[\prod_{i=1}^{C_w^{(c)}}\tr\left(M_c^{\sum_{a\in C_{i,w}^{(c)}}n_c^{(a)}}\right) \right] \right\rangle.
\end{multline}
We now aim at expressing the quantity of the right hand side of \eqref{eq:complicatedintegral2} as a contour integral of some $W$. First notice a simple fact. Given a Wick pairing $w$ on a invariant $\mathcal{B}(T,\overline{T})$, we notice that the cycles of $s_w\circ\sigma_k^{(-1)}$ are nothing but the cycles made of edges of colors $0$ and $k$ alternatively in the corresponding graph. The set of cycles is nothing but the set of faces of colors $(0k)$ in the graph where $k$ is allowed to run from $1$ to $d$. For each face $f_k$ of color $(0k)$ we introduce a complex variable $x_{f_k}^{(k)} \in \hat{\C}_k$. Then each mean value of the right hand side of \eqref{eq:complicatedintegral2} can be expressed as a contour integral as follows
\begin{multline}\label{eq:expressioncontour}
 \left\langle \prod_{c=1}^d\left[\prod_{i=1}^{C_w^{(c)}}\tr\left(M_c^{\sum_{a\in C_{i,w}^{(c)}}n_c^{(a)}}\right) \right] \right\rangle = \\
  \prod_{c=1}^d\left(\oint_{\Gamma_c^{|F_c|}}\left(\prod_{f_c\in F_c }\frac{dx^{(c)}_{f_c}}{2i\pi }\right)\right)\left(\prod_{f_c\in F_c} (x_{f_c}^{(c)})^{\sum_{a \in f_c}n_c^{(a)}} \right)\overline{W}_{\sum_c |F_c|}(x_{\bigcup_c F_c}).
\end{multline}
Let us explain a bit the notations of \eqref{eq:expressioncontour}. First we denote $F_c$ the set of faces of color $(0c)$ of a given Wick pairing $w$ of our invariant. $|F_c|$ is its cardinal. Although these sets depend on the specific Wick pairing $w$ at play we do not make explicit this dependence in the notations. We recall that $\Gamma_c$ is the cut in $\hat{\C}_c$, while $\overline{W}_{\sum_c |F_c|}$ is the global counterpart of $\overline{W}_{\vk}$ introduced in equation \eqref{eq:coloredbarW}, in section \ref{sec:observablesnotations}, and $\sum_c |F_c|$ is the total number of faces of color $(0c)$, $c$ running from $1$ to $d$. In the exponent of $x_{f_c}^{(c)}$, the labels $a$ index edges of color $c$ that belong to a face $f_c$ and such that one of their ends is a white vertex labeled $a$. We now claim that,
\begin{multline}\label{eq:Wickamplitude}
\frac{[N^{d-1}(1+iG)]^{p}}{Z_{\textrm{tensor}}}\int \prod_{c=1}^d dX_c \ w(\cB(T,\overline{T})) e^{-\frac{1}{2}\sum_{c=1}^d\tr Y_c^2 - \tr \log\left(\mathbbm{1}^{\otimes d}+i\sum_{c=1}^dg_c Y_c\right)} \\
 = \prod_{c=1}^d \left[\oint_{\Gamma_c^{|F_c|}}\prod_{f\in F_c}\Biggl(\frac{dx^{(c)}_f}{2i\pi}\prod_{\substack{e\in E_0 \\ e\in f}}\oint_{\Gamma_c}\frac{dx_e^{(c)}}{2i\pi}\frac{1}{x_f^{(c)}-x_e^{(c)}}\Biggr)\right]\left[ \prod_{e\in E_0}\frac{1}{1+N^{(2-d)/2}\sum_{c=1}^d\alpha_cx_e^{(c)}}\right]\overline{W}_{\sum_c |F_c|}(x_{\bigcup_c F_c})\\
 = \sum_{J\vdash \bigcup_c F_c }\prod_{c=1}^d \left[\oint_{\Gamma_c^{|F_c|}}\prod_{f\in F_c}\Biggl(\frac{dx^{(c)}_f}{2i\pi}\prod_{\substack{e\in E_0 \\ e\in f}}\oint_{\Gamma_c}\frac{dx_e^{(c)}}{2i\pi}\frac{1}{x_f^{(c)}-x_e^{(c)}}\Biggr)\right]\left[ \prod_{e\in E_0}\frac{1}{1+N^{(2-d)/2}\sum_{c=1}^d\alpha_cx_e^{(c)}}\right]\prod_{J_i}W_{J_i}(x_{J_i}).
\end{multline}
We call here $E_0$ the set of edges of color $0$. In this equation, we attach $d$ complex variables $x_e^{(c)} \in U_{\Gamma_c}$ to each edge $e\in E_0$ of color $0$. Each $0$-colored edge $e$ comes with a weight $\frac{1}{[N^{d-1}(1+iG)]}\frac{1}{1+N^{(2-d)/2}\sum_{c=1}^d\alpha_cx_e^{(c)}}$. Moreover each $x_e^{(c)}$, attached to an edge $e$, needs to be identified with the $x_f^{(c)}$ attached to the face $e$ belongs to. This is the reason for the factor of the form $\oint_{\Gamma_c}\frac{dx_e^{(c)}}{2i\pi}\frac{1}{x_f^{(c)}-x_e^{(c)}}$.  \\
We also need to specify that $\forall e \in f$, $f\in F_c$, we have to make a choice of contours around the cuts $\Gamma_c$ such that $|x_e^{(c)}|<|x_f^{(c)}|$. This formula can be obtained by combining  \eqref{eq:complicatedintegral2} and \eqref{eq:expressioncontour}. This is a rather straightforward computation although notations can be heavy.   
\begin{remark}
The mean value associated to an observable is a formal series in $N, N^{-1}$, thus the "propagator" term $\frac{1}{[N^{d-1}(1+iG)]}\frac{1}{1+N^{(2-d)/2}\sum_{c=1}^d\alpha_cx_e^{(c)}}$ associated to any edge $e\in E_0$ has to be understood as a formal series in $N, N^{-1}$ as well.
\end{remark}
One can deduce some sort of Feynman rules to apply to get the exact expression associated to an observable.
\begin{itemize}
\item Given an observable $\mathcal{B}$ associated to its representing $d$-colored graph that we also call $\mathcal{B}$, perform all the Wick contraction on it by adding edges of color $0$ in all possible ways on $\mathcal B$. For each given Wick contraction construct the following weight:
\item To each edge $e$ of color $0$ associate $\{x_e^{(c)}\}_{c=1}^d$ a propagator $\frac{1}{[N^{d-1}(1+iG)]}\frac{1}{1+N^{(2-d)/2}\sum_{c=1}^d\alpha_cx_e^{(c)}}$.
\item To each face $f\in F_c$ of color $(0c)$ associate a variable $x^{(c)}_f$.
\item To each edge $e$ that belongs to a face $f$ of color $(0c)$ associate a weight $\frac{1}{x_f^{(c)}-x_e^{(c)}}$.
\item Multiply these local weights with $\overline{W}_{\sum_c|F_c|}(x_{\bigcup_c F_c})$. The choice of $\overline W$ depends on the number $\sum_c |F_c|$ of faces of color $(0c)$ of the given Wick contraction. 
\item Integrate variables $x_f^{(c)}, x_e^{(c)}$ around a cut $\Gamma_c$ respecting the prescription that $|x_e^{(c)}|<|x_f^{(c)}|$, $\forall e\in f$. 
\item Sum the obtained weights for all Wick contractions. 
\end{itemize}

In order to illustrate our result, in the next paragraph we construct the formula for a precise example. We restrict to the case $d=6$ since it is the case we have shown that the topological recursion\footnote{However our demonstration works almost without change in dimension $d=4\delta+2$.} enables us to compute the corresponding $W$. However formula \eqref{eq:Wickamplitude} applies whenever there exists $W$ functions which is the case in any dimension $d\ge 2$.

\begin{figure}
\begin{center}
 \includegraphics[scale=1.2]{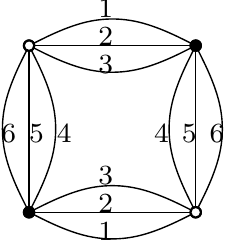}
 \caption{A six dimensional necklace.\label{fig:necklace}}
\end{center}
\end{figure}

We consider the observable described by the graph of Figure \ref{fig:necklace}.
We have, 
\begin{multline}
\Biggl\langle\raisebox{-5.0ex}{\includegraphics[scale=0.8]{necklace6d.pdf}}\Biggr\rangle_{\!\overline{T},T}= \Biggl\langle\raisebox{-5.0ex}{\includegraphics[scale=0.8]{necklace6d.pdf}}\Biggr\rangle_{\!\overline{T},T,\{M_c\}} = \Biggl\langle\raisebox{-10.0ex}{\includegraphics[scale=0.8]{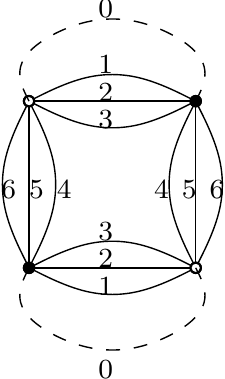}}\Biggr\rangle_{\!\{M_c\}}+\Biggl\langle\raisebox{-5.2ex}{\includegraphics[scale=0.8]{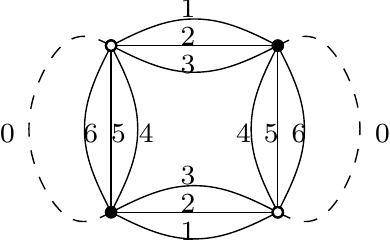}}\Biggr\rangle_{\!\{M_c\}} \\
=\Biggl(\prod_{c=1}^6\oint_{\Gamma_c}\frac{dx^{(c)}_1}{2i\pi}\Biggr) \Biggl(\prod_{c=1}^3\oint_{\Gamma_c} \frac{dx^{(c)}_2}{2i\pi} \Biggr)\left[\frac{1}{N^5(1+iG)}\right]^2\frac{\overline{W}_9(x_1^{(1)},x_1^{(2)},x_1^{(3)},x_1^{(4)},x_1^{(5)},x_1^{(6)},x_2^{(1)},x_2^{(2)},x_2^{(3)}) }{[1+N^{-2}\sum_{c=1}^6\alpha_c x_1^{(c)}][1+N^{-2}(\sum_{c=1}^3\alpha_c x^{(c)}_2+\sum_{c=4}^6\alpha_c x^{(c)}_1)]} \\
+\Biggl(\prod_{c=1}^6\oint_{\Gamma_c}\frac{dx^{(c)}_1}{2i\pi}\Biggr)\Biggl(\prod_{c=4}^6\oint_{\Gamma_c} \frac{dx^{(c)}_2}{2i\pi} \Biggr)\left[\frac{1}{N^5(1+iG)}\right]^2\frac{\overline{W}_9(x_1^{(1)},x_1^{(2)},x_1^{(3)},x_1^{(4)},x_1^{(5)},x_1^{(6)},x_2^{(4)},x_2^{(5)},x_2^{(6)})}{[1+N^{-2}\sum_{c=1}^6\alpha_c x^{(c)}_1][1+N^{-2}(\sum_{c=1}^3\alpha_cx^{(c)}_1+\sum_{c=4}^6\alpha_c x^{(c)}_2]}.
\end{multline}
In this formula we already have integrated out the edge variables, the variables left to integrate are the face variables. The second line comes from the first Wick pairing of the first line while the third line comes from the second Wick pairing in the first line. One can easily obtain from this formula that at leading order the observable scales like $N^{-1}$. Indeed at large $N$ the $\overline{W}_k$ scales like $N^k$. Then $\overline{W}_9\sim N^9$. There are two $N^{-5}$ coming from the edges of color $0$. Consequently the large $N$ scaling of these integrals is $N^{9-10}=N^{-1}$. The coefficient of $N^{-1}$ can also be computed as the propagator terms at large $N$ reduce to $1/(1+iG)^2=-G^2/(\sum_cg_c^2)^2$ and the resulting integral on $\overline{W}_9$ leads to $N^9$. This results is the expected result coming from the literature \cite{New1/N, SigmaReview}.

\section*{Conclusion}

Thanks to a Hubbard-Stratonovich transformation, quartic tensor models can be turned into matrix models \cite{QuarticModels}. In the case of quartic melonic interactions, the matrix eigenvalues all fall in the potential well \cite{IntermediateT4}, as expected from Gurau's universality theorem \cite{Universality} and the direct analysis of the tensor model's Schwinger-Dyson equations \cite{SDLargeN}. This suggests to study the fluctuations around the potential well, giving rise to another matrix model whose large $N$ limit is ($d$ copies) of the GUE \cite{IntermediateT4}.

In the present article, this matrix model has been analyzed at all orders of the $1/N$ expansion. It is shown that its correlation functions satisfy the blobbed topological recursion of \cite{BlobbedTR, BlobbedTR2}, where the key point was to set as the spectral curve the union of the $d$ GUE spectral curves. The loop equations then take exactly the same form as in \cite{BlobbedTR}. The correlation functions are split into a polar part, determined by the usual formula for the topological recursion, and a holomorphic part. For completeness, we further gave the graphical expansion of \cite{BlobbedTR2} which in particular is important to evaluate the holomorphic part. While possible in principle, this is still an intricate procedure and using this blobbed topological recursion for explicit computations is challenging.

Our results constitute a proof of principle that tensor models can be solved at all orders in the $1/N$ expansion and feature similar structures as matrix models, in particular the topological recursion. This opens numerous questions. A natural question is that of the integrability of our model. While the ordinary topological recursion is a modern way to study the integrability in matrix models, it is unclear that its extension to the blobbed recursion still features integrability. This is a question for the blobbed recursion as a whole rather than our specific situation.

We have studied the case $d=6$ which is part of the sequence $d = 4\delta +2$ for which the matrix model has a topological expansion in the sense of an expansion in powers of $1/N^2$. For other values of $d$, the non-topological corrections come from $1/\sqrt{N}$ expansions of the coupling constants. One solution is thus to perform the blobbed topological recursion just like we did and eventually expand the coupling constants in the correlation functions. It may be interesting to look for another approach, like finding a recursion for the non-topological corrections to the correlation functions.

Finally, an exciting direction is to transfer our results directly into the framework of tensor models. Indeed, the blobbed topological recursion we found applies to a matrix model which is obtained from the quartic melonic tensor model after several transformations. It thus gives a solution to the loop equations of the so-obtained matrix model. The tensor model however has its own Schwinger-Dyson equations, which as explained in the Introduction are quite challenging.

We now possess two instances of tensor models for which a topological recursion applies. One is the case of matrix-like interactions in tensor model, where one defines multi-indices, say $(a_1 a_2) = A, (a_3 a_4) = B$, sets $M_{AB} = T_{a_1 a_2 a_3 a_4}$ and considers as tensor model any matrix model on $M$. The second instance is the result from the present article. Both instances thus make use of matrix models. It is important for the development of tensor models to solve their Schwinger-Dyson equations directly. A first approach is to try and transfer or translate the expressions found here for the matrix correlators to expressions for the tensorial observables.

Solving the Schwinger-Dyson equations of tensor models directly also would help to compare our results with existing results on the full $1/N$ expansion. There is indeed a classification due to Gurau and Schaeffer \cite{GurauSchaeffer} of all the Feynman graphs of tensor models with respect to Gurau's degree. This classification expresses the free energy and the 2-point function at each order of the $1/N$ expansion as a finite sum over objects called schemes and it gives the corresponding singularities. The quartic melonic tensor model which was our starting point generates a subset of the graphs studied in \cite{GurauSchaeffer} with the same exponents of $N$. In fact, an approach similar to \cite{GurauSchaeffer} was used in \cite{DoubleScalingDartois} suggesting a similar classification. Therefore, a relation between our work and \cite{GurauSchaeffer} should exist and it would be particularly interesting to understand it.

\section*{Acknowledgements}

Both authors would like to thank Ga\"etan Borot and Bertrand Eynard for numerous discussions on the topic of this paper and the topological recursion in general. Their help was of the utmost importance for the realization of this project.

Stephane Dartois would like to thank several institutions that provided support and offices during parts of this work. Among them, the Laboratoire de Physique Th\'eorique d'Orsay, the Laboratoire d'Informatique de Paris-Nord, the Albert Einstein Institute of Potsdam-Golm and the Max Planck Institute for Mathematics in Bonn. He also acknowledges the COST Action MP1405 for Quantum Structure of Space-time for providing financial support during part of this work.

This research was supported by the ANR MetACOnc project ANR-15-CE40-0014.


\end{document}